\documentclass{aa}  

\makeatletter
\let\linenumbers\relax
\let\nolinenumbers\relax
\makeatother
\usepackage{url}
\usepackage{graphicx}
\usepackage{txfonts}
\usepackage{subcaption}         
                                
\usepackage{hyperref}

\usepackage{xcolor}

\begin{document}

\title{Unraveling the mysteries of Jets in peculiar NLSy1 galaxies through multi-wavelength variability}

\author{Vineet Ojha\inst{1}\fnmsep\thanks{Corresponding author: vineetojhabhu@gmail.com, wuxb@pku.edu.cn}
        \and Xue-Bing, Wu\inst{2, 1}
        \and Luis C. Ho \inst{1, 2}
        \and Raj Prince \inst{3}
        \and Joysankar Majumdar \inst{4}
        \and Hum Chand \inst{5}
        \and Chi-Zhuo Wang \inst{2, 1}
        }

\institute{Kavli Institute for Astronomy and Astrophysics, Peking University, Beijing \it{100871}, China
      \and
    Department of Astronomy, School of Physics, Peking University, Beijing \it{100871}, China
     \and
    Department of Physics, Institute of Science, Banaras Hindu University, Varanasi, \it{221005}, Uttar Pradesh, India
    \and
    Astronomical Observatory, University of Warsaw, Al. Ujazdowskie 4, \it{00-478}, Warsaw, Poland
    \and
    Department of Physics and Astronomical Science, Central University of Himachal Pradesh, Dharamshala, \it{176215}, India
     }

\date{Received February 30, 20XX}

  \abstract
{Radio-quiet narrow-line Seyfert 1 galaxies (RQ-NLSy1s) are generally considered to be dominated by thermal emission from the accretion disk. However, detections of recurring flaring at 37 GHz by the Mets{" a}hovi Radio Observatory from seven RQ-NLSy1s suggest that non-thermal processes may also contribute to their observed emission.}
{We performed a systematic optical and mid-infrared (MIR) variability study along with broadband Spectral Energy Distribution (SED) modeling of them to investigate the origin of their flux variations and assess the relative contributions of the accretion disk and potential jet-related components.}
{High-cadence optical light curves in the \emph{g}, \emph{r}, and \emph{i} bands were obtained from the \emph{Zwicky Transient Facility (ZTF)}, and long-term MIR light curves in the \emph{W1} and \emph{W2} bands from the \emph{Wide Field Infrared Survey Explorer}. Optical variability was quantified using the $F_{\mathrm{AGN}}$-test, the peak-to-peak variability amplitude ($\psi_{\mathrm{pp}}$), and fractional variability ($F_{\mathrm{var}}$), while MIR variability was characterized by the redshift-corrected intrinsic variability amplitude. Optical variability was examined on timescales from intra-night to long-term, and MIR variability on long-term timescales. Color–magnitude behavior was analyzed using bias-resistant representations, and correlations with the physical parameters of active galactic nuclei (AGNs) were explored.}
{All seven sources exhibit statistically significant long-term optical variability, with variability amplitudes systematically increasing toward shorter wavelengths on the investigated timescales. In three sources, exhibiting pronounced bluer-when-brighter trends, both $\overline{\psi_{\mathrm{pp}}}$ and $\overline{F_{\mathrm{var}}}$ increase across the \emph{ZTF} bands, a behavior that can also arise from accretion-disk variability, but, together with the observed sub-timescale variability, supports a non-thermal contribution. Intrinsic MIR variability is detected in three of the four sources. Significant optical–MIR and MIR intra-band lags are observed, while intra-band optical lags are not statistically significant in the sample. Optical variability amplitudes are anti-correlated with the Eddington ratio and emission-line ratios and positively correlated with black hole mass. }
{A subset of RQ-NLSy1s exhibits variability patterns similar to those observed in jet-dominated AGNs, suggesting that weak or intermittent jets contribute to their optical and MIR emission. Independently of these variability diagnostics, a full-baseline \emph{Fermi}-LAT analysis yields significant ($\geq 5\sigma$) GeV detections for all six SED-modeled sources, while the corresponding broadband SED modeling requires a non-thermal, jet-related emission component in these objects. Although the inferred jet conditions are similar to those of blazars, the faint GeV emission and absence of persistent radio jets indicate low-power and/or intermittent jets. Coordinated multi-wavelength monitoring is nevertheless required to further constrain the physical origin of the observed variability.}

\keywords{surveys -- galaxies: active -- galaxies: jets -- $\gamma$-ray-galaxies: photometry -- galaxies: Seyfert – radio continuum: galaxies}

\maketitle

\section{Introduction}
 \label{sec1.0}
Active galactic nuclei (AGNs) are among the most luminous and variable objects in the Universe, powered by accretion onto supermassive black holes with masses of $\sim10^{6}$–$10^{10}$ M$_\odot$~\citep{Lynden-Bell1969Natur.223..690L, Rees1984ARA&A..22..471R, Woo2002ApJ...579..530W, Bischetti2017A&A...598A.122B}. Variability observed across the electromagnetic spectrum, on timescales ranging from minutes to decades, provides a powerful probe of the physical processes operating in the immediate vicinity of the black hole, including accretion dynamics and relativistic outflows~\citep{Urry1995PASP..107..803U, Ulrich1997ARA&A..35..445U, Czerny2008MNRAS.386.1557C}. However, the origin of observed optical variability remains debated, as classical disk models predict characteristic timescales that are often longer than those inferred from observations~\citep{Czerny2008MNRAS.386.1557C, MacLeod2010ApJ...721.1014M}. This has motivated alternative interpretations involving disk instabilities, localized temperature fluctuations, reprocessing of high-energy emission, or additional non-thermal contributions from relativistic jets~\citep{Kawaguchi1998, Lawrence2018, NodaDone2018}. Although only a small fraction of AGNs are known to host powerful relativistic jets, traditionally identified through strong radio emission~\citep{Kellermann1989AJ.....98.1195K, Urry1995PASP..107..803U}, the jet power and radiative output are expected to scale non-linearly with black hole mass ~\citep{Heinz2003MNRAS.343L..59H, Merloni2003MNRAS.345.1057M}. As a result, jet signatures in low-mass, rapidly accreting AGNs may be intrinsically weak at radio frequencies, rendering radio loudness an incomplete tracer of jet activity in such systems, particularly in radio-quiet sources with narrow emission lines~\citep{Heinz2003MNRAS.343L..59H, Foschini2014IJMPD, Caccianiga2015A&A, Jarvela2017A&A...606A...9J, Padovani2017FrontAA}.\par
Narrow-line Seyfert 1 (NLSy1) galaxies represent one such particular subclass of AGN. These are traditionally defined by narrow permitted emission lines with FWHM(H$\beta$) $<$ 2000 km s$^{-1}$, weak  [O~{\sc iii}] emission relative to H$\beta$, and often strong Fe~{\sc ii} multiplets~\citep{Osterbrock1985ApJ...297..166O, Goodrich1989ApJ...342..908G, VeronCetty2001A&A...372..730V}. NLSy1s typically host low-mass black holes relative to quasars, while accreting at high Eddington ratios~\citep[e.g.,][]{Peterson2000ApJ...542..161P}, and are therefore suggested to represent a phase of rapid supermassive black hole growth in AGN evolution~\citep{Mathur2000MNRAS.314L..17M, Grupe2004ApJ...606L..41G, Zhou2006ApJS..166..128Z, Komossa2018rnls.confE..15K, Paliya2019JApA...40...39P, Ojha2020ApJ...896...95O}. Although the majority of NLSy1s are radio-quiet, the discovery of radio-loud and especially $\gamma$-ray-detected NLSy1s ($\gamma$-NLSy1s) has demonstrated that relativistic jets can be launched even in systems with comparatively lower black hole masses than those traditionally associated with powerful jets in elliptical galaxies hosting massive SMBHs ($M_{\rm BH} > 10^8 M_{\odot}$)~\citep{Urry1995PASP..107..803U, Chiaberge2011MNRAS.416..917C, 2016MNRAS.460.3202O}, despite their high accretion rates~\citep{Abdo2009ApJ...699..976A,  Abdo2009ApJ...707..727A, Abdo2009ApJ...707L.142A, Foschini2011nlsg.confE..24F, Foschini2015A&A...575A..13F, Ojha2022JApA...43...25O}. The presence of relativistic jets in NLSy1s has profound implications for our understanding of jet formation in low-luminosity AGNs, as these sources occupy a region of parameter space traditionally thought to be unfavorable for jet launching~\citep{Mathur2000MNRAS.314L..17M, Foschini2011nlsg.confE..24F, Foschini2017FrontAA}. Unlike BL Lac objects, where low jet power is often attributed to radiatively inefficient accretion~\citep{Heckman2014ARA&A..52..589H}, NLSy1s exhibit radiatively efficient accretion flows and systematically high Eddington ratios~\citep{Brandt1997MNRAS.285L..25B, Grupe2004ApJ...606L..41G, Zhou2006ApJS..166..128Z, Ojha2020ApJ...896...95O}. Their jet properties therefore challenge earlier paradigms and suggest that jet production is not suppressed at high accretion rates~\citep{Foschini2015A&A...575A..13F, Padovani2017FrontAA}. Recent numerical simulations further support this view, suggesting that powerful, collimated jets can form even at near- or super-Eddington accretion rates under suitable magnetic conditions~\citep{McKinney2017MNRAS, Liska2022ApJS}. These developments imply that previous observational biases, particularly the emphasis on radio-loud systems, may have obscured a broader diversity of jet phenomena in AGNs~\citep{Berton2017AandA, Padovani2017FrontAA}.\par
In this context, time-domain investigations provide a particularly sensitive means of probing the presence and influence of relativistic jets, especially when traditional radio-based diagnostics become ambiguous and link rapid optical fluctuations with jet activity in AGNs~\citep{Wagner1995ARA&A..33..163W, Ulrich1997ARA&A..35..445U, Gopal-Krishna2003ApJ...586L..25G, Ojha2024MNRAS.529L.108O}. Strong and frequent intra-night optical variability (INOV) is commonly observed in jet-dominated sources such as blazars and high-polarization quasars, with duty cycles exceeding 30\%~\citep{Jang1995ApJ...452..582J, deDiego1998, Stalin2005MNRAS.359.1022S, Goyal2012A&A...544A..37G}. In contrast, radio-quiet quasars generally show weaker and less frequent INOV, consistent with variability driven by accretion-disk processes or weak, misaligned jets~\citep{Chakrabarti1993ApJ...411..602C, Mangalam1993ApJ...406..420M, Goyal2013MNRAS.435.1300G}. Among NLSy1s, $\gamma$-ray-detected and radio-loud sources display blazar-like INOV characteristics, suggesting the connection between rapid optical variability and relativistic jet emission~\citep{Paliya2013MNRAS.428.2450P, Ojha2022MNRAS.514.5607O, Ojha2024MNRAS.529L.108O, Singh2025ApJ...990...79S}.\par
Moreover, color variability behavior of AGNs provides a powerful diagnostic of dominant emission processes, since color–magnitude trends encode the relative contributions of thermal accretion-disk emission and non-thermal synchrotron radiation from relativistic jets, with the direction and strength of the trend depending on the source type and underlying physical conditions~\citep{Vagnetti2003ApJ...590..123V, Gu2011A&A...534A..59G, Negi2022MNRAS.510.1791N, Ojha2024MNRAS.529L.108O}. For instance, in jet-dominated AGNs such as BL Lac objects, optical color variability is frequently characterized by a bluer-when-brighter (BWB) trend, commonly attributed to changes in synchrotron emission from the relativistic jet~\citep[e.g.,][]{Villata2002A&A...390..407V, Ikejiri2011PASJ...63..639I} although similar BWB trends are also commonly observed in accretion-disk dominated AGNs and quasars~\citep[e.g., ][]{Giveon1999MNRAS.306..637G, 2004ApJ...601..692V, Ruan2014ApJ...783..105R}. Within the shock-in-jet framework, freshly accelerated high-energy electrons at the shock front preferentially enhance short-wavelength emission and on shorter timescales than long-wavelength emission, resulting in larger variability amplitudes at higher frequencies~\citep[see][]{Kirk1998A&A...333..452K, Mastichiadis2002PASA...19..138M, 2011MmSAI..82..104T}. Alternative mechanisms, such as energy injection through internal shocks and variations in the Doppler beaming factor, can also naturally cause BWB behavior~\citep[e.g.,][]{Villata2004A&A...421..103V, Papadakis2007A&A...470..857P, Larionov2010A&A...510A..93L}.\par
A major challenge to this framework has emerged with the discovery of seven radio-quiet (or radio-silent) NLSy1 galaxies exhibiting recurrent, high-amplitude flaring at 37 GHz~\citep{2018A&A...614L...1L}. These sources show dramatic radio variability on timescales of days, reaching flux densities typically associated with powerful jetted AGNs. However, follow-up interferometric observations  with the Karl G. Jansky Very Large Array and the Very Long Baseline Array have revealed either very steep radio spectra or non-detections, with no evidence for persistent large-scale relativistic jets ~\citep{Berton2020A&A...636A..64B,Jarvela2024MNRAS.532.3069J}. Several physical explanations have been proposed, including compact and self-absorbed jets, strong free-free absorption by dense ionized gas, jet-interstellar medium interactions, or transient energy-release processes such as magnetic reconnection near the black hole~\citep{Antonucci1988ApJ...332L..13A, Bicknell1997ApJ...485..112B, Jarvela2024MNRAS.532.3069J}. Apart from constraints from radio studies, the intra-night optical variability detected in a subset of the sources in this sample has been suggested to originate not only from relativistic jets but also from magnetic reconnection events in the black hole magnetosphere~\citep[see ][]{Ojha2024MNRAS.529L.108O}. Regardless of the interpretation, these objects challenge the traditional radio-based classification of AGNs and raise questions about the reliability of radio loudness as a diagnostic for jets. Furthermore, optical spectroscopic studies have shown that, despite their extraordinary radio behavior, the emission-line properties of these seven sources are largely consistent with those of typical NLSy1s. They generally exhibit strong Fe~{\sc ii} emission and black hole masses that are not exceptional within the class, suggesting that their extreme radio variability is not simply driven by usual accretion-related properties~\citep{Crepaldi2025A&A...696A..74C}. This apparent disparity between radio variability and optical spectral properties emphasizes the importance of time-domain and multi-wavelength diagnostics.\par
In this context, optical and infrared variability studies provide a crucial and complementary probe of the physical components of AGNs. Optical variability primarily traces emission from the accretion disk, with a possible contribution from any jet component extending to these wavelengths, whereas mid-infrared (MIR) emission is generally associated with thermal radiation from dust in the circumnuclear torus~\citep{Antonucci1993ARA&A..31..473A, Urry1995PASP..107..803U}. Dust reverberation mapping studies have shown that MIR variability commonly lags optical variations, thereby providing direct constraints on the size, geometry, and structure of the dusty region~\citep{Suganuma2006ApJ...639...46S, Koshida2014ApJ...788..159K, Mandal2018MNRAS.475.5330M}. Moreover, the presence or absence of optical intra-band and optical–MIR inter-band time lags serves as a powerful diagnostic of the dominant variability mechanism. For instance, significant lags are expected when variability is disk-dominated and driven by reprocessing~\citep[e.g.,][]{Cackett2007MNRAS.380..669C, McHardy2014MNRAS.444.1469M}, whereas jet-dominated variability is expected to exhibit weak or negligible lags owing to its compact and non-thermal origin~\citep[e.g.,][]{Villata2002A&A...390..407V, Marscher2008Natur.452..966M}.\par
Despite the availability of high-cadence optical and infrared data from modern time-domain surveys such as the Zwicky Transient Facility~\citep[\emph{ZTF},][]{Bellm2019PASP..131a8002B} and the \emph{Wide Field Infrared Survey Explorer}~\citep[\emph{WISE},][]{Wright2010AJ....140.1868W}, detailed multi-wavelength variability studies of radio-quiet NLSy1 galaxies remain scarce. This is particularly true for sources that exhibit indirect yet compelling evidence for relativistic jet activity. The seven radio-quiet NLSy1s displaying extreme flaring at 37 GHz therefore occupy a uniquely informative parameter space, as they blur the traditional distinction between radio-quiet and jetted AGNs, challenge standard unification schemes, and provide a rare opportunity to investigate the interplay between accretion processes, jet activity, and circumnuclear structures in low- to intermediate-mass black hole systems. In addition, broadband spectral energy distribution (SED) modeling of these RQ-NLSy1s enables an assessment of the relative contributions of thermal emission from the accretion disk and dusty torus versus non-thermal, jet-related components, thereby offering an independent and complementary consistency check on the physical interpretation inferred from variability diagnostics, along with robust constraints on key SED parameters.
Motivated by the above considerations, we undertake a comprehensive optical and mid-infrared variability study of seven RQ-NLSy1s that exhibit extreme and recurring 37 GHz radio flaring. By combining analyses of optical variability on timescales from hours to years, long-term mid-infrared variability, optical and mid-infrared color variability, optical–infrared time-lag measurements, and broadband SED modeling, we aim to probe the relative contributions of accretion-driven and jet-related processes in these sources. We also examine how the variability amplitudes of these RQ-NLSy1s relate to their fundamental AGN properties, aiming to identify which physical parameters most strongly influence their variability amplitude. This article is organized as follows. Section~\ref{section_2.0} outlines the compilation of the data sets used in this study. The analysis techniques and methodological framework are presented in Sect.~\ref{sec_3.0}. Our results of the present work are reported in Sect.~\ref{sec_4.0}. Discussions are interpreted in Sect.~\ref{sec_5.0}, while the main conclusions are summarized in Sect.~\ref{sect_6.0}.

\section{Data compilation}
\label{section_2.0}
\subsection{\emph{ZTF} \emph{g}, \emph{r}, and \emph{i} bands optical photometric data}
\label{Optical_data}
Optical photometric light curves for the seven RQ-NLSy1s were obtained from \emph{ZTF} in the \emph{g}, \emph{r}, and \emph{i} bands using the 23$^{rd}$ public data release~\citep[see][]{Masci2019PASP..131a8003M}. Data were retrieved via the NASA/IPAC Infrared Science Archive (IRSA\footnote{https://www.ztf.caltech.
edu/ztf-public-releases.html}) application programming interface within an arcsec radius of the optical source positions. The availability of \emph{ZTF} data for individual sources in the different bands of \emph{ZTF} \emph{g}, \emph{r}, and \emph{i}, together with their basic properties, is reported in Table~\ref{tab:source_info}.

To minimize artificial variability introduced by independent photometric calibrations across different \emph{ZTF} fields and CCD quadrants, we retained, for each source and filter, only the light curve corresponding to the observation ID containing the maximum number of data points. The standard quality cuts were then applied following the recommendations of the \emph{ZTF} Science Data System\footnote{https://irsa.ipac.caltech.edu/data/ZTF/docs/ztf explanatory supplement.pdf}. Specifically, only measurements with catflags = 0 were selected, photometric points with magnitude uncertainties greater than 10\% were excluded, and a global 3$\sigma$ clipping was applied to remove outliers. In addition, a minimum of five valid data points per band was imposed. This has resulted in \emph{ZTF} \emph{g-}, \emph{r-}, and \emph{i-}band data for all except the \emph{i-}band data for RQ-NLSy1 J102906.69$+$555625.2 (see Table~\ref{tab:source_info}). The resulting light curves are characterized by cadences reaching a few hours (sub-day timescales) in densely sampled epochs and maximum temporal baselines of up to $\sim$2200 days, depending on the source and bands. These data are suitable for probing optical variability on shorter timescales, ranging from months and weeks down to intranight variations in the current sample of RQ-NLSy1s. A representative example of long-term light curves in the \emph{ZTF} \emph{r} band for these seven RQ-NLSy1s is presented in Fig.~\ref{fig: OP_flux_variability}.

\subsection{\emph{WISE} \emph{W1} and \emph{W2} bands mid-infrared data}
\label{MIR_data}
The mid-infrared photometric data for the seven RQ-NLSy1s were retrieved from the \emph{WISE} Multiepoch Photometry (MEP) database and the \emph{NEOWISE-R} single-exposure (L1b) source catalog via IRSA\footnote{https://irsa.ipac.caltech.edu/Missions/wise.html}. The combined data set spans the period from 2010 to 2024, with a gap between 2011 and 2014 associated with the cryogen depletion phase of the original \emph{WISE} mission. The availability of \emph{WISE} data for each source in \emph{WISE} \emph{W1} and \emph{W2} bands is summarized in Table~\ref{tab:source_info}. Our analysis was restricted to the \emph{W1} and \emph{W2} bands, as the \emph{W3} and \emph{W4} bands are sparsely sampled and largely confined to the early mission phase. To ensure robust photometry, we applied strict selection criteria to the single-exposure measurements. We selected reduced $\chi^{2}$ values of profile-fit photometry to be less than 5 in both bands, fewer than three PSF components in the fit, the highest image quality flag, the absence of known artifacts, and no active deblending. In addition, a minimum of five valid data points per band was imposed, and a global 3$\sigma$ clipping was also used to remove outliers. This has resulted in a sample of four RQ-NLSy1s with reliable measurements in \emph{WISE} \emph{W1} and \emph{W2} bands (see Table~\ref{tab:source_info})

\begin{figure*}
    \begin{minipage}[]{1.0\textwidth}
\includegraphics[width=0.24\textwidth]{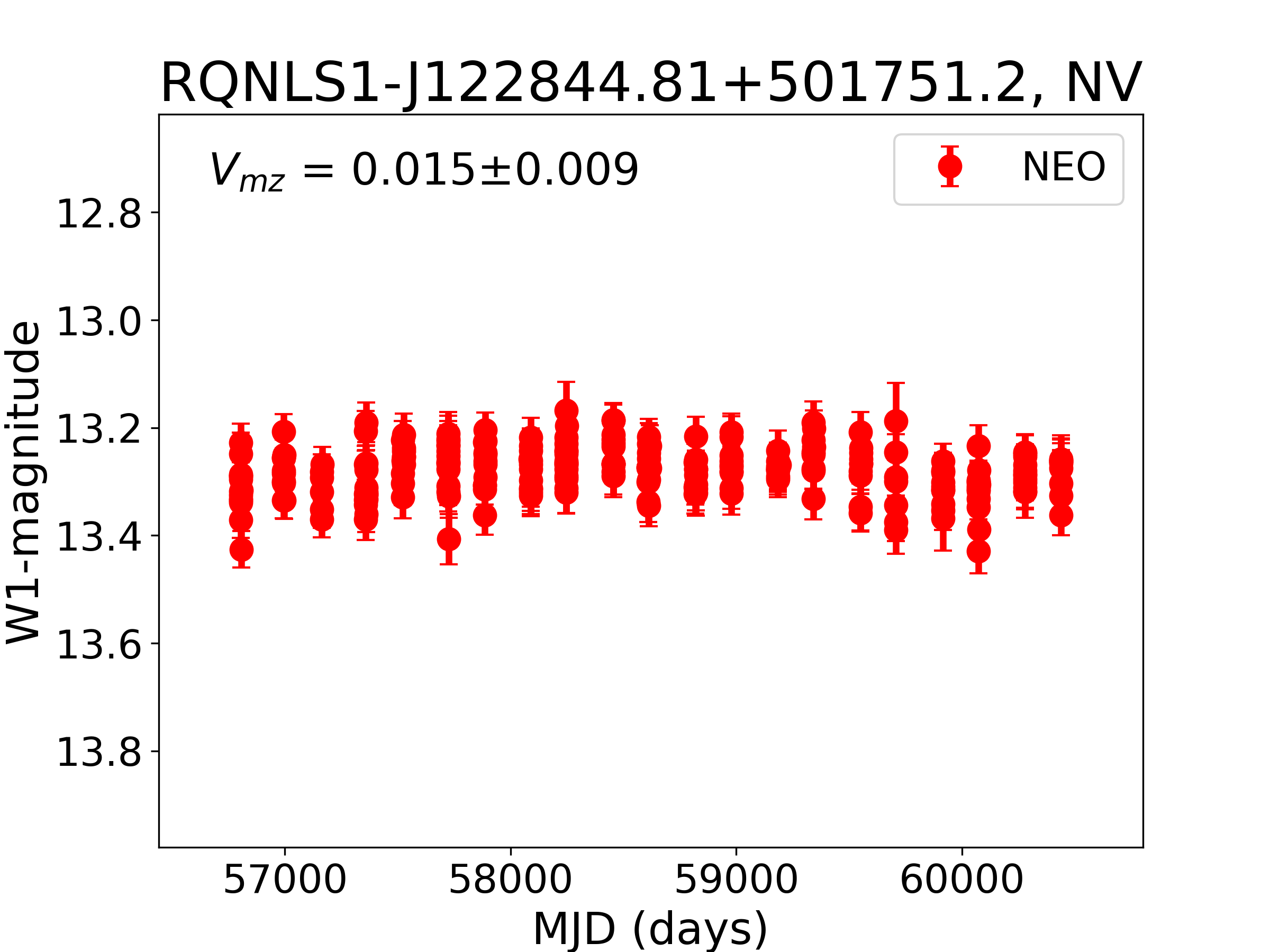}  
\includegraphics[width=0.24\textwidth]{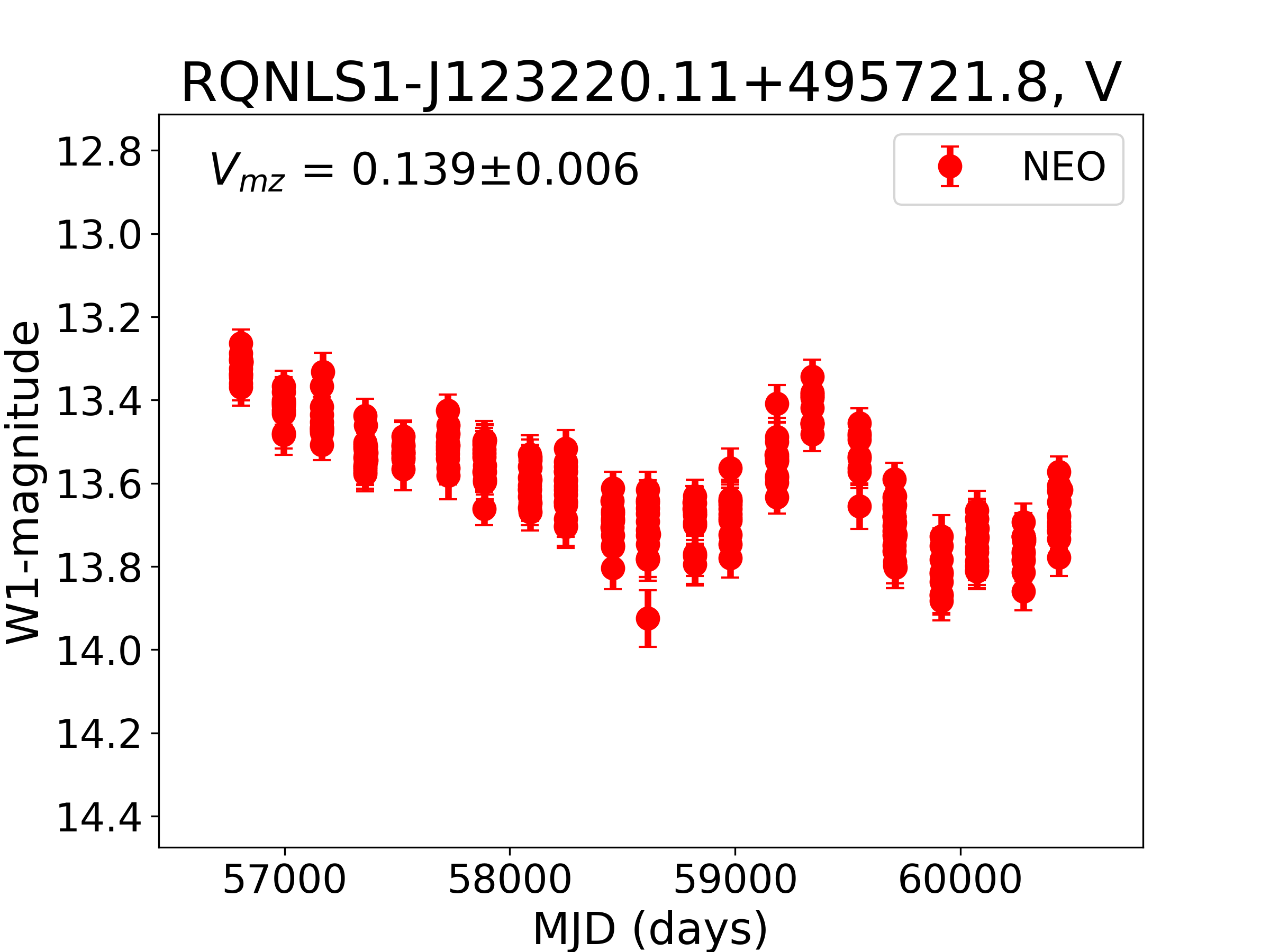} 
\includegraphics[width=0.24\textwidth]{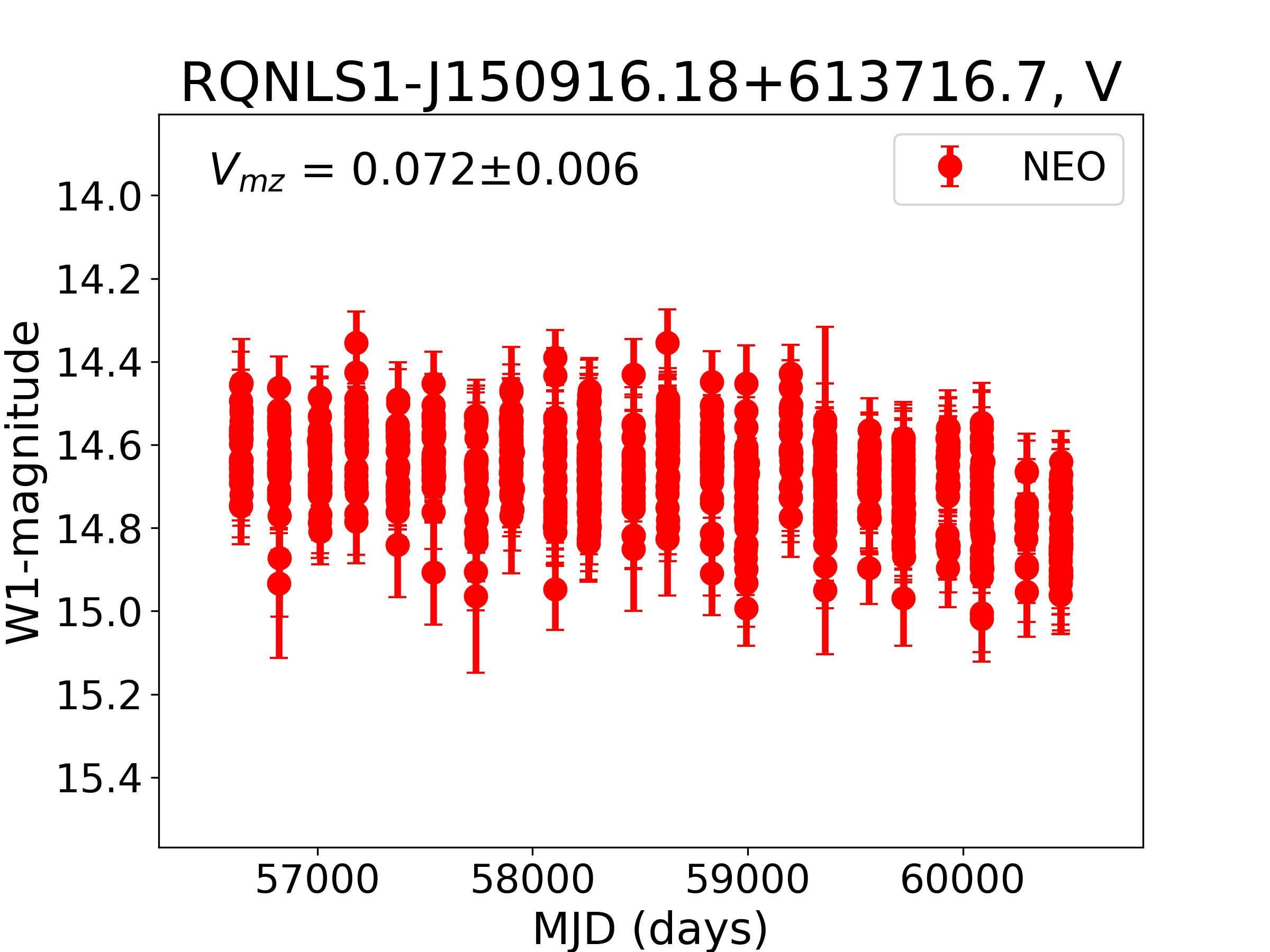}  
\includegraphics[width=0.24\textwidth,]{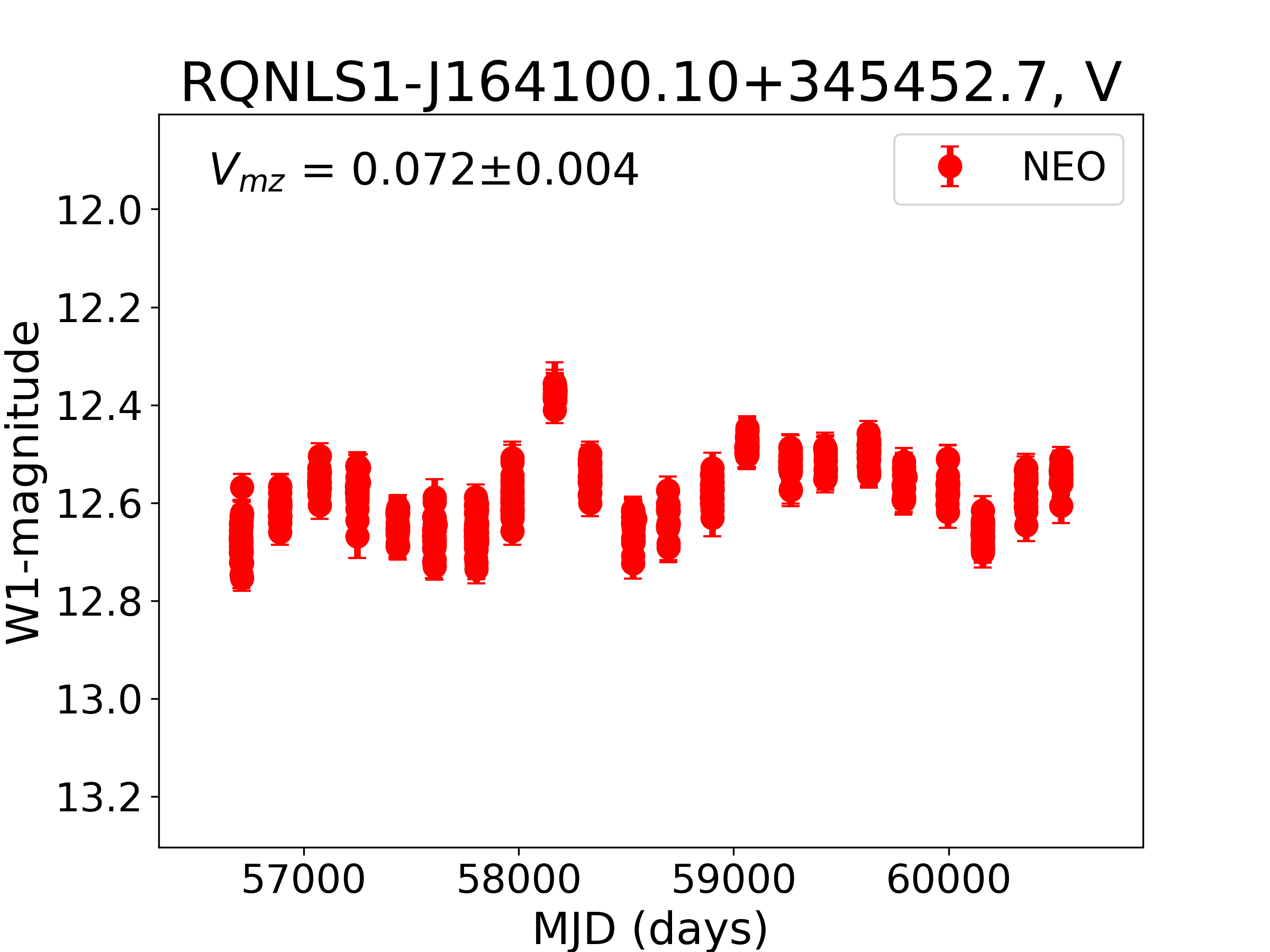}  
\includegraphics[width=0.24\textwidth,]{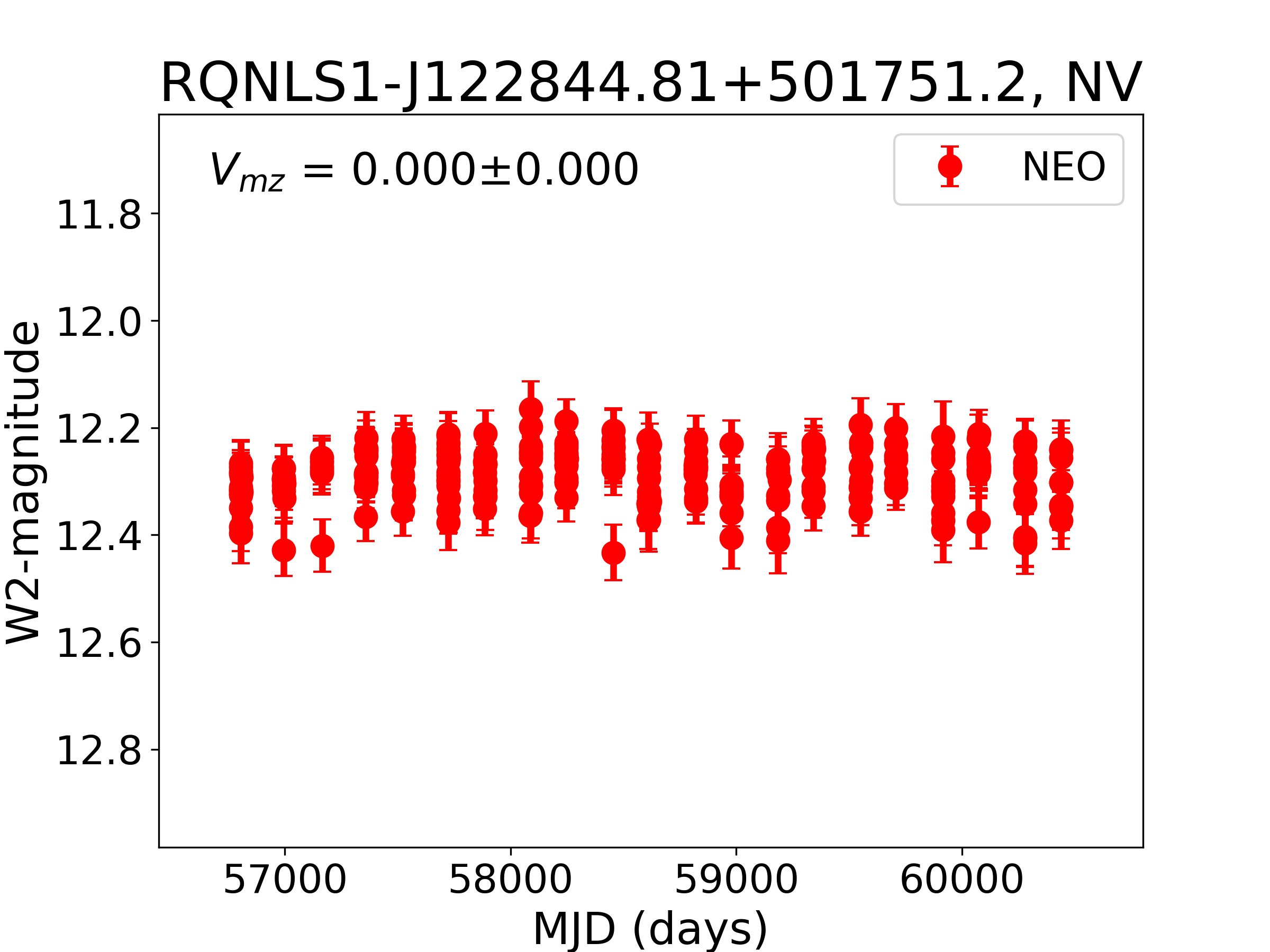} 
\includegraphics[width=0.24\textwidth,]{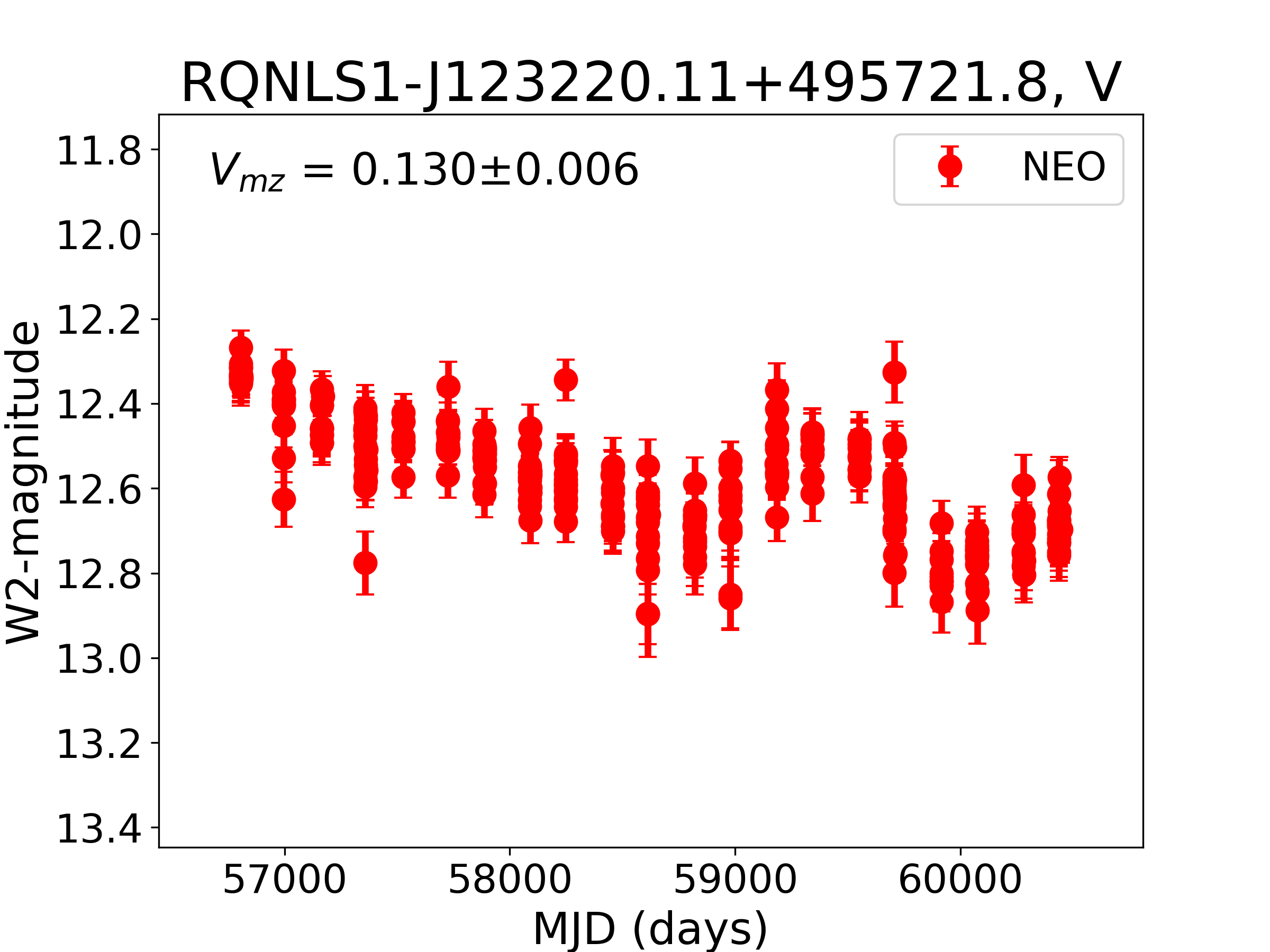}  
\includegraphics[width=0.24\textwidth,]{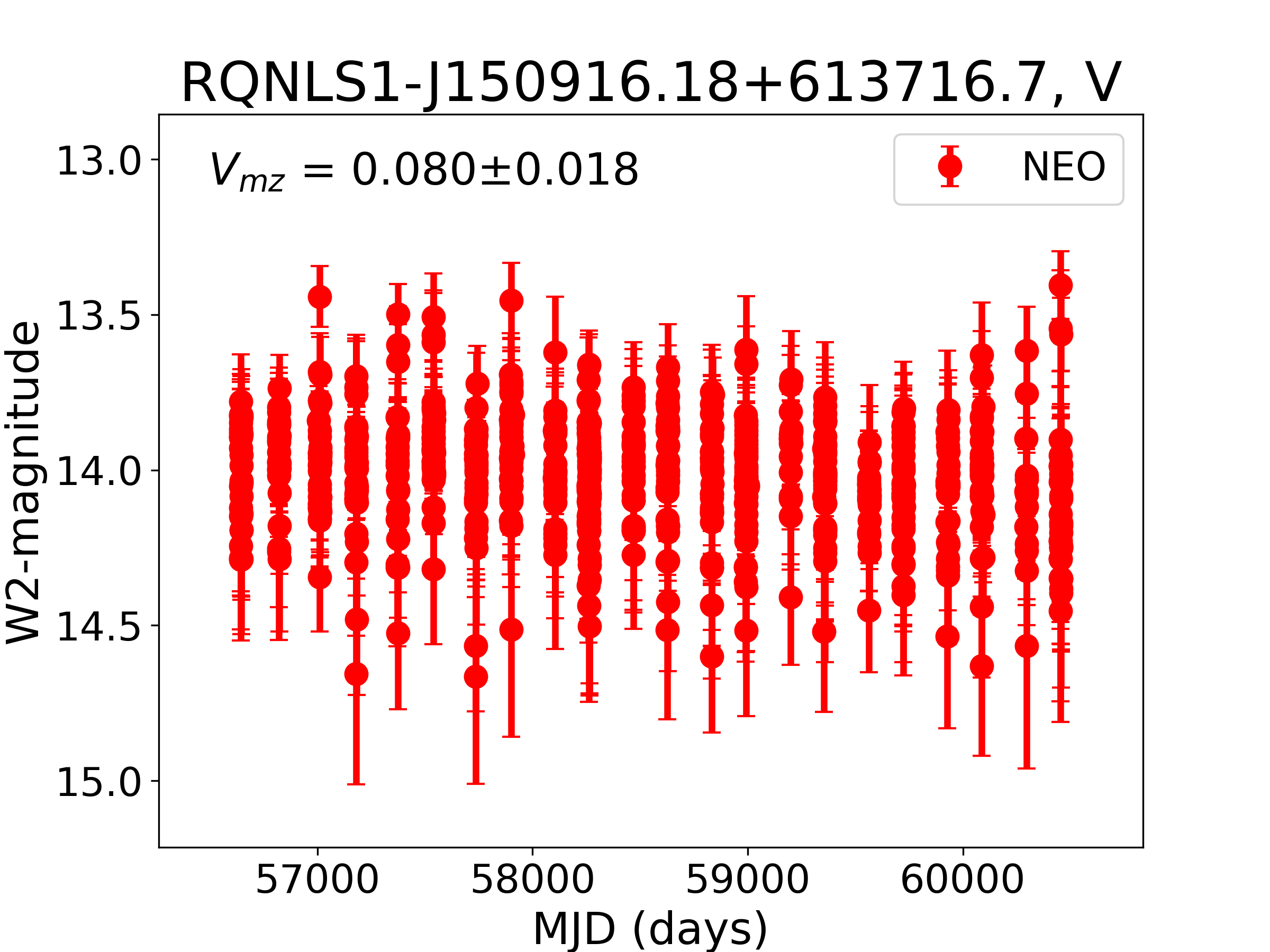}  
\includegraphics[width=0.24\textwidth,]{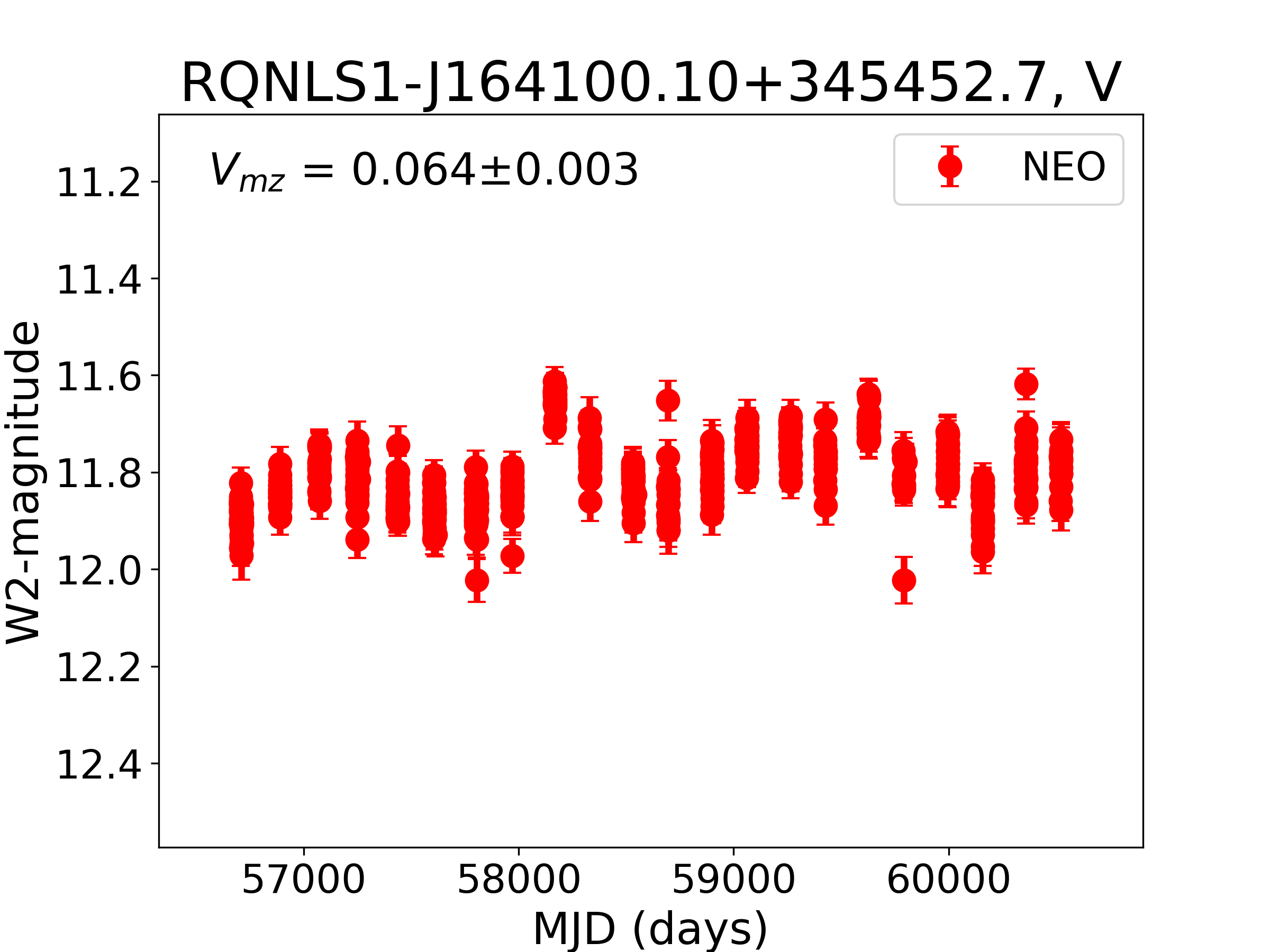} 
   
    \end{minipage}
    \caption{Long-term \emph{WISE} \emph{W1} and \emph{W2} light curves for the RQ-NLSy1s \emph{J122844.81$+$501751.2}, \emph{J123220.11$+$495721.8}, \emph{J150916.18$+$613716.7}, and \emph{J164100.10$+$345452.7} from the current sample, which have reliable MIR measurements. All sources exhibit significant variability in both bands, except for \emph{J122844.81$+$501751.2}, which shows no clear long-term variation. The name and long-term variability status of each source are indicated in the title of each panel, and the estimated redshift-corrected intrinsic variability amplitude, $V_{mz}$, is shown in the upper-left corner.}
    \label{fig: IR_flux_variability}
\end{figure*}

Inspection of the resulting light curves revealed closely spaced measurements separated by approximately 11 seconds during individual \emph{WISE} visits. Given the \emph{WISE} orbital period of about 1.5 hours, such measurements were averaged within each visit to construct the final \emph{W1} and \emph{W2} light curves. The processed mid-infrared data provide long-term temporal coverage suitable for investigating MIR variability in RQ-NLSy1s and for comparison with variability observed at other wavelengths. We have shown long-term MIR light curves for the available targets RQ-NLSy1-J122844.81$+$501751.2, RQ-NLSy1-J123220.11$+$495721.8, RQ-NLSy1-J150916.18$+$613716.7, and RQ-NLSy1-J164100.10$+$345452.7 in the \emph{WISE} \emph{W1} and \emph{W2} bands of the current sample in Fig.~\ref{fig: IR_flux_variability}.

\subsection{Compilation of multi-wavelength data for SED fitting}
\label{SED_data}
For the SED fitting of the seven RQ-NLSy1s, we compiled data from the NASA/IPAC Extragalactic Database\footnote{https://ned.ipac.caltech.edu/}. X-ray data were obtained with \emph{Swift}-XRT ~\citep{2005SSRv..120..165B} and reduced using the standard automatic products generator provided by the UK \emph{Swift} Science Data Centre~\citep{Evans2007A&A...469..379E,2009MNRAS.397.1177E}. The observation log, the number of observations available for each source, and the results of the default spectral fits are given in Sect.~\ref{Xray_data} and Table\ref{tab_xrt_log} of Appendix~\ref{app-xrt}. In addition, a full-baseline ($\sim$2008–2026) $\gamma$-ray dataset from \emph{Fermi}-LAT was analysed using \texttt{Fermipy}. A complete description of the $\gamma$-ray data reduction, the adopted instrument response functions and background models, the region-of-interest definition, and the source model is provided in Sect.~\ref{LAT_analysis}, together with a discussion of the limitations of the present $\gamma$-ray analysis. The analysis yielded individual detections with ${\rm TS}>25$ for all six SED-modelled sources (see Sect.~\ref{SED_Modeling}). The most recent radio measurements for these RQ-NLSy1s were adopted from Table3 of~\citet{Jarvela2024MNRAS.532.3069J} and included in the SEDs for comparison, although they were not used as constraints in the modeling. An initial inspection of the broadband SEDs suggests a double-humped structure similar to that observed in jet-dominated objects such as blazars. Since these objects have not been monitored in the past for the purpose of SED modeling, the data suffer from non-simultaneity. This is an important limitation: the radio, MIR, optical, X-ray, and $\gamma$-ray measurements represent averages over different and generally non-overlapping epochs. The resulting SEDs should therefore be regarded as time-averaged representations of the sources rather than snapshots of a single physical state. Accordingly, our objective is not to derive uniquely constrained or precise best-fit model parameters, but rather to assess qualitatively whether the broadband emission is consistent with the presence of jet-related non-thermal emission, as observed in jetted AGNs.

\begin{table*}
  \begin{minipage}{175mm} 
  \caption{Summary of multiwavelength data coverage and basic source properties of the seven radio-quiet narrow-line Seyfert 1 galaxies, including redshift, R-band magnitude, radio-loudness parameter, and black hole mass.}
\label{tab:source_info}
\begin{tabular}{lcccccccccccc}
 \hline
   &&&&&&& \multicolumn{3}{c}{\emph{ZTF}-data} & \multicolumn{2}{c}{\emph{WISE}-data}&\multicolumn{1}{c}{SWIFT-data} \\
   \hline
    {SDSS Name\footnote{\small The SDSS names of the source.}} & {RA.} & {DEC.} & {z\footnote{\small Redshift of the sources are taken from~\citet{2018A&A...614L...1L}.}} & {m$_R$\footnote{\small The apparent R-band magnitude of the sources are taken from~\citet{Monet1998AAS...19312003M}.}} & {RL\footnote{\small Radio-loudness parameter $RL\equiv F_{5.2~GHz}/F_{B-band}$ of current sources are taken from~\citet{Crepaldi2025A&A...696A..74C}}} & {$M_{BH}$\footnote{\small Average black hole masses of current sources reported in~\citet{Crepaldi2025A&A...696A..74C} are taken.}} & {g} & {r} & {i} & {W1} & {W2} & {X-ray} \\
    &&&&&& ($\times 10^{7} M_\odot$) &&&&&& \\
    \hline \\
    J102906.69+555625.2 & 157.27788 & 55.94033 & 0.451 & 19.1 & --  & 6.63$\pm$1.88  & \color{olive}\checkmark & \color{olive}\checkmark & \color{red}\texttimes & \color{red}\texttimes & \color{red}\texttimes & \color{olive}\checkmark \\
    J122844.81+501751.2 & 187.18671 & 50.29756 & 0.262 & 17.8 & 2.4  & 0.92$\pm$0.50 & \color{olive}\checkmark & \color{olive}\checkmark & \color{olive}\checkmark & \color{olive}\checkmark & \color{olive}\checkmark & \color{olive}\checkmark \\
    J123220.11+495721.8 & 188.08379 & 49.95606 & 0.262 & 16.9 & 0.1  & 2.73$\pm$0.88 & \color{olive}\checkmark & \color{olive}\checkmark & \color{olive}\checkmark & \color{olive}\checkmark & \color{olive}\checkmark & \color{olive}\checkmark \\
    J150916.18+613716.7 & 227.31742 & 61.62131 & 0.201 & 18.6 & --  & 3.02$\pm$0.73 & \color{olive}\checkmark & \color{olive}\checkmark & \color{olive}\checkmark & \color{olive}\checkmark & \color{olive}\checkmark & \color{olive}\checkmark \\
    J151020.06+554722.0 & 227.58358 & 55.78944 & 0.150 & 17.8 & 0.5  & 0.83$\pm$0.38 & \color{olive}\checkmark & \color{olive}\checkmark & \color{olive}\checkmark & \color{red}\texttimes & \color{red}\texttimes & \color{olive}\checkmark \\
    J152205.41+393441.3 & 230.52254 & 39.57814 & 0.077 & 13.1 & 1.7  & 0.31$\pm$0.05 & \color{olive}\checkmark & \color{olive}\checkmark & \color{olive}\checkmark & \color{red}\texttimes & \color{red}\texttimes & \color{olive}\checkmark \\
    J164100.10+345452.7 & 250.25042 & 34.91464 & 0.164 & 16.0 & 08  & 1.70$\pm$0.85 & \color{olive}\checkmark & \color{olive}\checkmark & \color{olive}\checkmark & \color{olive}\checkmark & \color{olive}\checkmark & \color{olive}\checkmark \\
    \hline
\end{tabular}
 \end{minipage}
\end{table*}

\begin{figure*}
    \begin{minipage}[]{1.0\textwidth}
    \includegraphics[width=0.24\textwidth]{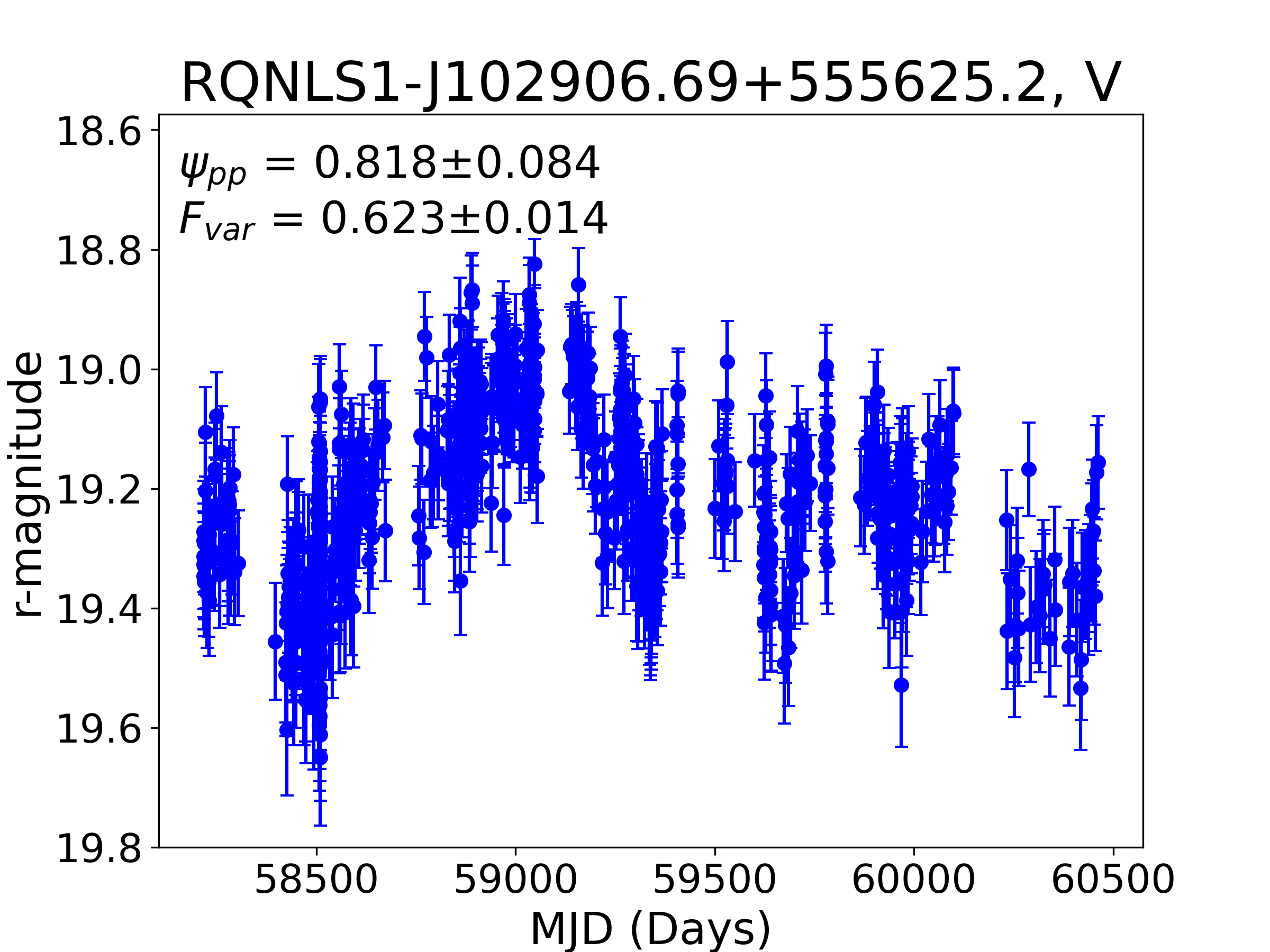}  
   \includegraphics[width=0.24\textwidth]{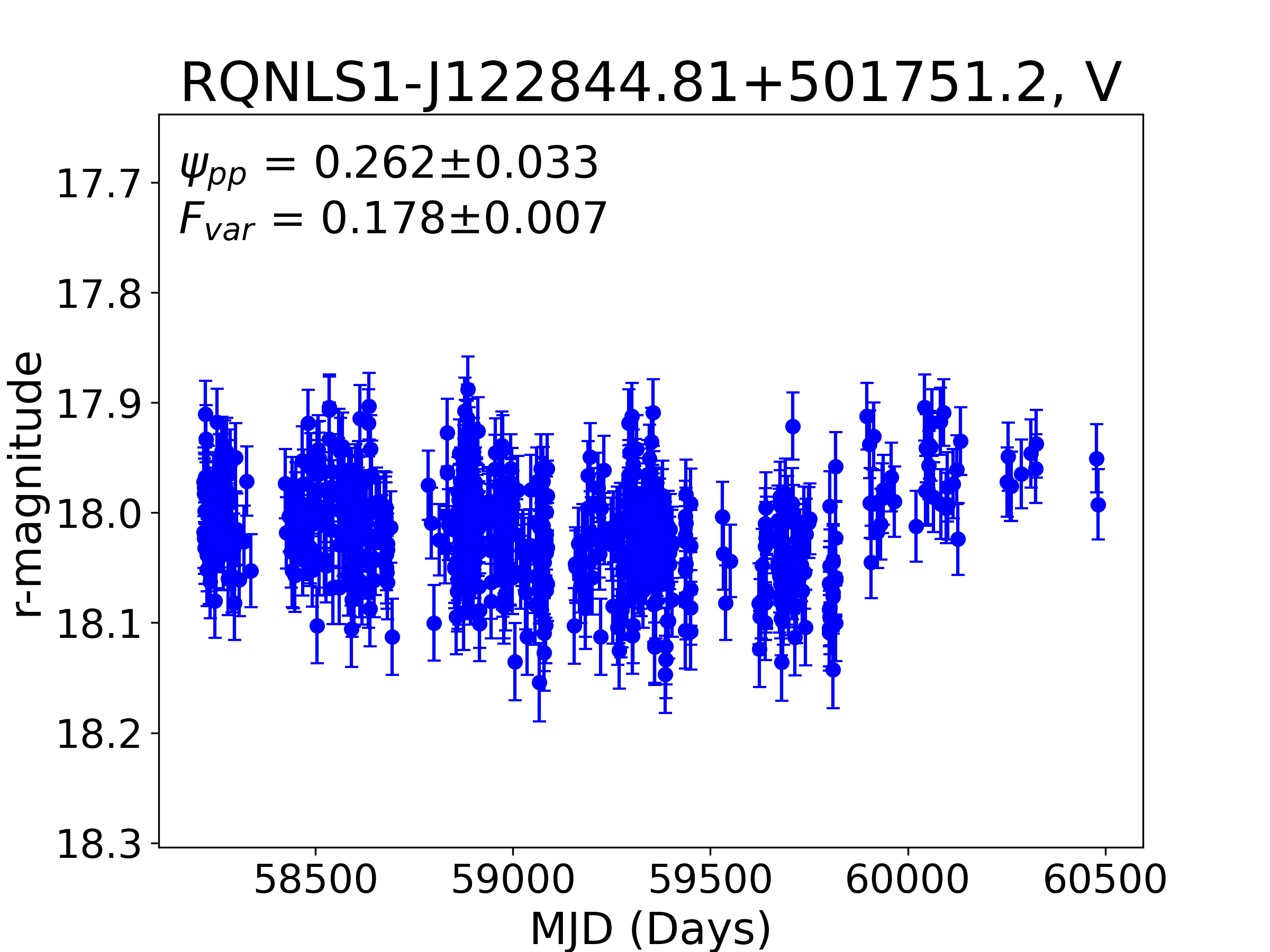} 
   \includegraphics[width=0.24\textwidth]{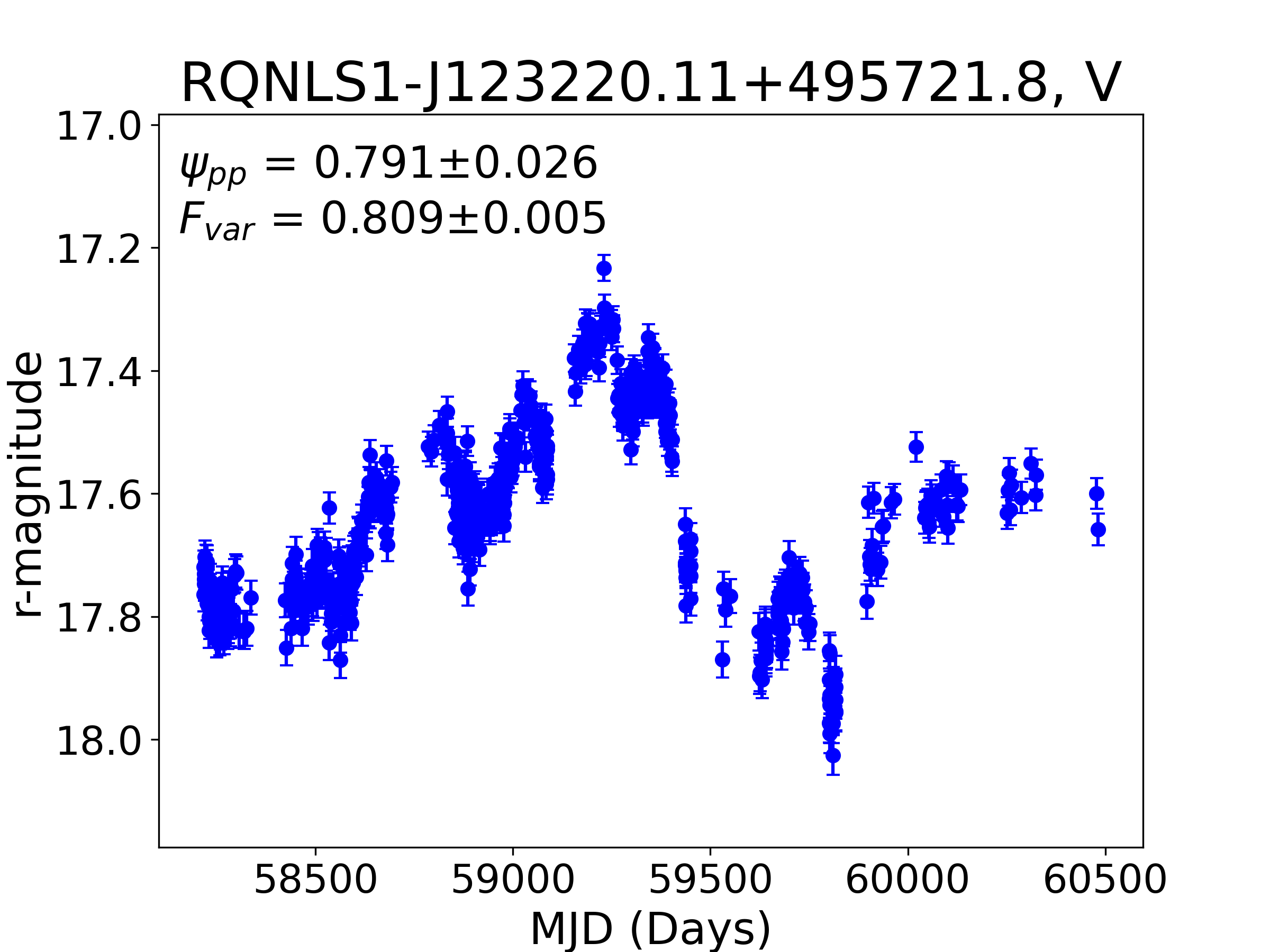}  
  \includegraphics[width=0.24\textwidth]{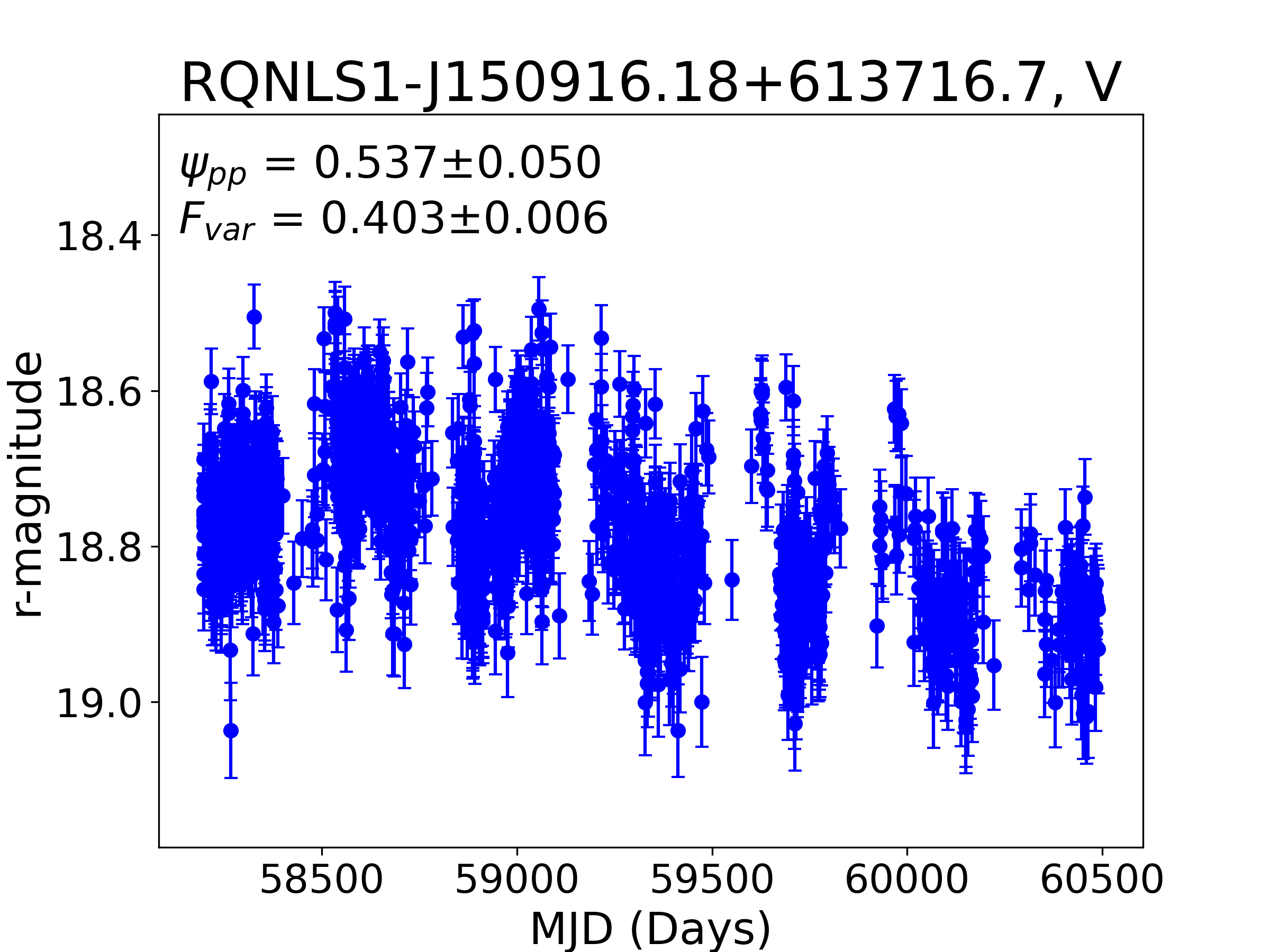}  
   \includegraphics[width=0.33\textwidth]{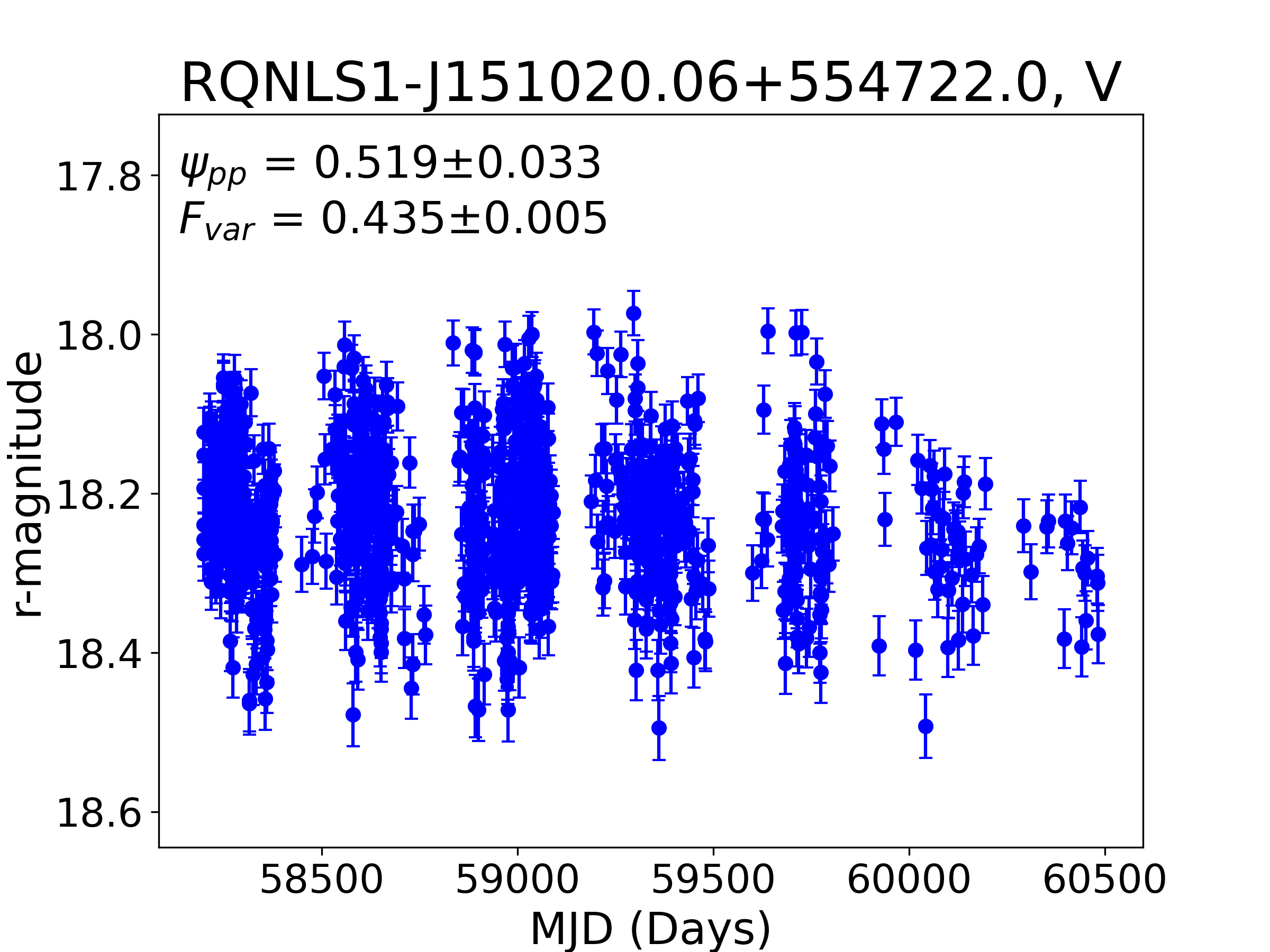} 
   \includegraphics[width=0.33\textwidth]{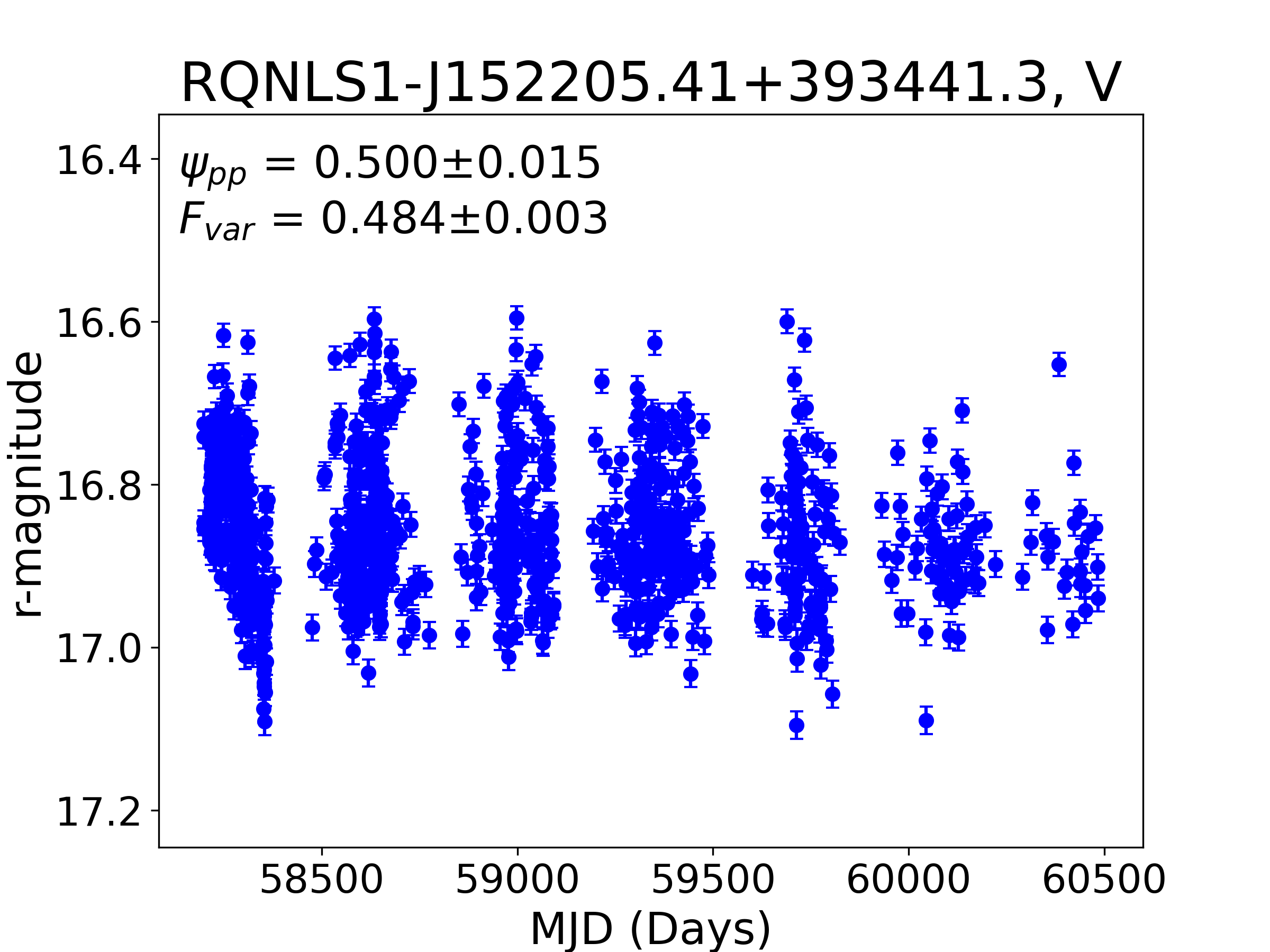}  
   \includegraphics[width=0.33\textwidth]{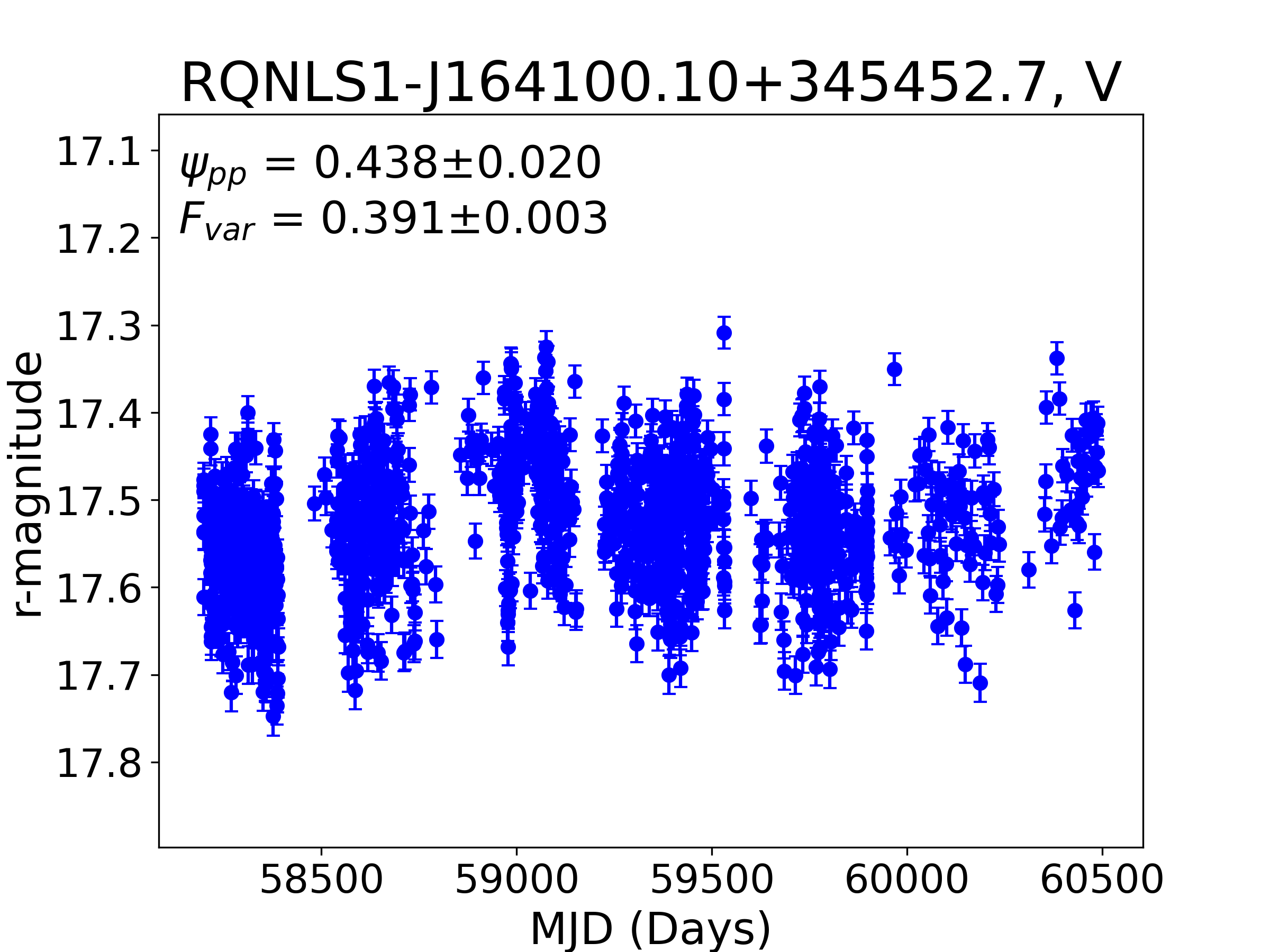}  
   
    \end{minipage}
    \caption{Long-term r-band \emph{ZTF} light curves of the seven RQ-NLSy1s from the current sample, showing variability (V) from all the sources on a year-like timescale. The name and long-term variability status of each RQ-NLSy1 galaxy are presented in the title of each panel, and estimated variability parameters $\psi_{\mathrm{pp}}$ and $F_\mathrm{var}$ are displayed in the upper-left corner of each panel.}
    \label{fig: OP_flux_variability}
\end{figure*}

\section{Methodology for analysis}
\label{sec_3.0}
Before performing the variability analysis, all observed \emph{ZTF} light curves were corrected for foreground Galactic extinction and absorption. For each source, the extinction value ($A_V$) was obtained from the NASA/IPAC Extragalactic Database (NED), and the correction was applied following the prescription of~\citet{Schlafly2011ApJ...737..103S}. Examples of extinction-corrected \emph{ZTF} $\emph{r}$-band light curves for the seven RQ-NLSy1s are presented in Fig.~\ref{fig: OP_flux_variability}.

\subsection{Variability analysis in Optical wavelength}
\label{section_3.1}

The optical variability properties of the seven RQ-NLSy1s were examined using their \emph{ZTF} $\emph{g}$, $\emph{r}$, and $\emph{i}$ band light curves. To assess whether the observed fluctuations are statistically significant, we applied the $F$-test, which is widely adopted in AGN variability studies~\citep{Goyal2012A&A...544A..37G}. The robustness and applicability of this approach for NLSy1s have been demonstrated in our previous work~\citep[see, ][]{Ojha2018BSRSL..87..387O, Ojha2024MNRAS.529L.108O}.

The variability statistic is defined as
\begin{equation}
\label{eq:Ftest_new}
F_{\mathrm{AGN}} = \frac{s_{\mathrm{lc}}^{2}}{\kappa^{2}\,\overline{\epsilon^{2}}}
\end{equation}
where $s_{\mathrm{lc}}^{2}$ represents the variance of the observed light curve, and
\[
\overline{\epsilon^{2}} = \frac{1}{N}\sum_{i=1}^{N} \epsilon_i^{2}
\]
is the mean squared formal photometric uncertainty for the $N^{th}$ data points. The factor $\kappa$ accounts for the systematic underestimation of photometric errors.\par
Reliable characterization of measurement uncertainties is essential for robust variability analyses, particularly when source magnitudes are derived using aperture photometry. It is well established that commonly used aperture-based photometric reduction packages, including IRAF\footnote{http://iraf.noao.edu/} and DAOPHOT II\footnote{http://www.astro.wisc.edu/sirtf/daophot2.pdf}, tend to underestimate formal magnitude errors, typically by factors of about 1.3–1.75~\citep{Sagar2004MNRAS.348..176S, Bachev2005MNRAS.358..774B}. To account for this systematic effect,~\citet{Goyal2013JApA...34..273G} introduced a correction factor of $\kappa = 1.54$, derived from an extensive set of intra-night AGN monitoring observations.

In contrast, the \emph{ZTF} photometric pipeline determines source magnitudes using a point-spread-function fitting at the target position~\citep{Masci2019PASP..131a8003M, Dekany2020PASP..132c8001D}. As a result, the associated rms photometric uncertainties are not expected to suffer from the same level of underestimation characteristic of aperture-based reductions. Therefore, in this work, we adopted $\kappa = 1$ for the \emph{ZTF} light curves, consistent with the approach followed in recent variability studies based on \emph{ZTF} data~\citep[e.g., see][]{Negi2023MNRAS.522.5588N, Chand2024PASA...41..106C}.

Thus, for each light curve, the computed value of $F_{\mathrm{AGN}}$ was compared with the critical value of the $F$-distribution at a significance level of $\alpha = 0.01$, corresponding to a confidence level of 99\% for a variability detection. A source was shortlisted as optically variable when $F_{\mathrm{AGN}}$ exceeded the critical threshold, thus rejecting the null hypothesis of non-variability with high confidence.

To characterize the amplitude of the detected optical variability and intrinsic fractional variability, we estimated two complementary quantities: the peak-to-peak variability amplitude and the fractional variability amplitude.

The peak-to-peak amplitude is defined as~\citep{Heidt1996A&A...305...42H}
\begin{equation}
\label{eq:Amp_pp_new}
\psi_{\mathrm{pp}} =
\sqrt{(m_{\max} - m_{\min})^{2} - 2\sigma_{\mathrm{corr}}^{2}}
\end{equation}
where $m_{\max}$ and $m_{\min}$ are the maximum and minimum magnitudes in the light curve, respectively, and
\[
\sigma_{\mathrm{corr}}^{2} = \kappa^{2}\,\overline{\epsilon^{2}}
\]
is the mean rms error for the
data points in the light curve.

The intrinsic fractional variability, which quantifies the source variability after removing the contribution of  measurement noise, was calculated following~\citet{Vaughan2003MNRAS.345.1271V} as
\begin{equation}
\label{eq:Fvar_new}
F_{\mathrm{var}} =
\sqrt{\frac{s_{\mathrm{obs}}^{2} - \overline{\epsilon^{2}}}{\langle f \rangle^{2}}},
\end{equation}
where $s_{\mathrm{obs}}^{2}$ is the sample variance of the flux measurements and $\langle f \rangle$ is the mean flux. The uncertainty associated with $F_{\mathrm{var}}$ was calculated using the prescription given by~\citet{Vaughan2003MNRAS.345.1271V}. For each of the seven RQ-NLSy1s, we list the full temporal baseline of the \emph{ZTF} light curves, the number of data points in each light curve, and all optical variability diagnostics ($F_{\mathrm{AGN}}$, variability status, $\psi_{\mathrm{pp}}$ and $F_{\mathrm{var}}$) in Table~\ref{tab:OP_IR_variability}. The quoted duration refers to the total observed time span of the light curves. However, Table~\ref{var_diff_time} summarizes, for each source, the number of variable epochs and the total number of available epochs on intranight, week-like, month-like and year-like timescales. In Fig.~\ref{fig: Intranight_optical_variability_1}, we present a selected set of light curves for each source, illustrating variability on intranight, week-like and month-like timescales. The term ``selected'' here indicates that only illustrative epochs are shown for visualization purposes and that these epochs are not necessarily representative of the mean variability amplitude, number of data points, or observational cadence of the complete dataset.  The quantitative variability analysis is instead based on all independent sub-epochs satisfying the data-quality criterion described in Sect.~\ref{results_op_mir_var}. The remaining light curves can be made publicly available upon request.

\subsubsection{Duty cycle of variability}
Since AGN variability is not continuous in time, we quantified its occurrence using the duty cycle (DC), which measures the fraction of time during which a source exhibits statistically significant variability. The DC was calculated following the formalism introduced by~\citet{Romero1999A&AS..135..477R} and subsequently applied by~\citet{Stalin2004JApA...25....1S}.\par
The monitoring intervals adopted for the DC calculation correspond to the intra-night, week-like, and month-like sub-epochs as defined in the second paragraph of  Sect.~\ref{results_op_mir_var}

The duty cycle is expressed as
\begin{equation}
\label{eq:DC_new}
\mathrm{DC} = 100 \,
\frac{\sum_{j=1}^{M} V_j (\Delta t_j)^{-1}}
{\sum_{j=1}^{M} (\Delta t_j)^{-1}}
\end{equation}
where $\Delta t_j = \Delta t_{\mathrm{obs},~j}/(1+z)$ is the rest-frame duration of the $j^{th}$ monitoring interval. The indicator $V_j$ is set to unity when variability is detected and zero otherwise.\par
 The estimated DC depends on the adopted interval length because the probability of detecting variability generally increases with the duration of the monitored timescale. Variability signatures that may remain undetected within short intra-night intervals can become statistically significant when examined over longer week-, month-, or year-like timescales. Consequently, the DC is expected to increase with increasing interval length, as also observed in our analysis, where substantially higher DC values are found on long-term timescales than on intra-night timescales (see Table~\ref{DC_diff_time}~\&~\ref{tab:OP_IR_variability}) The selected interval lengths therefore allow us to probe variability across distinct temporal regimes while maintaining sufficient temporal sampling and statistical robustness.

The resulting DC values, together with the mean values of $\psi_{\mathrm{pp}}$ ($\overline\psi_{\mathrm{pp}}$) in the $\emph{g}$, $\emph{r}$, and $\emph{i}$ bands for each RQ-NLSy1 galaxy for the investigated timescales, are tabulated in Table~\ref{DC_diff_time}. Note that only light curves classified as variable were included in the calculation of $\overline\psi_{\mathrm{pp}}$.

\subsection{Variability analysis in mid-infrared wavelength}
\label{section_3.2} 
The mid-infrared variability of the seven RQ-NLSy1s was quantified using the intrinsic variability amplitude, a metric commonly employed in MIR variability studies~\citep[e.g.][]{Rakshit2019MNRAS.483.2362R, Anjum2020MNRAS.494..764A, Wang2020RAA....20...21W}. This approach measures the scatter in the observed magnitudes after accounting for the measurement uncertainties~\citep{Sesar2007AJ....134.2236S, Ai2010ApJ...716L..31A}. We adopt this estimator for the MIR analysis because \emph{WISE} light curves are generally sparsely sampled and are affected by larger photometric and systematic uncertainties compared to optical data, which limits the robustness of multi-parameter variability estimators commonly applied at optical wavelengths. The adopted formalism explicitly accounts for measurement errors and incorporates systematic uncertainties associated with the \emph{WISE} \emph{W1} and \emph{W2} bands, as characterized by~\citet{Jarrett2011ApJ...735..112J} for the mission~\citep{Wright2010AJ....140.1868W}. This approach has been widely used to reliably quantify intrinsic MIR variability in AGN and related sources~\citep[e.g.][]{Sesar2007AJ....134.2236S, Ai2010ApJ...716L..31A, Rakshit2019MNRAS.483.2362R}. Therefore, the use of this estimator ensures a homogeneous and robust assessment of MIR variability and facilitates a direct comparison with previous studies. \par

The observed dispersion of the MIR light curve is given by
\begin{equation}
\label{eq:sigma_MIR}
\sigma_{\mathrm{MIR}} =
\sqrt{\frac{1}{N-1}
\sum_{i=1}^{N}
(m_i - \langle m \rangle)^2}
\end{equation}
where $m_i$ denotes the magnitude at $i^{th}$ epoch and $\langle m \rangle$ is the weighted mean magnitude.

The intrinsic variability amplitude is then defined as
\begin{equation}
V_{\mathrm{int}} =
\begin{cases}
\sqrt{\sigma_{\mathrm{MIR}}^{2} - \sigma_{\mathrm{err}}^{2}}, & \sigma_{\mathrm{MIR}} > \sigma_{\mathrm{err}} \\
0, & \text{otherwise}
\end{cases}
\end{equation}
where
\[
\sigma_{\mathrm{err}}^{2} =
\frac{1}{N}\sum_{i=1}^{N}\delta_i^{2} + \delta_{\mathrm{sys}}^{2}
\]
combines individual photometric uncertainties $\delta_i$ with systematic errors $\delta_{\mathrm{sys}}$, adopted to 0.024 mag for \emph{W1} and 0.028 mag for \emph{W2}~\citep{Jarrett2011ApJ...735..112J}.

Assuming independent and normally distributed measurements, the uncertainty in $\sigma_{\mathrm{MIR}}$ can be approximated to $\sigma_{\mathrm{MIR}}/\sqrt{2(N-1)}$~\citep[e.g., see][]{Ojha2025ApJ...994...84O}. To correct for the redshift of the object, we computed the rest-frame $V_{\mathrm{mz}}$ by multiplying $V_{\mathrm{int}}$ with $\sqrt{1+z}$ following~\citet{Rakshit2019MNRAS.483.2362R}

\begin{equation}
\label{eq:mir_variability}
V_{\mathrm{mz}} = V_{\mathrm{int}} \sqrt{1+z}.
\end{equation}

A source was considered variable in the MIR if $V_{\mathrm{mz}} \gtrsim 0.1$. The MIR variability amplitudes of the rest-frame and their uncertainties for individual sources in \emph{WISE} \emph{W1} and \emph{W2} are tabulated in Table~\ref{tab:OP_IR_variability}.

\begin{table*}
\caption{Variability statistics for the seven radio-quiet narrow-line Seyfert 1 in the optical \emph{g}, \emph{r}, and \emph{i} bands, along with in the mid-infrared \emph{W1} and \emph{W2} bands.}
    \centering
    \resizebox{\textwidth}{!}{
    \begin{tabular}{ccccccc c|ccc} 
    \hline
    \multicolumn{8}{c|}{Optical wavelength} &  \multicolumn{3}{c}{Mid-infrared wavelength} \\ \hline
      RQ-NLSy1 name  &   \emph{ZTF}    & Time & Light&F$_{AGN}$  &Status & $\psi_{\mathrm{pp}}\pm\Delta(\psi_{\mathrm{pp}})$ & $F_{\mathrm{var}}\pm\Delta(F_{\mathrm{var}})$ & \emph{WISE} &$V_{mz}\pm\Delta(V_{mz})$  & Status\\
                    & band & (in days)& {curve} & values             & \multicolumn{1}{c}{type}    &             &           &  band    &  &type  \\
                   & &    & dpts$^{\dagger}$ &  &     &             &           &      &   \\               
      \hline 
J102906.69+555625.2 & g & 2259.963 &  610 &  4.650 & V  & 1.026$\pm$0.087 & 0.858$\pm$0.018 &   &  &     \\                    
J102906.69+555625.2 & r & 2244.980 &  927 &  3.056 & V  & 0.818$\pm$0.084 & 0.623$\pm$0.014 &   &  &     \\                    
J122844.81+501751.2 & g & 2285.926 &  675 &  1.353 & V  & 0.337$\pm$0.046 & 0.146$\pm$0.009 & W1 &0.015$\pm$0.009& NV\\        
J122844.81+501751.2 & r & 2266.055 &  754 &  1.968 & V  & 0.262$\pm$0.033 & 0.178$\pm$0.007 & W2 &0.000$\pm$0.000& NV\\        
J122844.81+501751.2 & i & 1737.042 &  217 &  1.642 & V  & 0.234$\pm$0.031 & 0.142$\pm$0.012 &    &     &\\                     
J123220.11+495721.8 & g & 2285.926 &  695 & 35.168 & V  & 0.840$\pm$0.028 & 0.907$\pm$0.006 & W1 &0.139$\pm$0.006& V\\         
J123220.11+495721.8 & r & 2266.055 &  760 & 31.541 & V  & 0.791$\pm$0.026 & 0.809$\pm$0.005 & W2 &0.130$\pm$0.006& V\\         
J123220.11+495721.8 & i & 1737.042 &  221 & 22.238 & V  & 0.558$\pm$0.027 & 0.708$\pm$0.010 &     &     &\\                    
J150916.18+613716.7 & g & 2286.953 & 1948 &  3.682 & V  & 0.872$\pm$0.077 & 0.656$\pm$0.009 &  W1 &0.072$\pm$0.006& V\\       
J150916.18+613716.7 & r & 2292.040 & 2161 &  3.319 & V  & 0.537$\pm$0.050 & 0.403$\pm$0.006 &  W2 &0.080$\pm$0.018& V\\       
J150916.18+613716.7 & i & 2205.954 &  755 &  3.382 & V  & 0.411$\pm$0.039 & 0.402$\pm$0.008 &     &     &\\                    
J151020.06+554722.0 & g & 2287.829 & 1098 &  2.304 & V  & 0.609$\pm$0.064 & 0.384$\pm$0.010 &     &     &\\                    
J151020.06+554722.0 & r & 2284.894 & 1145 &  6.695 & V  & 0.519$\pm$0.033 & 0.435$\pm$0.005 &     &     &\\                    
J151020.06+554722.0 & i & 1638.737 &  380 &  9.256 & V  & 0.465$\pm$0.026 & 0.426$\pm$0.008 &     &     &\\                    
J152205.41+393441.3 & g & 2286.829 & 1003 &  9.183 & V  & 0.529$\pm$0.026 & 0.422$\pm$0.005 &     &     &\\                    
J152205.41+393441.3 & r & 2285.922 & 1029 & 28.898 & V  & 0.500$\pm$0.015 & 0.484$\pm$0.003 &     &     &\\                    
J152205.41+393441.3 & i & 1767.048 &  235 & 32.125 & V  & 0.514$\pm$0.016 & 0.546$\pm$0.006 &     &     &\\                    
J164100.10+345452.7 & g & 2285.864 & 1345 &  4.281 & V  & 0.510$\pm$0.039 & 0.382$\pm$0.006 &  W1  &0.072$\pm$0.004& V\\      
J164100.10+345452.7 & r & 2290.946 & 1387 & 12.938 & V  & 0.438$\pm$0.020 & 0.391$\pm$0.003 &  W2  &0.064$\pm$0.003& V\\      
J164100.10+345452.7 & i & 2210.971 &  229 & 14.877 & V  & 0.383$\pm$0.018 & 0.396$\pm$0.007 &     &                    &\\

\hline
\multicolumn{10}{l}{$^{~~}$Data is not available. $^{\dagger}$Number of data points in the optical band light curve. \emph{V:} Variable; \emph{NV:} Non-variable.}\\
\end{tabular}
}
\label{tab:OP_IR_variability}
\end{table*}

 \begin{table*}
  \caption{Number of optically variable epochs and total number of epochs of seven radio-quiet narrow-line Seyfert 1 galaxies on different time scales.}
\label{var_diff_time}
\resizebox{\textwidth}{!}{
\begin{tabular}{ccccc}
 \hline
    Sources             & Intranight     & Week-like           & Month-like         & Year-like  \\ \hline
    J102906.69+555625.2 & NO (g, r)      & NO (g, r)           & YES (g[1/41], r[0/41])          & YES (g[1/1], r[1/1])   \\
    J122844.81+501751.2 & NO (g, r, i)   & NO (g, r, i)        & YES (g[0/41], r[0/35], i[1/21]) & YES (g[1/1], r[1/1], i[1/1]) \\
    J123220.11+495721.8 & NO (g, r, i)   & NO (g, r, i)        & YES (g[2/41], r[1/35], i[0/21]) & YES (g[1/1], r[1/1], i[1/1]) \\
    J150916.18+613716.7 & NO (g, r, i)   & YES (g[1/74], r[4/76], i[0/20])        & YES (g[1/48], r[3/49], i[2/26])  & YES (g[1/1], r[1/1], i[1/1]) \\
    J151020.06+554722.0 & NO (g, r, i)   & YES (g[3/50], r[14/50], i[6/16])        & YES (g[6/43], r[23/39], i[6/22]) & YES (g[1/1], r[1/1], i[1/1]) \\
    J152205.41+393441.3 & YES (g[1/10], r[7/13]) & YES (g[11/47], r[40/48], i[5/6]) & YES (g[35/44], r[37/42], i[16/21]) & YES (g[1/1], r[1/1], i[1/1])  \\
    J164100.10+345452.7 & YES (g[3/13], r[4/20])     & YES (g[6/75], r[29/71], i[2/3])          & YES (g[11/49], r[43/53], i[15/25]) & YES (g[1/1], r[1/1], i[1/1])  \\
    \hline
    \multicolumn{5}{l}{'[]' represents [variable sessions/total number of sessions]}\\
\end{tabular}
}
 \end{table*}

 \begin{table*}
  \caption{Duty cycle (DC) and mean peak-to-peak amplitude of variability ($\overline\psi_{pp}$) of seven radio-quiet narrow-line Seyfert 1 galaxies on different time scales and in different bands.}
\label{DC_diff_time}
\resizebox{\textwidth}{!}{
\begin{tabular}{cccc}
 \hline
 Sources             & Intranight(DC(\%)/$\overline\psi_{pp}$)     & Week-like (DC(\%)/$\overline\psi_{pp}$)  & Month-like (DC(\%)/$\overline\psi_{pp}$)    \\ 
                        \hline
    J102906.69+555625.2 & 0\% (g, r)      & 0\% (g, r)           &  (g[2.02/69.40], r[0/0])           \\
    J122844.81+501751.2 & 0\% (g, r, i)   & 0\% (g, r, i)        &  (g[0/0], r[0/0], i[3.96/25.10])  \\
    J123220.11+495721.8 & 0\% (g, r, i)   & 0\% (g, r, i)        &  (g[4.37/16.90], r[2.35/16.30], i[0/0])  \\
    J150916.18+613716.7 & 0\% (g, r, i)   &  (g[1.40/45.00], r[5.35/31.30], i[0/0])                 &  (g[1.96/30.60], r[5.70/29.17], i[7.05/20.45])   \\
    J151020.06+554722.0 & 0\% (g, r, i)   &  (g[7.31/47.23], r[28.43/25.41], i[37.04/19.71])        &  (g[13.34/43.27], r[55.06/27.05], i[26.65/21.52])  \\
    J152205.41+393441.3 &  (g[10.91/63.00], r[56.05/17.87]) &  (g[23.79/24.68], r[84.44/20.48], i[78.23/21.08])       &  (g[75.66/26.67], r[87.39/25.30], i[78.17/27.88])   \\
    J164100.10+345452.7 &  (g[20.27/37.57], r[14.60/26.68]) &  (g[8.58/35.57], r[39.92/19.01], i[61.17/12.00])        &  (g[21.49/33.76], r[79.69/18.86], i[57.21/20.45])   \\
    \hline
    \multicolumn{4}{l}{'[]' represents [DC(\%)/$\overline\psi_{pp}$], where $\overline\psi_{pp}$ was computed by considering only the $\psi_{pp}$ values of variable sub-epochs,  computed following Eq.~\ref{eq:Amp_pp_new}.}\\
\end{tabular}
}
 \end{table*}

\subsection{Color behavior}
\label{Color_variability}
The optical radiation detected from AGNs is generally understood as a superposition of two physically distinct components: quasi-thermal emission associated with the accretion disk and non-thermal synchrotron emission produced by relativistic jets. Investigating color variability, therefore, provides a useful diagnostic in probing the dominant contributions between these components to the observed flux. In a similar context, infrared emission, which primarily traces optical and ultraviolet radiation reprocessed by circumnuclear dust, offers complementary insight into the coupling between thermal and non-thermal processes in AGNs~\citep[see, e.g.,][]{Storchi-Bergmann1992ApJ...395L..73S, Lu2016MNRAS.458..575L}. Since AGNs can exhibit variability on hour-like time scales, this can bias color measurements if non-simultaneous data are combined. To ensure a reliable characterization of spectral variations, we therefore required observations in different bands to be quasi-simultaneous. Following the criterion adopted by \citet{Ojha2025ApJ...994...84O}, only data points obtained within a 30-minute interval in the relevant bands were used for the color analysis, thus minimizing the influence of intrinsic flux variability on the derived colors. In addition, we required at least five quasi-simultaneous measurements per source to ensure statistical robustness. The same selection criteria were consistently applied to the mid-infrared data. Consequently, the optical and MIR color variability analysis presented here is based on a subset of available light-curve data, restricted to observations that satisfy these temporal and sampling constraints.  For the optical data, this requirement results in the use of approximately 6–10\% of the total observations, depending on the source, whereas nearly all available WISE measurements satisfy the adopted criteria and are therefore included in the MIR analysis. Although the temporal matching requirement reduces the number of optical data points, it minimizes artificial color variations caused by intrinsic source variability between non-simultaneous observations. Given that some of these sources have been found to exhibit significant intranight variability~\citep[see ][]{Ojha2024MNRAS.529L.108O}, using full optical light curves without such constraints could lead to misleading color variability measurements. Note that INOV was detected in at least one observing epoch for all sources except J150916.17+613716.6~\citep{Ojha2024MNRAS.529L.108O}.

Given that these sources are known to exhibit significant intranight variability~\citep[see ][]{Ojha2024MNRAS.529L.108O}, using the full optical light curves without such constraints could lead to misleading color variability measurements.\par
For the seven RQ-NLSy1s that meet the above criterion, we constructed color–magnitude diagrams using quasi-simultaneous data in the optical and MIR bands. Specifically, we examine the relations $\emph{g-r}$ versus $\emph{r}$, $\emph{r-i}$ versus $\emph{i}$, and $\emph{W1-W2}$ versus $\emph{W2}$. In these diagrams, the magnitude at the longer wavelength was initially placed on the abscissa, whereas the color index, defined as the difference between the magnitudes at the shorter and longer wavelengths, was plotted on the ordinate. This choice ensures a uniform approach across the optical and MIR wavelengths. However, during the analysis, we found that the inferred color–magnitude trend can reverse when the shorter-wavelength (bluer) magnitude is used on the abscissa, an effect that is particularly noticeable in the color-magnitude diagrams $\emph{g-r}$ versus $\emph{r}$ and $\emph{g-r}$ versus $\emph{g}$ (see Fig.~\ref{fig: OP_color_variability_gr_1_RQ-NLSy1}). To account for this behavior and to test the robustness of the trends, we therefore constructed three alternative representations for each source: $(m_1 - m_2)$ versus $m_2$, $(m_1 - m_2)$ versus $m_1$, and $(m_1 - m_2)$ versus $(m_1 + m_2)$. Representative examples of these long-term color–magnitude relations in optical and MIR wavelengths are shown in Figs.~\ref{fig: OP_color_variability_gr_1_RQ-NLSy1},~\ref{fig: OP_color_variability_ri_1_RQ-NLSy1}, ~\ref{fig: IR_color_variability_w1w2_RQ-NLSy1}. For each diagram, the dependence between color and magnitude was quantified using orthogonal distance regression, which accounts for uncertainties in both variables. In addition, the Pearson correlation coefficient, $\rho_r$, was calculated to characterize the strength and direction of the correlation.\par
To assess the presence of systematic color trends, we classified sources based on the value of the Pearson rank correlation coefficient ($\rho_r$). A positive correlation with $\rho_r \geq 0.5$ was interpreted as a BWB trend, while a negative correlation with $\rho_r \leq -0.5$ was taken to indicate a redder-when-brighter (RWB) trend. Sources with intermediate values of $-0.5 < \rho_r < 0.5$ were considered to show no statistically significant color–magnitude correlation (denoted as NOT). All color–magnitude representations from optical to MIR are presented in Figs.~\ref{fig: OP_color_variability_gr_1_RQ-NLSy1},~\ref{fig: OP_color_variability_ri_1_RQ-NLSy1},~\ref{fig: IR_color_variability_w1w2_RQ-NLSy1}, and in the left corner of each sub-panel, the best-fit slope and $\rho_r$ (Pearson-$\rho$) values are shown for each source.

\subsection{Lag Measurement}
\label{Lag_measurement}
The multi-wavelength light curves exhibit pronounced variability and a high degree of coherence, with recurring features such as maxima and minima appearing at the corresponding epochs. This strong correlation enables a reliable estimation of intra-band and inter-band time delays. We therefore computed the intra-band and inter-band lags using the interpolated cross-correlation function (ICCF) in conjunction with flux randomization and random subset sampling (FR/RSS), following the implementation of~\citet{Peterson2004ApJ...613..682P}. For optical  intra-band lag measurements, the \emph{ZTF} light curves were binned using a \emph{1-}day window and subsequently used for the ICCF and FR/RSS analysis.  The 1-day binning was adopted to reduce the effect of uneven sampling and photometric scatter while preserving the short-timescale variability relevant for inter-band lag measurements. For optical-infrared lag measurements, we adopted a \emph{180-}day binning window for the \emph{WISE} light curves and a \emph{10-}day binning window for the \emph{ZTF} light curves.  The larger binning window for the \emph{WISE} data was motivated by its sparse sampling pattern, where observations are clustered within individual observing epochs separated by approximately six months. This binning improves the signal-to-noise ratio and provides more reliable cross-correlation measurements between the optical and infrared bands.
All optical intra-band lags were measured with respect to the $\emph{g}$ band, as it lies closest to the thermal emission peak of the accretion disk among the \emph{g}, \emph{r}, and \emph{i} bands. The time lag was derived from the centroid of the cross-correlation function (CCF), considering only correlation coefficients that exceeded 80\% of the peak value. Uncertainties on the lag measurements were quantified using the FR/RSS method. For each band pair, 1000 FR/RSS realizations were generated by randomly resampling the light curves with replacement and perturbing the fluxes according to their measurement uncertainties. The centroid lag was measured for each realization, yielding a centroid cross-correlation distribution (CCCD). The final lag was taken as the median of the CCCD, with the 1$\sigma$ uncertainties defined by the 16$^{th}$ and 84$^{th}$ percentiles of the distribution. For optical-optical lag analysis, the CCF was evaluated in a lag range of -30 to 30 days. In the case of optical-MIR lags,  we adopted a broader lag search range from -200 to 600 days. For estimating lags between optical and MIR wavelengths, the \emph{ZTF} \emph{r}-band light curve was adopted as the reference because it contains the largest number of data points among the optical bands. In addition, lags between the \emph{WISE} \emph{W1} and \emph{W2} bands were measured with a lag search range of -600 to 600 days. Lag measurements for a representative RQ-NLSy1, J122844.81+501751.2, are shown in Figs.~\ref{fig: Lag_optical_bands_J1232},~\ref{fig: Lag_optical_MIR_bands_J1232}, while the corresponding plots for the remaining sources are presented in the Appendix (see Figs.~\ref{fig: Lag_optical_bands_A1} to~\ref{fig: Lag_optical_bands_A3}).

\subsection{Broadband Spectral Energy Distribution}
\label{SED_Modeling}
Modeling the broadband SED helps us to understand the main physical mechanism responsible for the emission. We have compiled the spectral points from all available bands for all objects in our sample. All sources are detected in near-IR, optical–UV, and X-rays. The reduction and analysis of the two datasets that provide the most diagnostic information, namely the \emph{Swift}-XRT and \emph{Fermi}-LAT observations, are described in detail in Sects.~\ref{Xray_data} and \ref{LAT_analysis}, respectively, while the SED modelling procedure is described in Sect.~\ref{jetset_modelling}.

\subsubsection{\emph{Swift}-XRT observations and spectral analysis}
\label{Xray_data}
All seven RQ-NLSy1s have been observed at least once with \emph{Swift}-XRT~\citep{2005SSRv..120..165B} in photon-counting mode. Table~\ref{tab_xrt_log} in Appendix~\ref{app-xrt} lists the observation identifiers, observation dates, the number of observations combined for each target, and the corresponding total on-source exposures.
The spectra were generated using the standard automatic products generator hosted by the UK \emph{Swift} Science Data Centre (the ``Build \emph{Swift}-XRT products\footnote{\url{https://www.swift.ac.uk/user_objects/index.php}}’’ interface;~\citealt{Evans2007A&A...469..379E,2009MNRAS.397.1177E}), which performs event screening, selection of the source and background extraction regions, assessment and correction of pile-up, and generation of exposure maps, ancillary response files, and grouped spectra following a documented and reproducible procedure. When more than one observation is available for a target, the tool combines them into a single time-averaged spectrum, which this study uses.

The products generator also performs a default spectral fit using an absorbed power-law model, with the absorbing column fixed to the Galactic value along the line of sight. The resulting photon indices, $\Gamma_{\rm X}$, and observed 0.3–10keV fluxes are listed in Table~\ref{tab_xrt_log} of Appendix~\ref{app-xrt}. We stress that these fitted parameters are \emph{not} used to constrain the broadband modelling. The SEDs shown in Fig.~\ref{fig: sed_figures} include the unfolded spectral data points returned by the tool, and these points, rather than the fitted power-law parameters, are used in the \texttt{JETSET} fits.

\subsubsection{\emph{Fermi}-LAT data reduction and analysis}
\label{LAT_analysis}
We analysed $\sim$17.9~yr of \emph{Fermi}-LAT data (mission elapsed times 239557417--801878405, i.e. 2008 August 4 to 2026 June 16), retrieved from the \emph{Fermi} Science Support Center data server on 2026 June 19 and processed with \texttt{Fermipy v1.3}~\citep{2017ICRC...35..824W}. We selected Pass~8 \texttt{SOURCE}-class events (\texttt{evclass=128}, \texttt{evtype=3}) between 100~MeV and 500~GeV, applying a zenith-angle cut of $<90^{\circ}$ and the standard good-time-interval filter $\mathrm{DATA\_QUAL}>0$ \&\& $\mathrm{LAT\_CONFIG}==1$, and adopted the \texttt{P8R3\_SOURCE\_V3} instrument response functions with energy dispersion enabled for all components except the two diffuse templates. The region of interest (ROI) is a $20^{\circ}\times20^{\circ}$ square centred on the optical (SDSS) position of each target, binned with $0.1^{\circ}$ pixels and eight logarithmic energy bins per decade. The background model comprises the Galactic and isotropic templates \texttt{gll\_iem\_v07.fits} and \texttt{iso\_P8R3\_SOURCE\_V3\_v1.txt}, together with all sources within $10^{\circ}$ of the ROI centre taken from 4FGL-DR4~\citep{2022ApJS..260...53A,2023arXiv230712546B} with their catalogued spectral shapes and parameters. None of our targets is present in 4FGL-DR4, so each was added as a point source at its optical position with a power-law spectrum, $dN/dE\propto E^{-\Gamma_{\gamma}}$, with normalisation and $\Gamma_{\gamma}$ as free parameters.

After a preliminary optimization with command \texttt{optimize()}, we freed the normalisations of the catalogued sources within $3^{\circ}$ of the target, the two diffuse normalisations, and both spectral parameters of the target, while fixing all the parameters of sources beyond $3^{\circ}$ at their 4FGL-DR4 values. A binned maximum-likelihood fit over the full energy range yields a Test Statistic, ${\rm TS}= 2(\ln\mathcal{L}_{1}-\ln\mathcal{L}_{0})$, above 25 for all six SED-modelled sources. The $\gamma$-ray spectral points in Fig.~\ref{fig: sed_figures} were obtained with the command \texttt{sed()}, which refits the target normalisation in each of nine logarithmic energy bins with the in-bin shape fixed to a power law of index~2 and all other parameters fixed at their global best-fit values. Bins reaching ${\rm TS} > 25$ are plotted as spectral points, while the remainder are shown as 95\% confidence upper limits, which are not used to constrain the models.

\subsubsection{Broadband SED modelling with \texttt{JETSET}}
\label{jetset_modelling}
Since the standard thermal AGN components, namely the accretion disk, X-ray corona, and dusty torus, are not expected to produce GeV emission, a significant $\gamma$-ray detection at the source position necessarily implies the presence of an additional non-thermal inverse-Compton component. Furthermore, the overall broadband SED shapes resemble those of blazars and $\gamma$-NLSy1 galaxies, suggesting a similar underlying physical origin. This motivated us to model the broadband SEDs by including non-thermal jet components.

We modeled the broadband SED with a publicly available leptonic code, \texttt{JETSET}\footnote{https://jetset.readthedocs.io/en/1.3.0/index.html}.
The models presented here comprise a multicolour Shakura--Sunyaev accretion disk together with a one-zone leptonic jet that produces synchrotron, synchrotron self-Compton (SSC), and external Compton (EC) emission through interactions with the disk and BLR photon fields.

In our models, the XRT band falls in the low-flux valley between the synchrotron and inverse-Compton humps, where the jet contribution is least well constrained; the SSC component provides a good description of the X-ray points. The GeV detections alone establish the need for a jet-related non-thermal component in these objects. The X-ray points constrain the relative strength of the two SED humps only if the X-ray emission is assumed to be jet-dominated, and the corresponding SSC parameters should therefore be regarded as upper limits on the jet contribution in this band.

\begin{figure*}
    \begin{minipage}[]{1.0\textwidth}

\includegraphics[width=0.33\textwidth,height=0.24\textheight,angle=00]{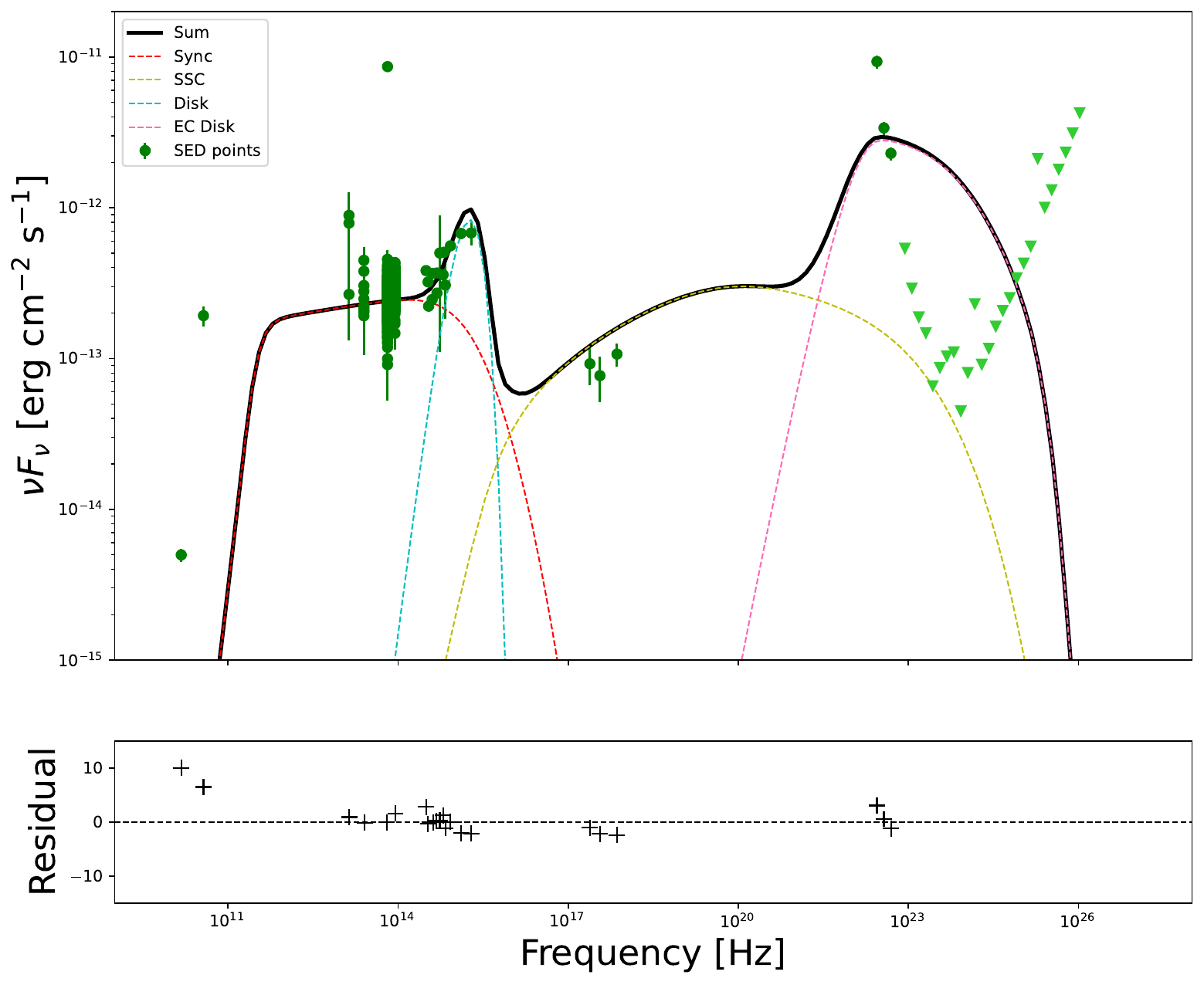}  
\includegraphics[width=0.33\textwidth,height=0.24\textheight,angle=00]{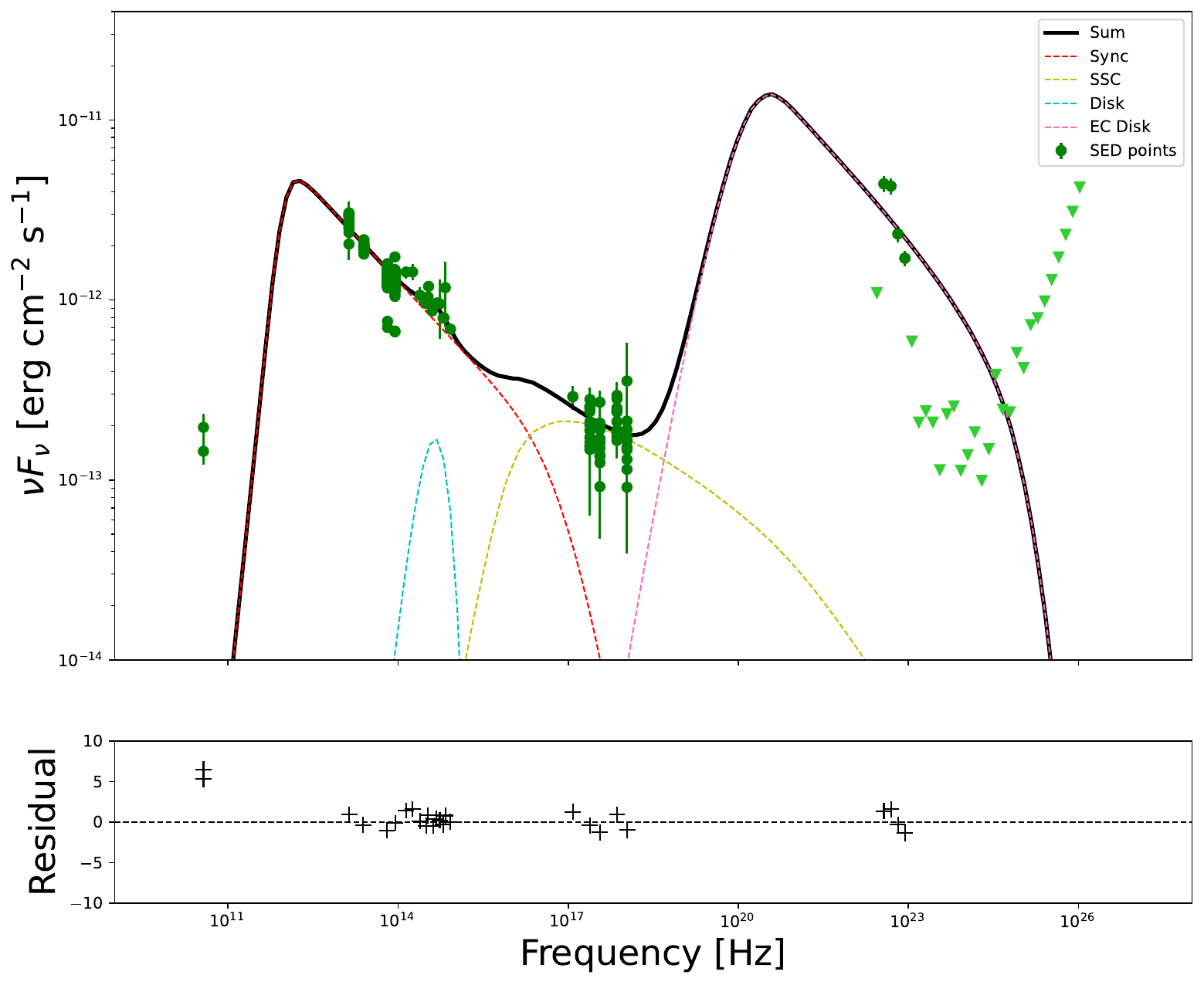}   
\includegraphics[width=0.33\textwidth,height=0.24\textheight,angle=00]{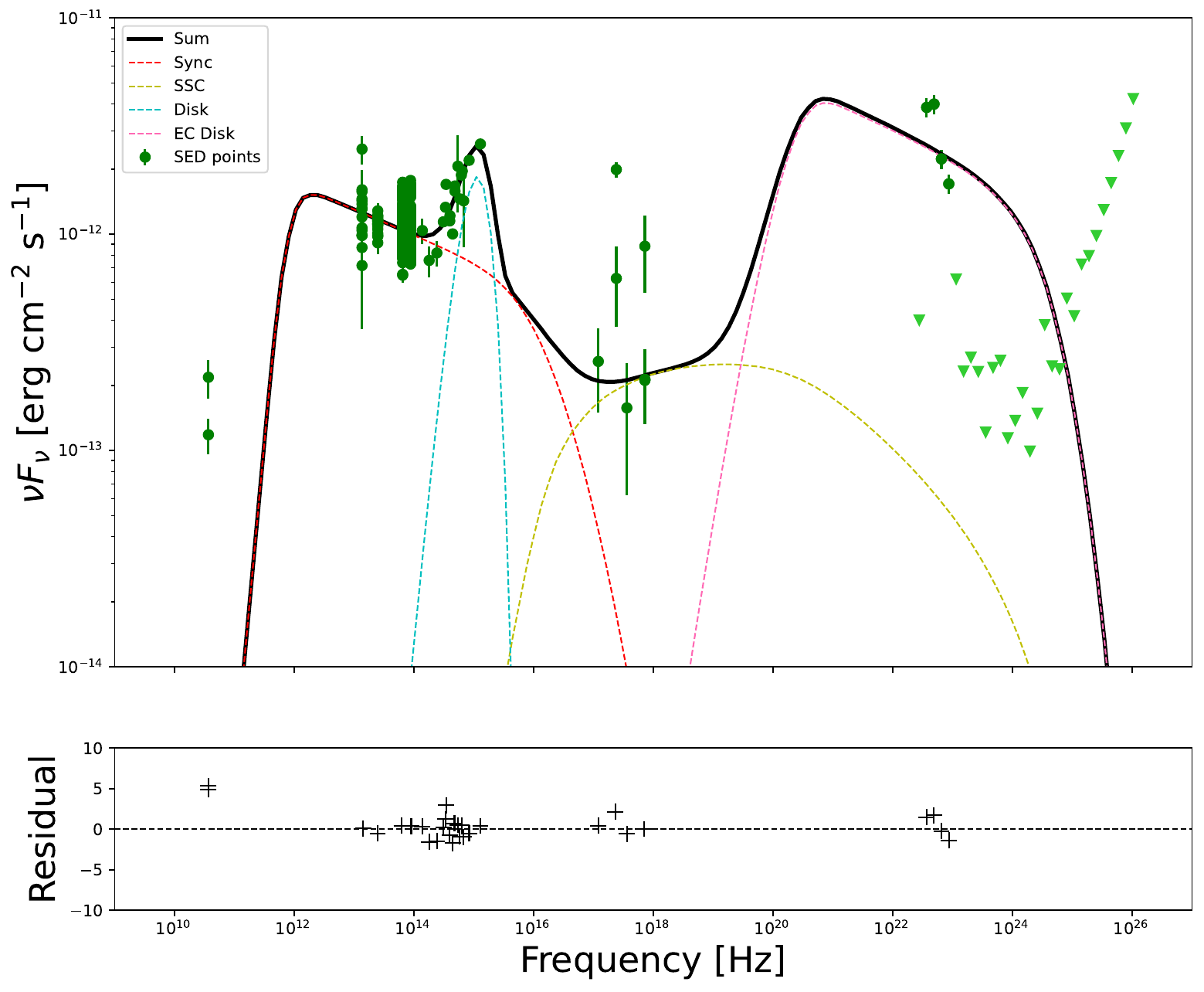}  
\includegraphics[width=0.33\textwidth,height=0.24\textheight,angle=00]{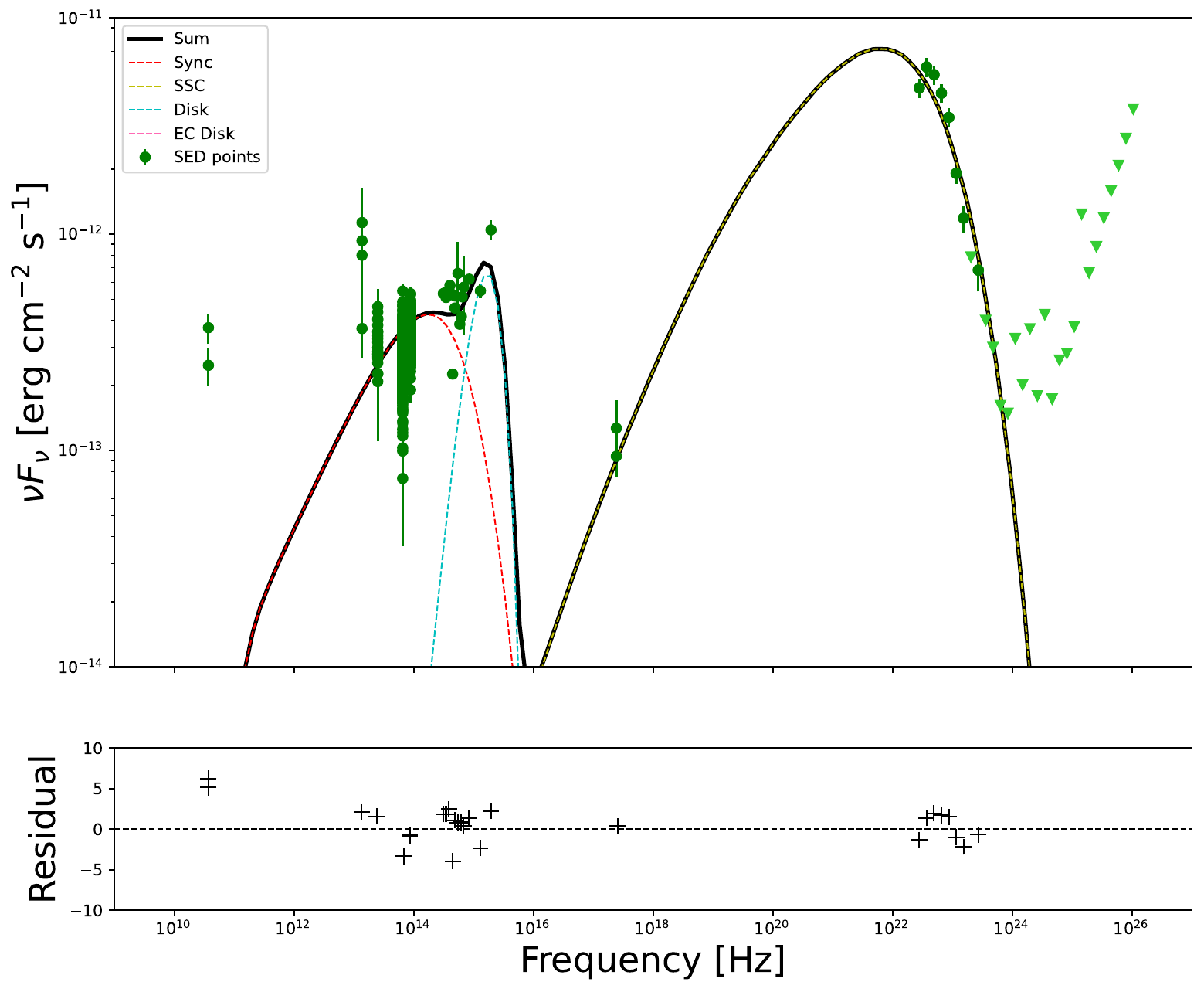}  
\includegraphics[width=0.33\textwidth,height=0.24\textheight,angle=00]{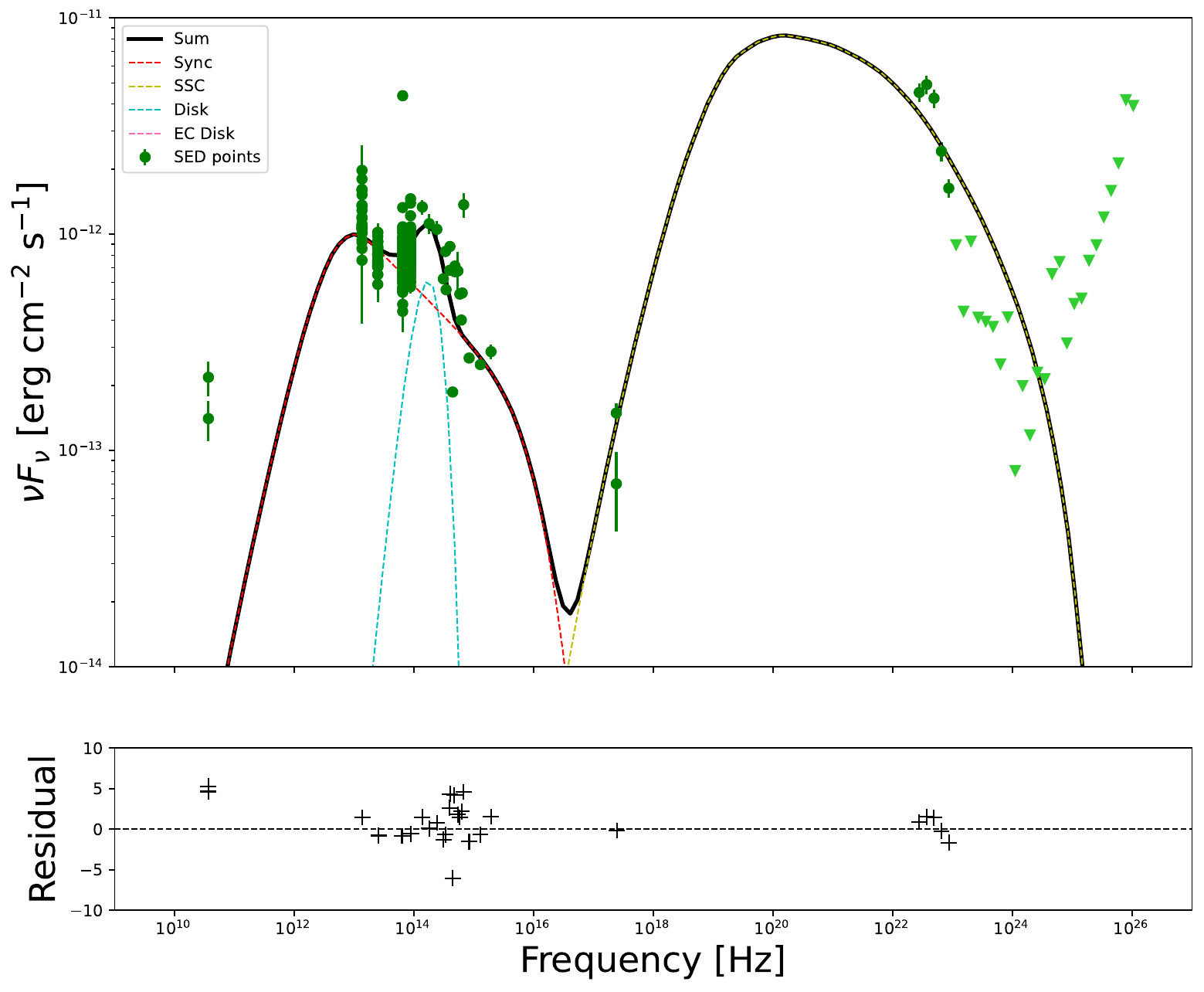}  
\includegraphics[width=0.33\textwidth,height=0.24\textheight,angle=00]{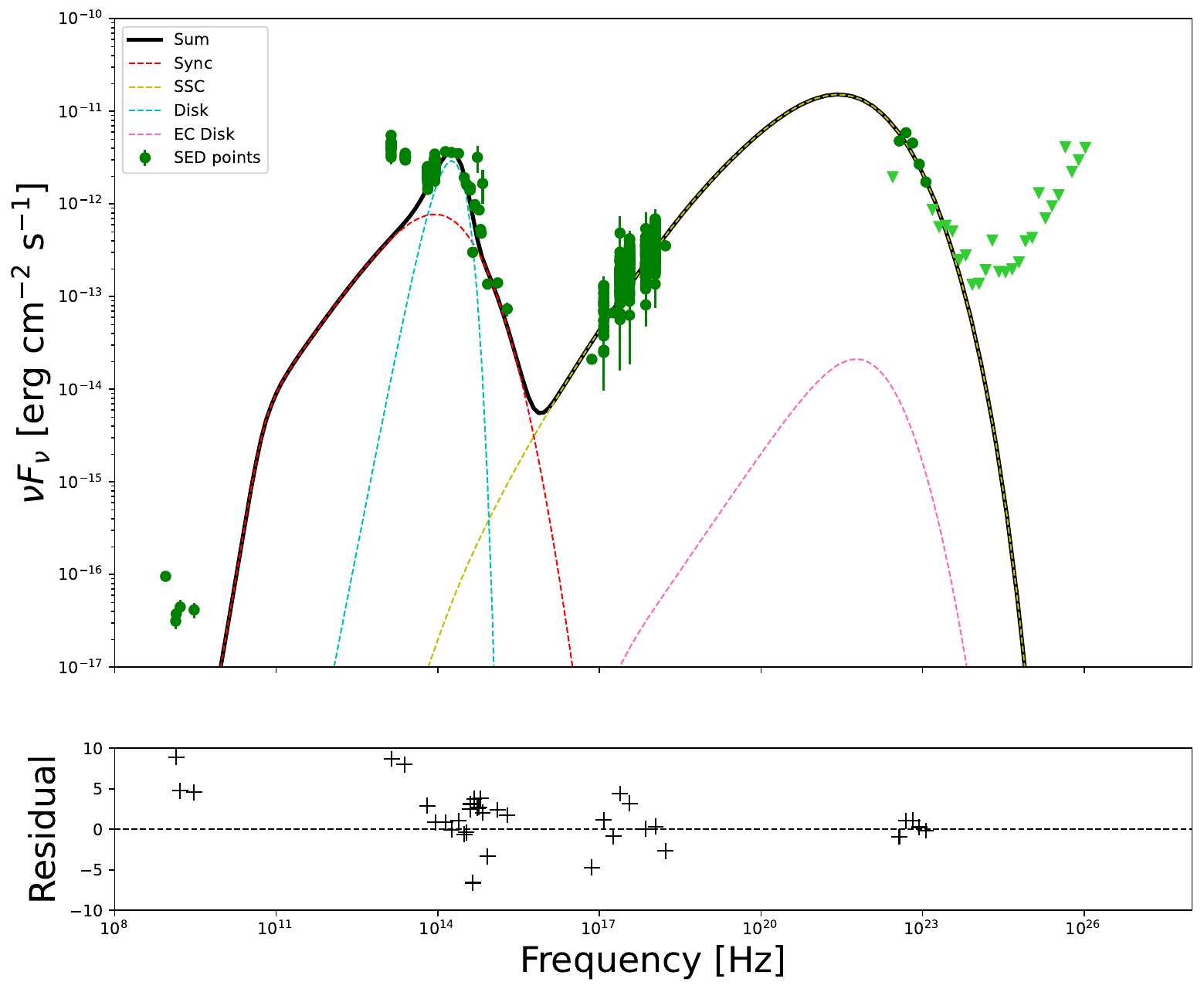}  
    \end{minipage}
    \caption{Spectral Energy Distribution (SED) of six RQ-NLSy1s, and all are modeled with \texttt{JETSET}, \emph{Top left:}{SDSS J102906.69$+$555625.2}, \emph{Top middle:}{SDSS J122844.81$+$501751.2}, \emph{Top right:}{SDSS J123220.11$+$495721.8}, \emph{Bottom left:}{SDSS J150916.18$+$613716.7}, \emph{Bottom middle:}{SDSS J151020.06$+$554722.0} \emph{Bottom right:}{SDSS J164100.10$+$345452.7}. The best fit parameters are shown in Table~\ref{tab_SED_Modeling_data}. All model components contributing within the plotted flux range are shown. \emph{Sync} (red dashed) denotes the jet synchrotron emission; \emph{SSC} (yellow dashed), the synchrotron self-Compton emission; \emph{Disk} (cyan dashed), the thermal multicolour accretion-disk continuum; \emph{EC Disk} (magenta dashed), the external Compton emission from up-scattered accretion-disk photons; and \emph{Sum} (solid black), the total model compared with the data. The external Compton component associated with the BLR photon field (EC~BLR) is included in the fits of all six sources but is negligible in every case, and therefore lies below the plotted range and is not shown in the legend. Filled green circles represent the SED data points; downward green triangles represent \emph{Fermi}-LAT 95\% confidence upper limits, which are shown for illustration only and were not used in the fits. A dusty-torus component and an X-ray coronal component are \emph{not} included in these models; the reasons for their omission and the implications for the interpretation of the X-ray band are discussed in Sect.~\ref{jetset_modelling}. The lower panels show the residuals, $(F_{\rm obs}-F_{\rm model})/\sigma_{\rm obs}$ of the fit.}
    \label{fig: sed_figures}
\end{figure*}

For clarity, the component acronyms used in the legend of Fig.~\ref{fig: sed_figures} are defined as follows: \emph{Sync} denotes the synchrotron emission from the jet; \emph{SSC} denotes the synchrotron self-Compton emission; \emph{Disk} denotes the thermal multicolour accretion-disk continuum; \emph{EC Disk} denotes the external Compton emission produced by up-scattering of accretion-disk photons; and \emph{Sum} denotes the sum of all components, which is the curve compared with the data. The lower panel of each figure shows the residuals, $(F_{\rm obs}-F_{\rm model})/\sigma_{\rm obs}$, of the fit.

Fig.~\ref{fig: sed_figures} shows the broadband SED fit, and Table~\ref{tab_SED_Modeling_data} lists the corresponding parameters. Out of \emph{7} objects in our sample, only \emph{6} have available broadband SED data, and hence the modeling is done only for those objects. The BLR external photon field was also included for all sources, but in every case its contribution was negligible relative to that of the disk photon field, so that the EC~BLR curve lies below the plotted range; the high-energy hump is produced by SSC and EC-disk alone (see Fig.~\ref{fig: sed_figures}).
For modeling, a logarithmic parabola (lp) was used with a low-energy power-law branch particle distribution with minimum and maximum energy between $\gamma_{min}$ and $\gamma_{max}$, and has $r$ and $s$ parameters as the low-energy spectral index and spectral curvature. The parameter $\gamma_{0-lp}$ is the reference energy~\citep{2006A&A...448..861M,2025ApJ...991...78X}.
In most cases, $\gamma_{min}$ is found to be close to \emph{50} and $\gamma_{max}$ close to \emph{10$^{6}$}, which is very commonly seen in blazars, indicating that the particle energy distribution in the emitting region is similar to that of blazar jets. This similarity refers to the physical conditions in the emitting region rather than the jet power: the GeV emission is detected only after integrating over $\sim$17.9 yrs, is not significant in 4FGL-DR4, and no persistent relativistic radio jet has been detected\citep{Berton2020A&A...636A..64B,Jarvela2024MNRAS.532.3069J}, suggesting jets of low power and/or low duty cycle. The size of the emission region $R$ is kept free in all cases. In an ideal situation, $R$ can be constrained by the variability time scale. The size of the emission region in all cases is found to be on the order of \emph{10$^{15}$} cm, suggesting a compact emission region. For a conical jet with a half-opening angle of $\sim1/\Gamma$, the jet cross-sectional radius at a distance $R_H$ from the black hole is approximately $R_H/\Gamma$. For the fitted parameters ($\Gamma\approx5$–$24$ and $R_H\approx10^{16}$–$10^{18}$ cm), this yields a characteristic jet radius of $\sim6\times10^{14}$–$2\times10^{17}$ cm. The fitted emission-region size, $R\sim10^{15}$ cm, is comparable to or smaller than this value in most sources, supporting a scenario in which the radiating region is a compact dissipation zone embedded within the conical jet.
Similarly, we also optimized the location of the emission region ($R_H$). In our modelling, the BLR radius was fixed at $R_{\rm BLR}=10^{17}$ cm ($\approx0.03$ pc), consistent with the BLR sizes expected for these low-luminosity, low-black-hole-mass NLSy1 galaxies from the radius–luminosity relation ($R_{\rm BLR}\propto L^{1/2}$; see also~\citealt{Crepaldi2025A&A...696A..74C}). This value is also well within the dust/torus scale of $\approx10^{17}$–$10^{18}$ cm ($\approx0.05$–$0.3$ pc) inferred from the optical-to-MIR lag measurements (Sect.~\ref{sec_4.0}) and therefore provides a more physically motivated choice than the canonical BLR sizes ($\sim0.1$–$1$ pc) typically adopted for powerful blazars. The fitted dissipation distance, $R_H$, spans $\approx10^{16}$ cm ($\approx0.003$ pc) to $\approx10^{18}$ cm ($\approx0.32$ pc). Thus, in sources with $R_H < R_{\rm BLR}$, the emitting region is located within the BLR, whereas in sources with $R_H \gtrsim R_{\rm BLR}$, it lies at or beyond the BLR boundary. In all cases, however, the external-Compton emission is dominated by the accretion-disk photon field (EC-disk; Fig.~\ref{fig: sed_figures}), while the BLR contribution remains negligible.
In sources where $R_H$ is of the order of $10^{17}$--$10^{18}$~cm, the high-energy hump is produced by a combination of SSC and external Compton scattering of disk photons, whereas in J102906.69$+$555625.2 and J122844.81$+$501751.2 the EC-disk component alone accounts for the GeV hump, as is commonly observed in FSRQ-type blazars. In a few sources, the thermal disk component also contributes prominently in the optical--UV range (the \emph{Disk} curve in Fig.~\ref{fig: sed_figures}), another feature commonly observed in FSRQs. 
We also observed varying magnetic fields ($B$) in all the cases, and these lie in the range of what is observed in blazars. The viewing angle of all these objects is not known, and hence it was optimized during the fitting. We observed a range between \emph{1} and \emph{7} degrees. The bulk Lorentz factor ($\Gamma$) is also optimized to obtain the best fit, and is found between {\emph{5} and \emph{24}}, suggesting that the jet flow in RQ-NLSy1s is similar to blazars, in terms of its bulk motion, although its power is much lower.  The redshift ($z$) of all sources is fixed to their nominal value. We have also included the contribution of the accretion disk, which is controlled by the luminosity of the accretion disk ($L_{disk}$) and the temperature ($T_{disk}$). The accretion disk is modelled using the multicolour Shakura–Sunyaev disk implemented in \texttt{JETSET}, characterised by the disk luminosity, $L_{\rm disk}$, and the peak disk temperature, $T_{\rm disk}$, the latter being determined by $L_{\rm disk}$ and $M_{\rm BH}$. This approach follows the standard methodology adopted in previous studies of $\gamma$-ray-detected NLSy1 galaxies \citep{Paliya2013MNRAS.428.2450P}. For two sources, J151020.06+554722.0 and J164100.10+345452.7, the inferred disk temperatures ($T_{\rm disk}\sim2.5\times10^{3}$ K) are lower than typically expected for AGN accretion disks. We attribute this primarily to the absence of simultaneous UV observations, which results in weak constraints on the thermal disk component, together with the assumption of a standard thin-disk geometry. Therefore, the derived disk parameters should be regarded as indicative rather than definitive. Given the high Eddington ratios generally associated with NLSy1 galaxies, a slim or advective disk model may provide a more realistic description of the accretion flow in these sources.\par
The X-ray detections play a role in constraining the model, mostly the SSC part, and in placing the sources relative to the blazar synchrotron-peak sequence. In all six sources, the XRT band falls between the synchrotron and inverse-Compton humps rather than on the synchrotron peak itself, consistent with the behaviour characteristic of low- and intermediate-synchrotron-peaked objects; none of the sources is high-synchrotron-peaked. We stress once more, however, that this placement follows from models in which the X-rays are assigned to the jet by construction (Sect.~\ref{jetset_modelling}), and it would be modified if a substantial fraction of the observed 0.3--10~keV flux were of coronal origin, as allowed by the measured photon indices. The corresponding SSC normalisations are therefore best regarded as upper limits. The magnetic field in all the cases is optimized and found to be between \emph{0.04} and \emph{2.25} Gauss, as mostly seen in other classes of blazars. The viewing angle and $\Gamma$ of these objects are not known, and therefore they were kept free, but the modeled values come closer to the blazars (see Table~\ref{tab_SED_Modeling_data}). The accretion disk is included both as a direct thermal component in the optical--UV and as the seed photon field for external Compton scattering; among the external photon fields, it is the dominant one, while the BLR contribution is negligible in every source (Sect.~\ref{jetset_modelling}). The overall broadband SED modelling, combined with the ${\rm TS} > 25$ GeV detections that cannot be accounted for by thermal AGN components alone, indicates that all six SED-modelled sources require a non-thermal, jet-related emission component. We regard this as establishing the presence of $\gamma$-ray emission, and hence of a relativistic jet, in these objects, although a non-jet origin of the GeV emission cannot be fully excluded; we do not claim, however, that their jets dominate the emission in any band other than the GeV, nor that the specific jet parameters in Table~\ref{tab_SED_Modeling_data} are uniquely determined by the present, non-simultaneous data.

\begin{table*}
\centering
\caption{Broadband \emph{SED} fitting results. All sources are modeled with \emph{JETSET}, and their corresponding parameters are listed here. The corresponding best fit result is shown in Fig.~\ref{fig: sed_figures}.}
\label{tab_SED_Modeling_data}
\resizebox{\textwidth}{!}{
\begin{tabular}{llllllll} \hline 
Object & J102906.6+555625.2 & J122844.8+501751.2 & J123220.1+495721.8 & J150916.1+613716.7 & J151020.0+554722.0 & J164100.1+345452.7 \\ \hline 
$\gamma_{\text{min}}$ & {62.35} & {50.66} & {70.24} & {48.94} & {834.68} & {28.95} \\
$\gamma_{\text{max}}$ & {$9.69\times10^5$} &{$2.02\times10^6$} &{$2.04\times10^7$} &{$8.98\times10^6$} &{$2.68\times10^7$} &{$9.77\times10^5$}  \\
N [cm$^{-3}$] &{$1.41\times10^4$} &{$2.68\times10^3$} &{$3.58\times10^3$} &{$1.47\times10^4$} &{$9.07\times10^2$} &{$3.57\times10^3$} \\
$\gamma_{0-lp}$ &{$6.24\times10^3$} &{$9.87\times10^3$} &{$5.61\times10^3$} &{$5.16\times10^3$} &{$2.62\times10^4$} &{$3.06\times10^3$}  \\
s &{2.88} &{3.68} &{3.24} &{1.81} &{3.56} &{1.30} \\
r &{2.78} &{2.01} &{1.50} &{6.98} &{6.99} &{4.54}  \\
R [cm] &{$2.99\times10^{15}$} &{$4.95\times10^{15}$} &{$3.33\times10^{15}$} &{$2.82\times10^{15}$} &{$7.14\times10^{15}$} &{$5.99\times10^{15}$}  \\
R$_{\texttt{H}}$ [cm] &{$5.35\times10^{17}$} &{$1.01\times10^{16}$} &{$1.00\times10^{17}$} &{$9.98\times10^{17}$} &{$9.93\times10^{17}$} &{$7.72\times10^{17}$}  \\
B [G] &{0.12} &{1.85} &{1.23} &{0.20} &{0.17} &{0.25}  \\
$\theta$ [degree] &{0.28} &{2.98} &{1.29} &{4.99} &{6.77} &{6.93}  \\
$\Gamma$ &{17.25} &{16.26} &{9.13} &{5.100} &{15.22} &{24.01}  \\
z & 0.45 & 0.26 & 0.26 & 0.20 & 0.15 & 0.16  \\
L$_{\text{disk}}$ [erg/s] &{$9.06\times10^{44}$} &{$5.26\times10^{43}$} &{$5.68\times10^{44}$} &{$1.12\times10^{44}$} &{$5.45\times10^{43}$} &{$3.20\times10^{44}$}  \\
T$_{\text{disk}}$ [K] &{$3.29\times10^4$} &{$6.93\times10^3$} &{$1.79\times10^4$} &{$2.51\times10^4$} &{$2.48\times10^3$} &{$2.61\times10^3$}  \\ \hline 
\multicolumn{8}{l}{The tabulated 
parameters, $\gamma_{min}$, $\gamma_{max}$, $\gamma_{0-lp}$, $r$, and $s$ are  related to the particle distribution. $N$ is the particle number density. $R$ and $R_H$ are the size and the}\\
\multicolumn{8}{l}{ location of the emitting zone. $B$ is the magnetic field. $\theta$, $\Gamma$, and $z$ are the viewing angle, the bulk Lorentz factor, and the redshift, respectively.}\\
\multicolumn{8}{l}{$L_{\rm disk}$ and $T_{\rm disk}$ are the accretion disk luminosity and temperature, respectively.}
\end{tabular}
}
\end{table*}

\begin{figure*}
    \begin{minipage}[]{1.0\textwidth}
\includegraphics[width=1.0\textwidth,height=0.20\textheight,angle=00]{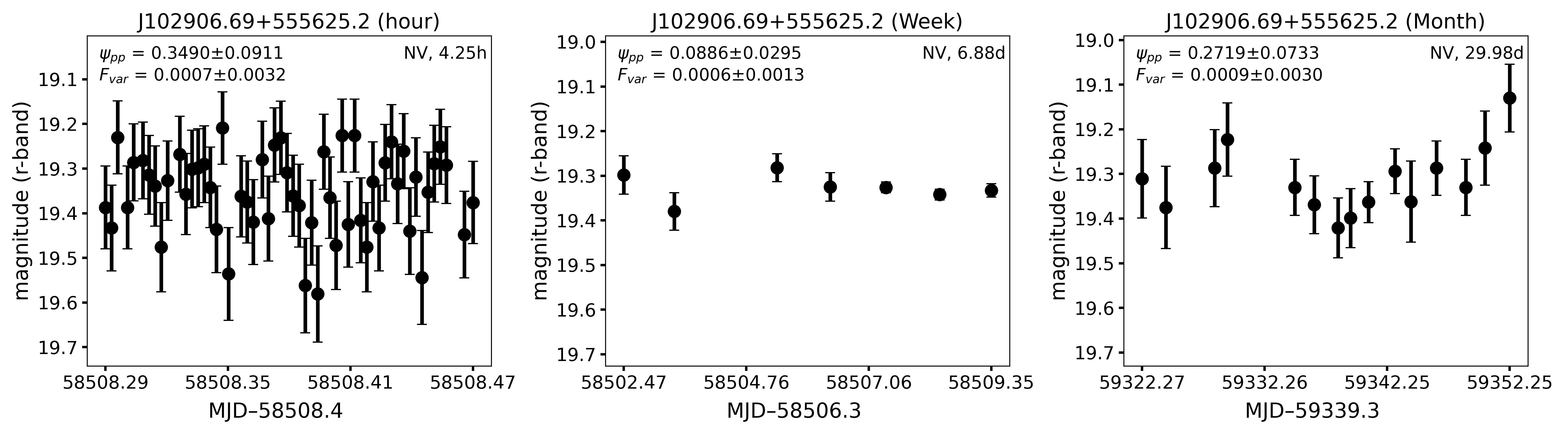}  
  \includegraphics[width=1.0\textwidth,height=0.20\textheight,angle=00]{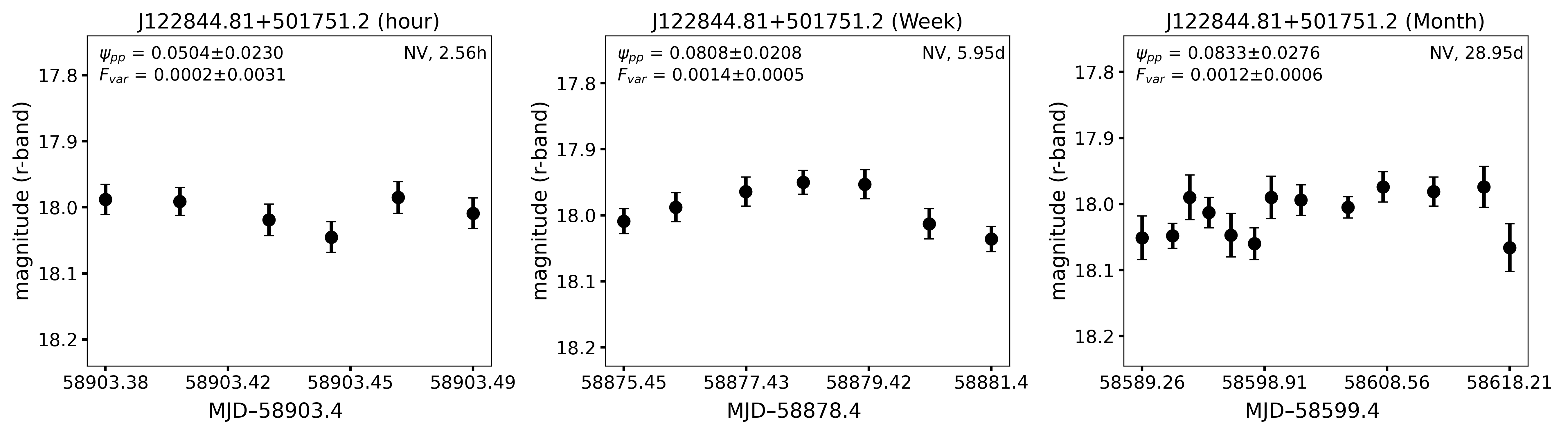}  
   \includegraphics[width=1.0\textwidth,height=0.20\textheight,angle=00]{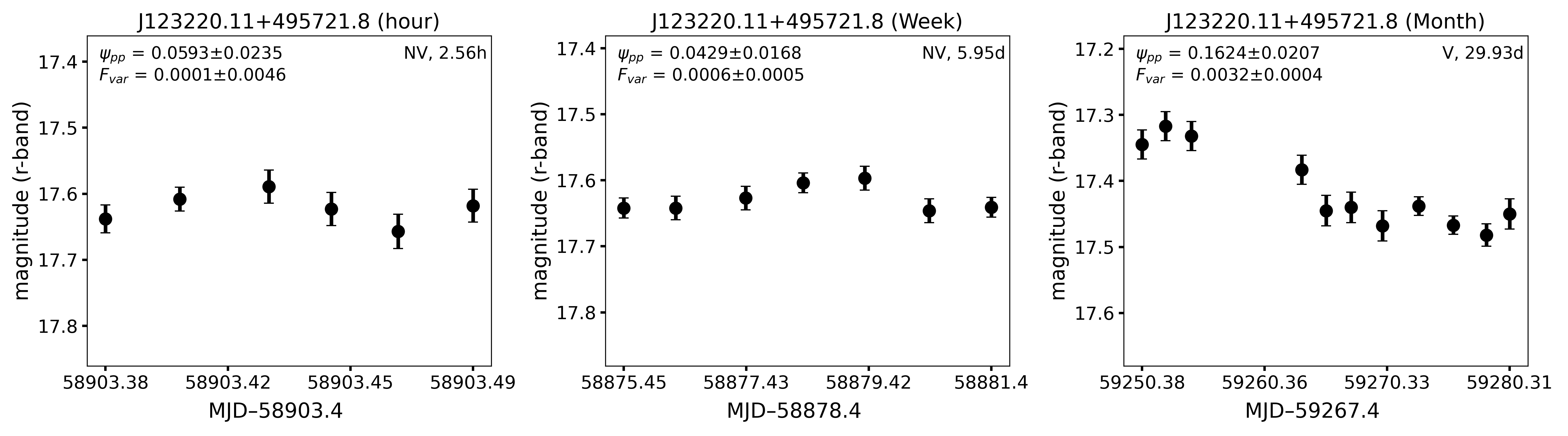} 
    \end{minipage}
    \caption{\emph{ZTF} \emph{r}-band light curves of the RQ-NLSy1s \emph{J102906.69$+$555625.2}, \emph{J122844.81$+$501751.2}, and \emph{J123220.11$+$495721.8} from the current sample, covering intra-night to month-like timescales. All sources exhibit significant non-variability (NV) on all timescales except for \emph{J123220.11$+$495721.8}, which shows variability only on month-like timescales. The name and analyzed timescale for each source are given in the panel title. Estimated variability parameters, $\psi_{\mathrm{pp}}$ and $F_\mathrm{var}$, along with the variability status and duration of each light curve, are indicated in the upper-left and upper-right corners of the respective panels.}
    \label{fig: Intranight_optical_variability_1}
\end{figure*}

\begin{figure*}
\ContinuedFloat
    \begin{minipage}[]{1.0\textwidth}
\includegraphics[width=1.0\textwidth,height=0.20\textheight,angle=00]{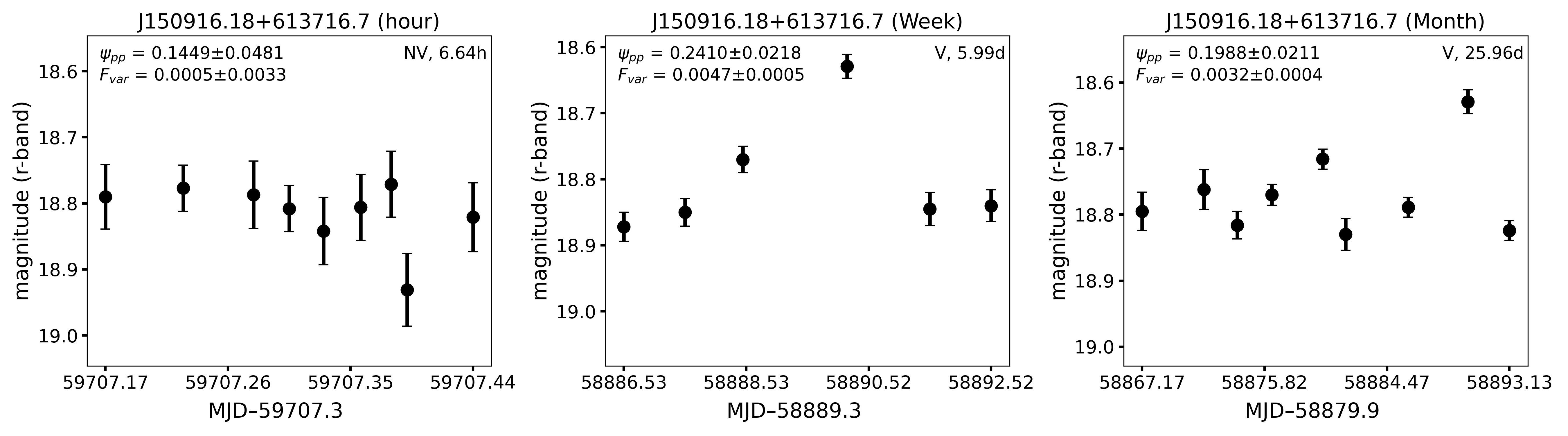}  
  \includegraphics[width=1.0\textwidth,height=0.20\textheight,angle=00]{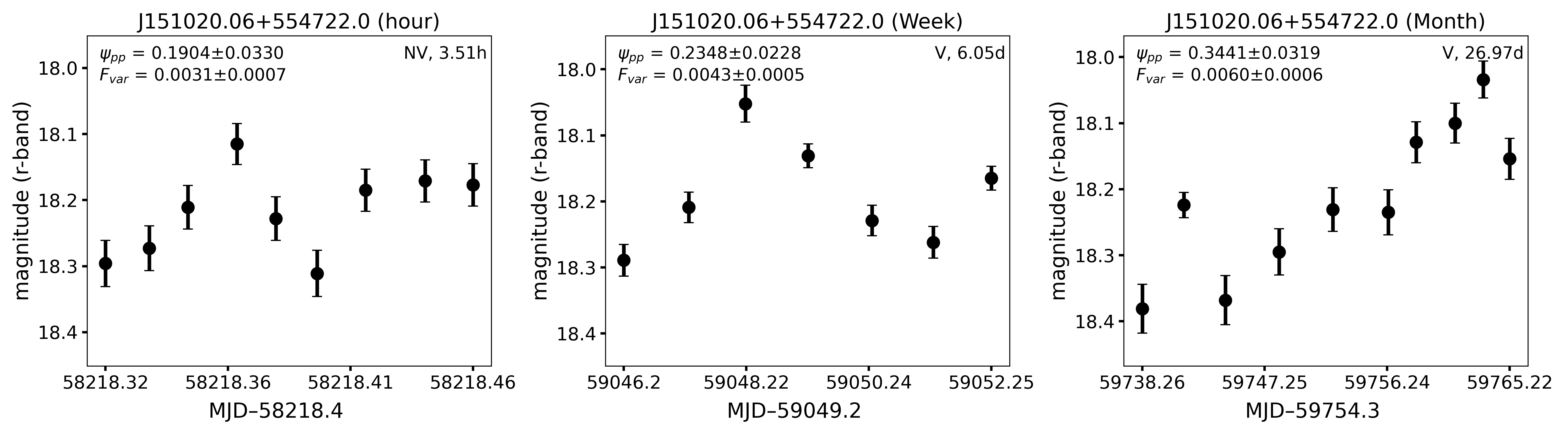}  
   \includegraphics[width=1.0\textwidth,height=0.20\textheight,angle=00]{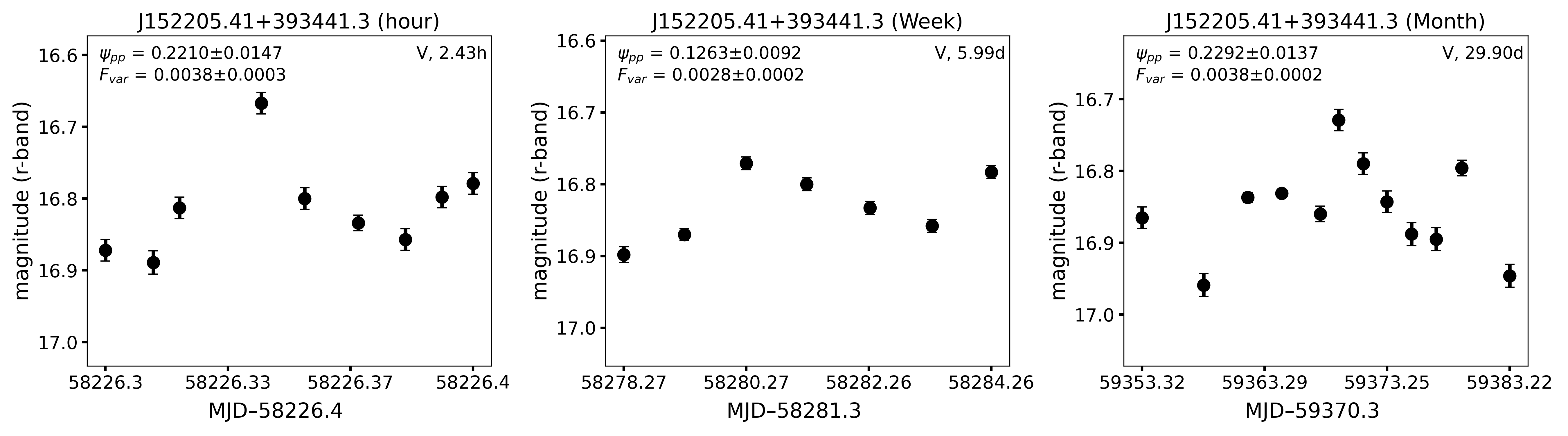} 
   \includegraphics[width=1.0\textwidth,height=0.20\textheight,angle=00]{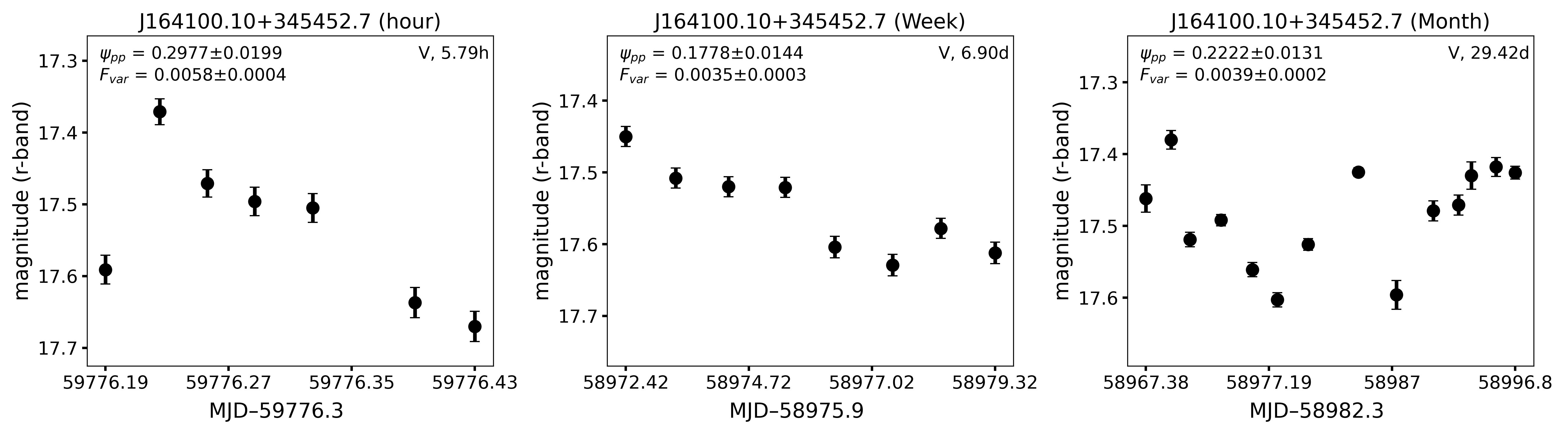} 
    \end{minipage}
    \caption{(Continued) Same as Fig.~\ref{fig: Intranight_optical_variability_1} but for RQ-NLSy1s \emph{J150916.18$+$613716.7}, \emph{J151020.06+554722.0}, \emph{J152205.41+393441.3}, and \emph{J164100.10$+$345452.7}, respectively. All four sources are exhibiting variability on all timescales except for  RQ-NLSy1s \emph{J150916.18$+$613716.7} and J151020.06+554722.0, which do not show variability on the intra-night timescale.}
\end{figure*}

\begin{figure*}
    \begin{minipage}[]{1.0\textwidth}
\includegraphics[width=1.0\textwidth,height=0.24\textheight,angle=00]{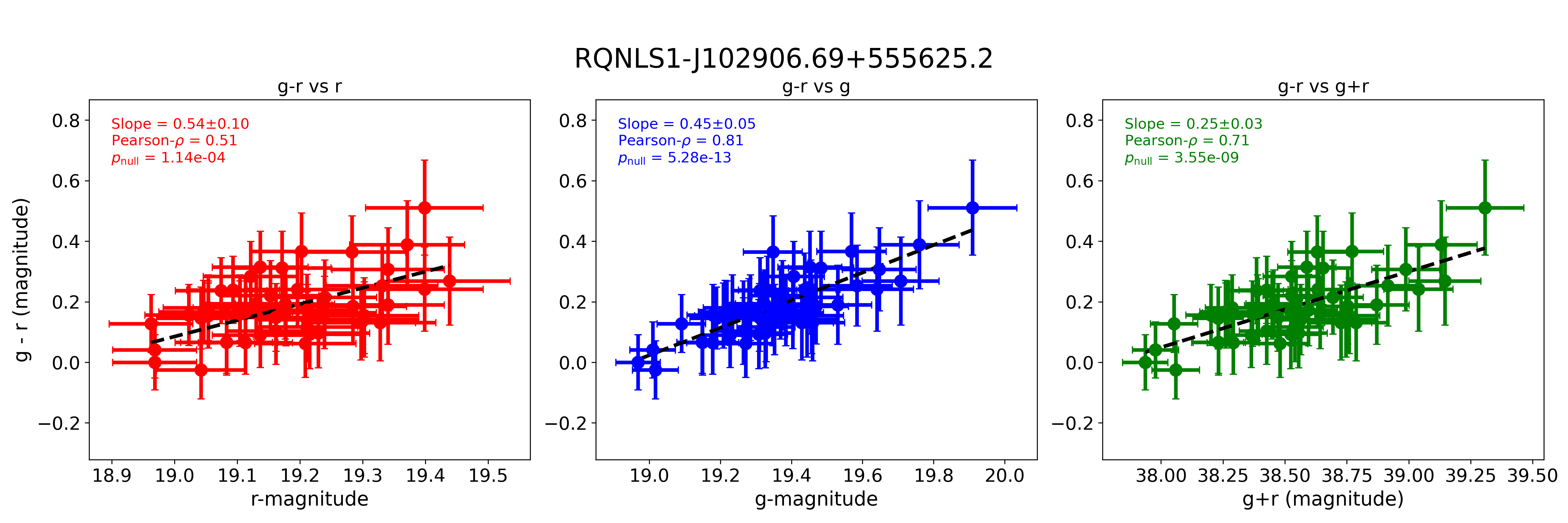}  
  \includegraphics[width=1.0\textwidth,height=0.24\textheight,angle=00]{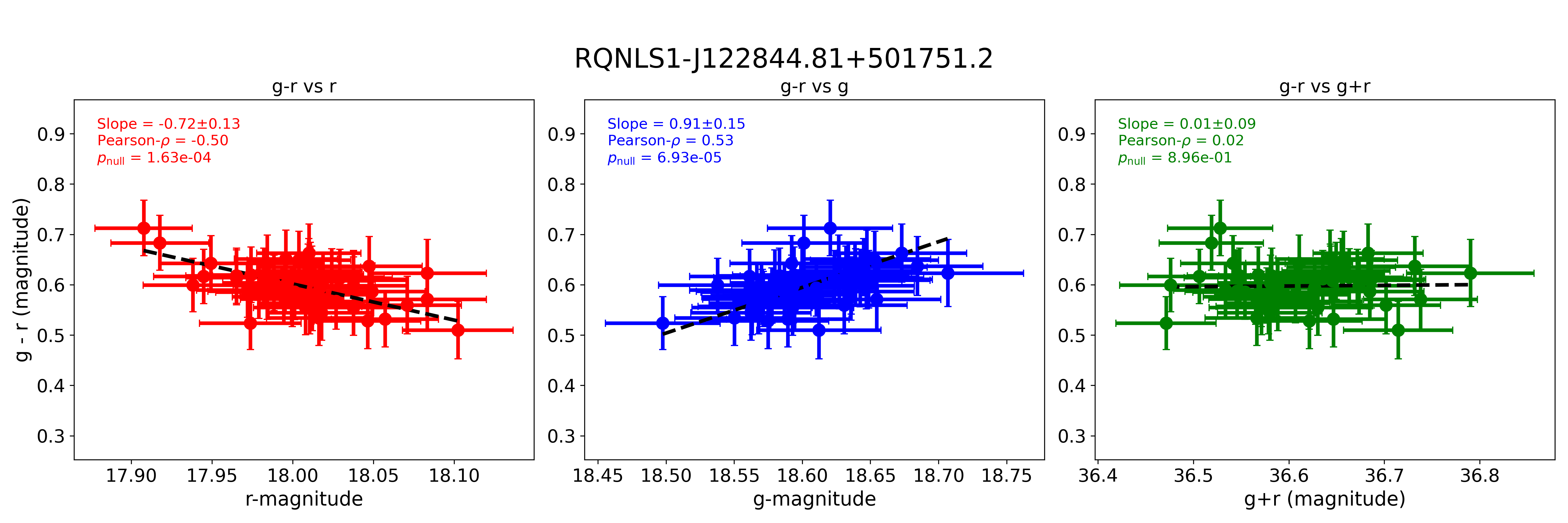}  
   \includegraphics[width=1.0\textwidth,height=0.24\textheight,angle=00]{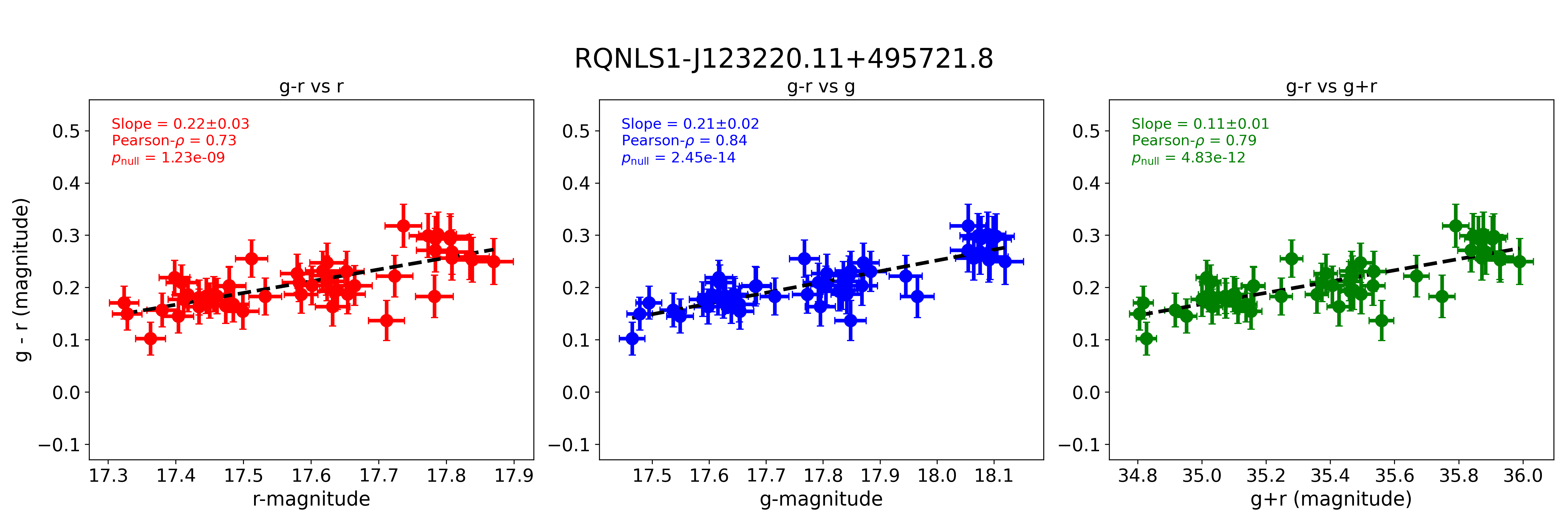} 
    \end{minipage}
    \caption{Long-term ($\emph{g-r}$) color variations versus $\emph{r}$-magnitude, $\emph{g}$-magnitude, and $\emph{(g+r)}$ for the RQ-NLSy1s \emph{J102906.69$+$555625.2} (top), \emph{J122844.81$+$501751.2} (middle), and \emph{J123220.11$+$495721.8} (bottom) from the current sample. Significant positive trends are observed in \emph{J102906.69$+$555625.2} and \emph{J123220.11$+$495721.8} across all color–magnitude combinations. The black dashed line shows the best fit obtained using orthogonal distance regression, accounting for uncertainties in both color and magnitude. The corresponding slope and Pearson-$\rho$ are indicated in the upper-left corner of each panel.}
    \label{fig: OP_color_variability_gr_1_RQ-NLSy1}
\end{figure*} 

\begin{figure*}
\ContinuedFloat
    \begin{minipage}[]{1.0\textwidth}
 \includegraphics[width=1.0\textwidth,height=0.238\textheight,angle=00]{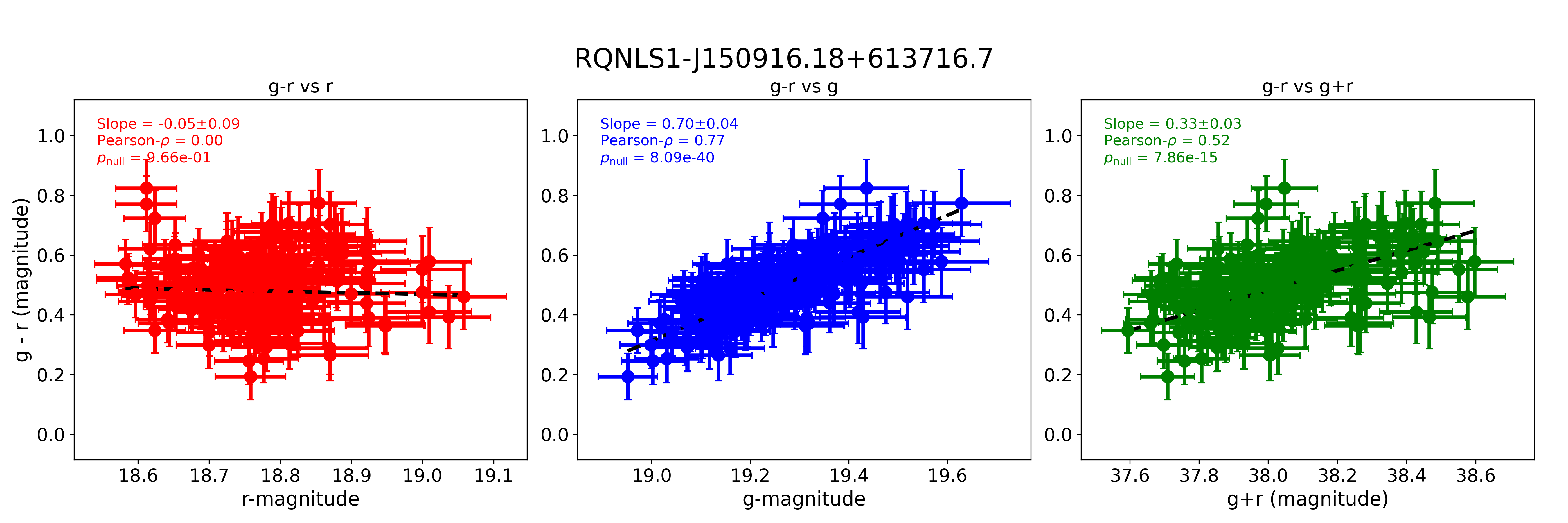}  
\includegraphics[width=1.0\textwidth,height=0.238\textheight,angle=00]{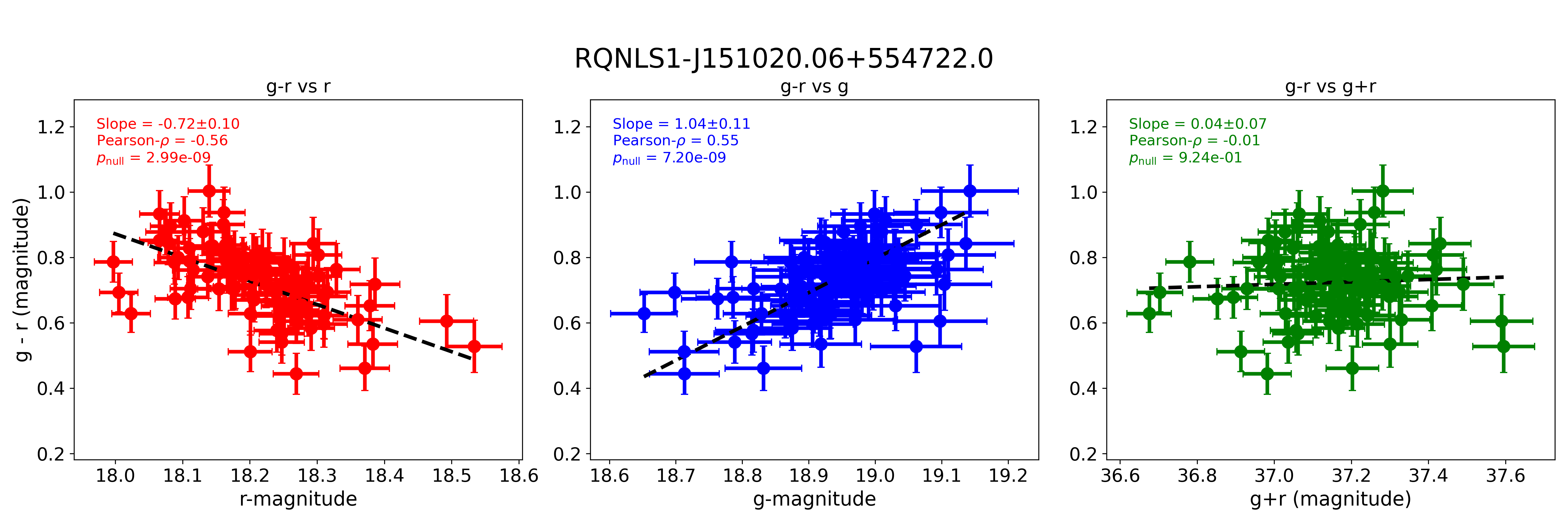}  
\includegraphics[width=1.0\textwidth,height=0.238\textheight,angle=00]{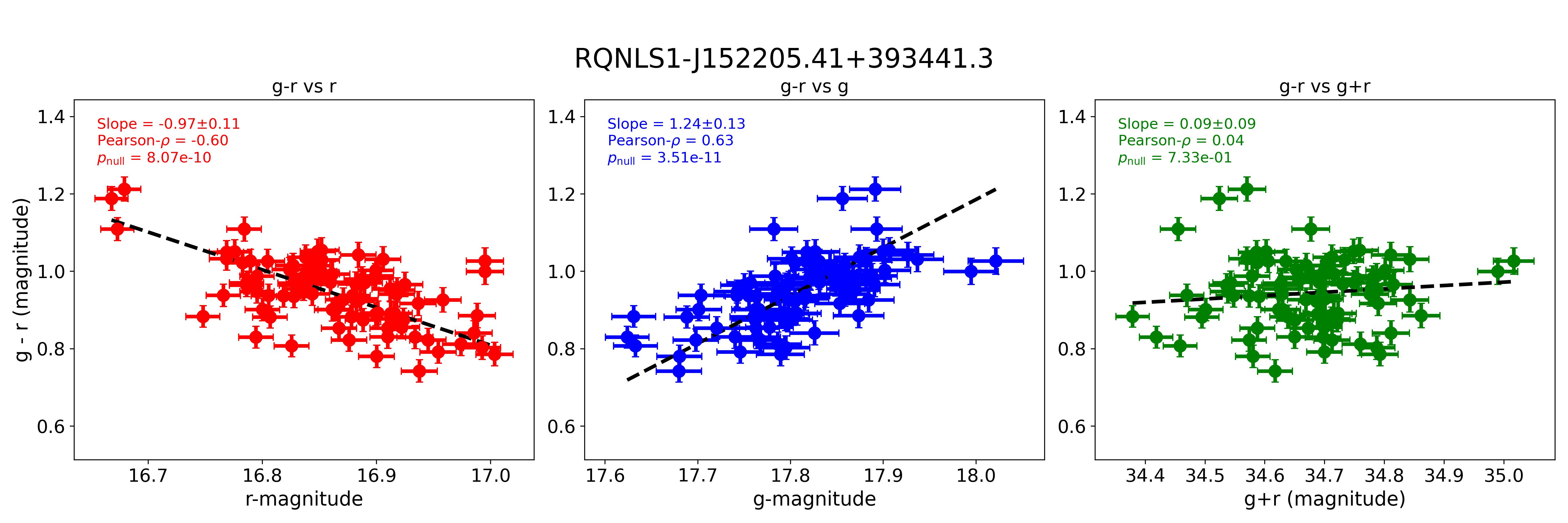} 
   \includegraphics[width=1.0\textwidth,height=0.238\textheight,angle=00]{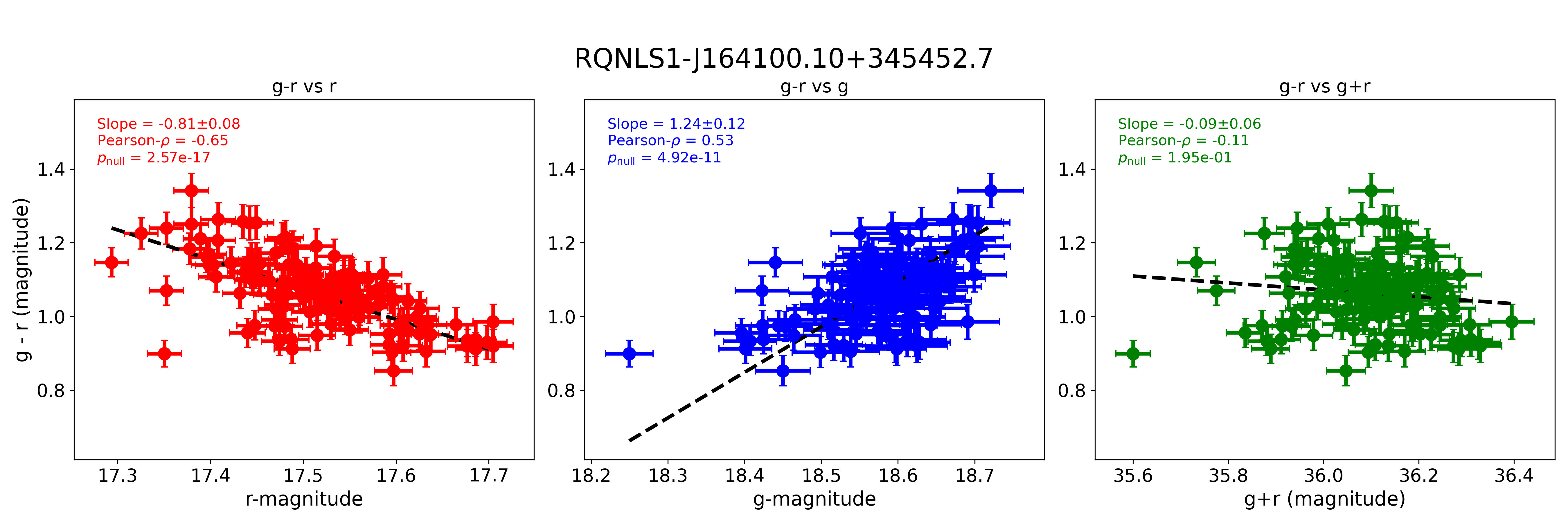} 
    \end{minipage}
    \caption{(Continued) Same as Fig.~\ref{fig: OP_color_variability_gr_1_RQ-NLSy1} but for RQ-NLSy1s \emph{J150916.18$+$613716.7}, \emph{J151020.06$+$554722.0}, \emph{J152205.41$+$393441.3}, and \emph{J164100.10$+$345452.7}, respectively, presented from the top to bottom. Only \emph{J150916.18$+$613716.7} showed a significant positive trend in its $\emph{g-r}$ versus $\emph{g+r}$ color-magnitude diagram.}
    \label{fig: OP_color_variability_gr_2_RQ-NLSy1}
\end{figure*}

\begin{figure*}
    \begin{minipage}[]{1.0\textwidth}
 \includegraphics[width=1.0\textwidth,height=0.24\textheight,angle=00]{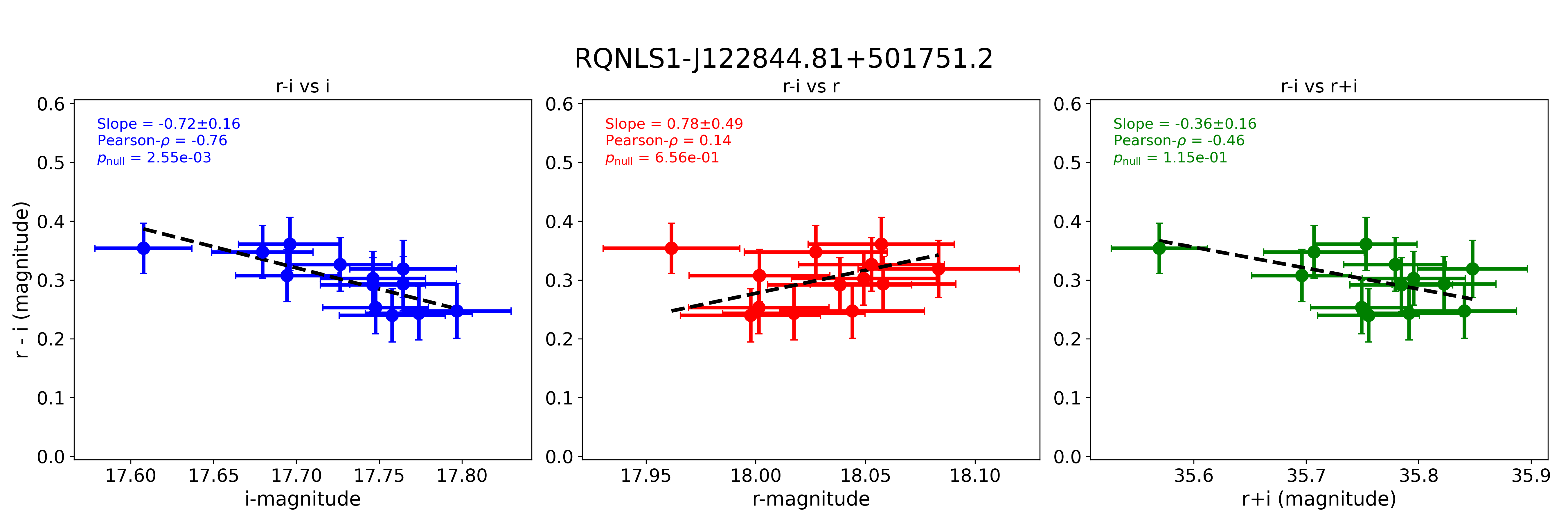}  
  \includegraphics[width=1.0\textwidth,height=0.24\textheight,angle=00]{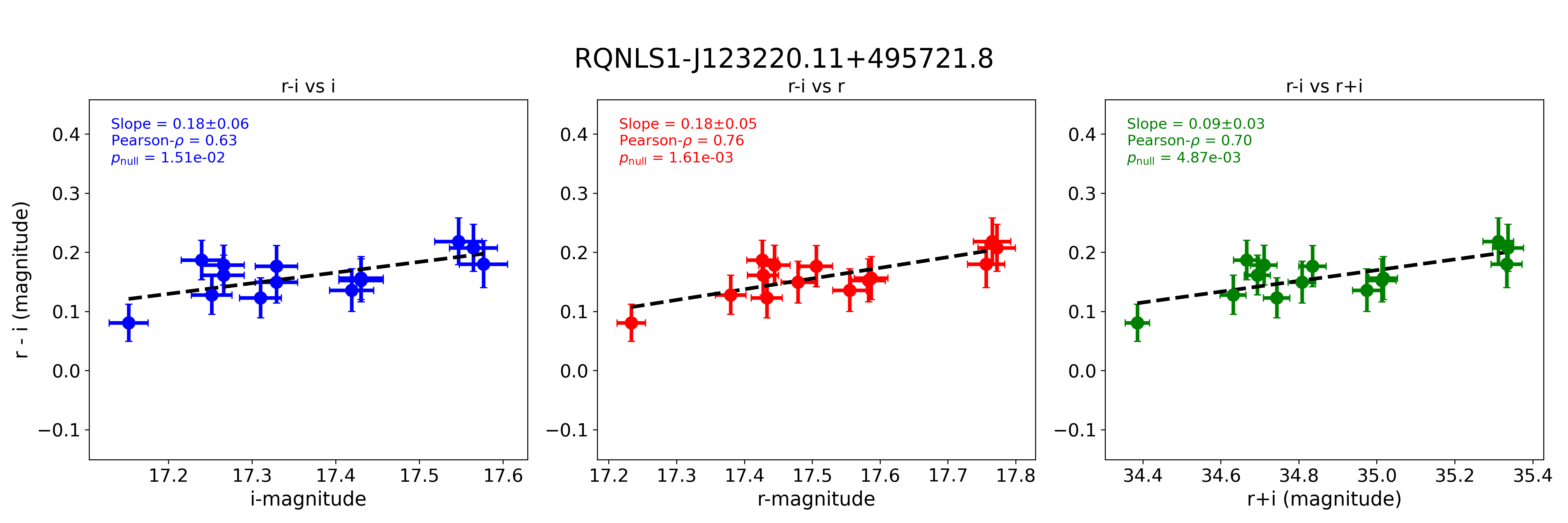}  
   \includegraphics[width=1.0\textwidth,height=0.24\textheight,angle=00]{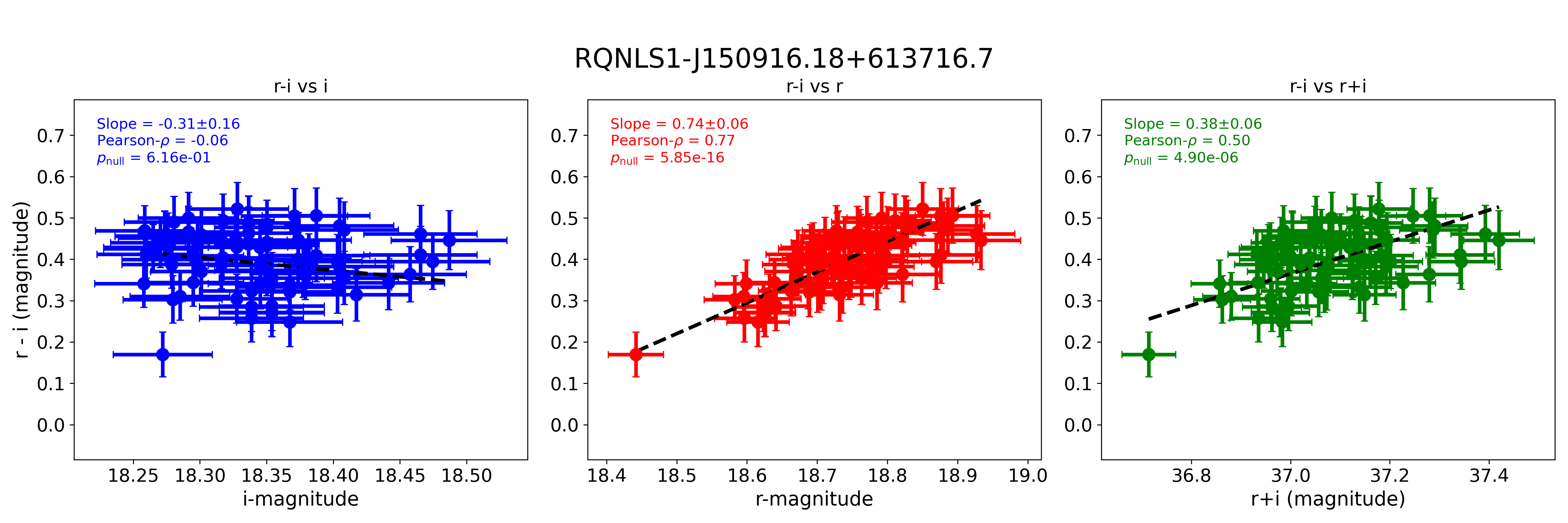} 
    \end{minipage}
    \caption{Same as Fig.~\ref{fig: OP_color_variability_gr_1_RQ-NLSy1} but for the long-term ($\emph{r-i}$) color variations versus $\emph{i}$-magnitude, $\emph{r}$-magnitude, and $\emph{(r+i)}$ for the RQ-NLSy1s \emph{J122844.81$+$501751.2} (top), \emph{J123220.11$+$495721.8} (middle), and \emph{J150916.18$+$613716.7} (bottom) from the current sample. Significant positive trends are observed in \emph{J123220.11$+$495721.8} and \emph{J150916.18$+$613716.7} from their $\emph{r-i}$ versus $\emph{r+i}$ color-magnitude diagrams.}
    \label{fig: OP_color_variability_ri_1_RQ-NLSy1}
\end{figure*}

\begin{figure*}
\ContinuedFloat
    \begin{minipage}[]{1.0\textwidth}
 \includegraphics[width=1.0\textwidth,height=0.24\textheight,angle=00]{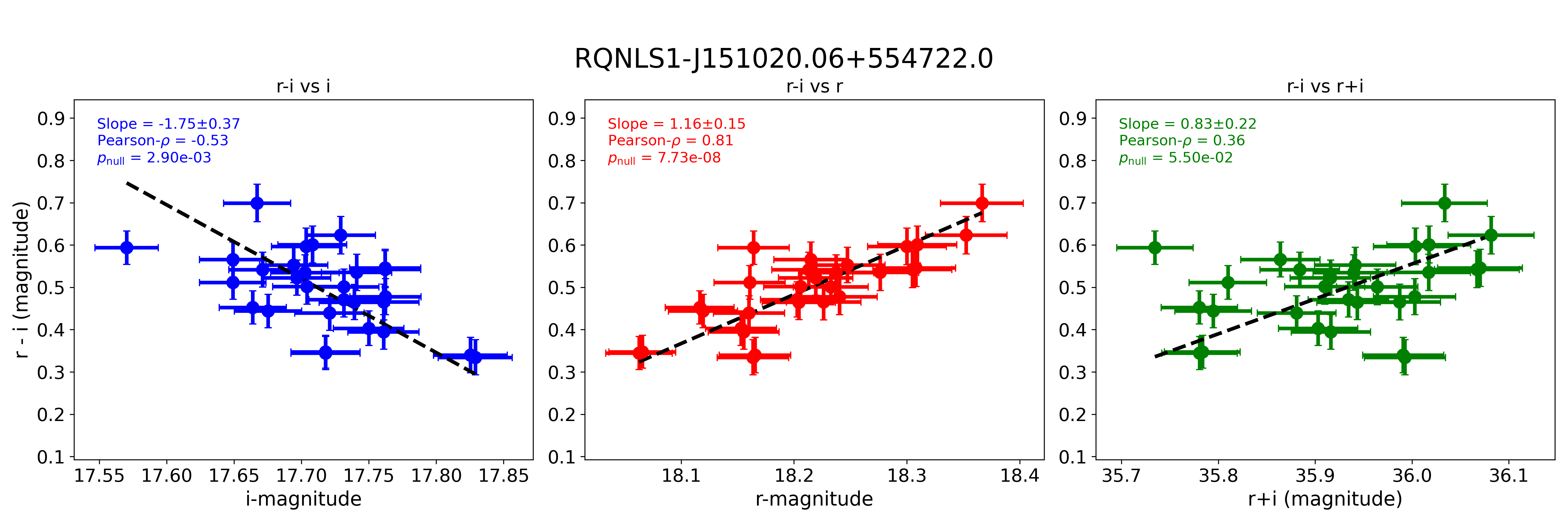}  
  \includegraphics[width=1.0\textwidth,height=0.24\textheight,angle=00]{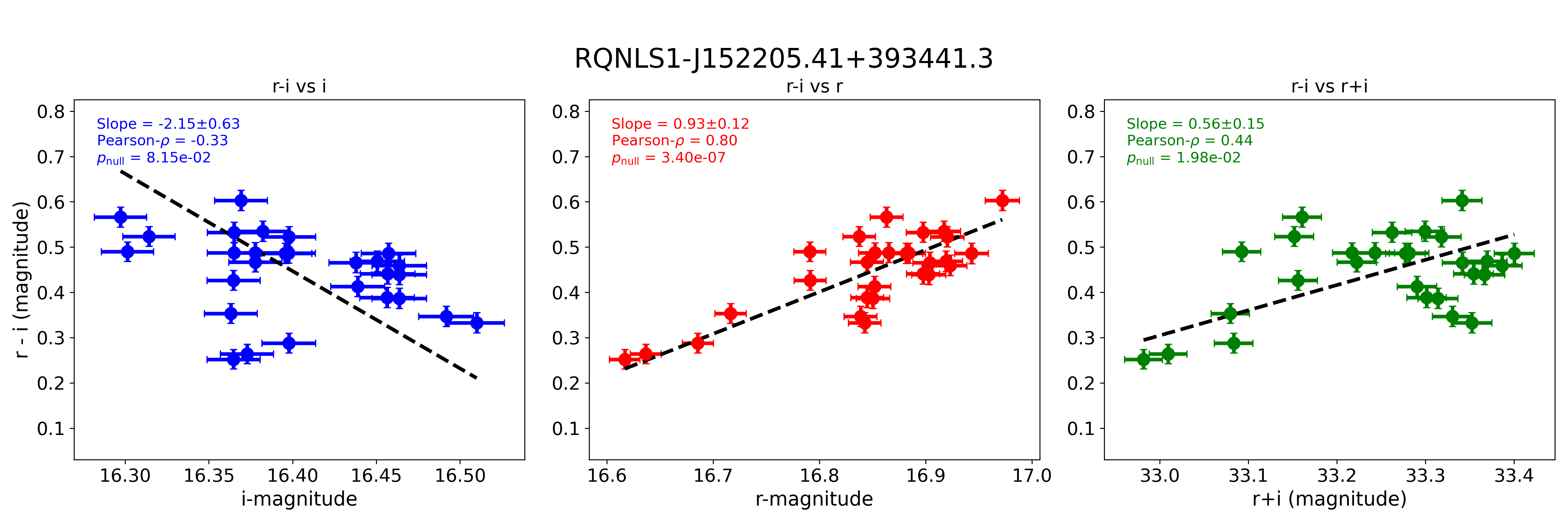}  
   \includegraphics[width=1.0\textwidth,height=0.24\textheight,angle=00]{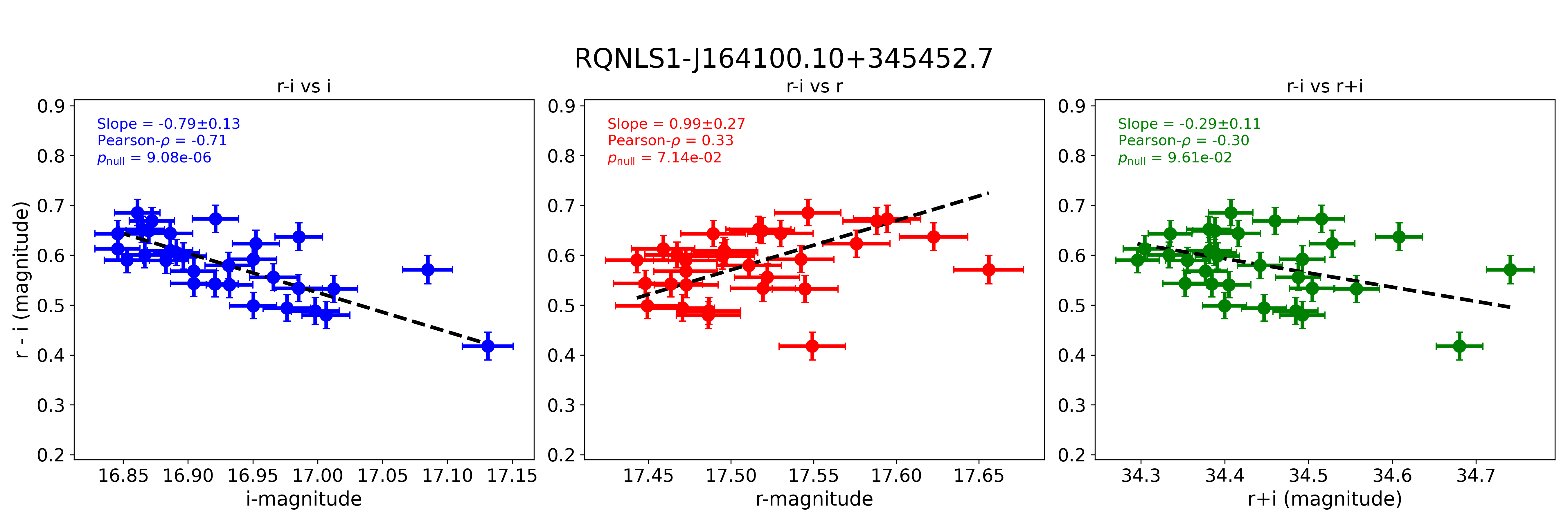} 
    \end{minipage}
    \caption{(Continued) Same as Fig.~\ref{fig: OP_color_variability_ri_1_RQ-NLSy1} but for RQ-NLSy1s \emph{J151020.06$+$554722.0} (top), \emph{J152205.41$+$393441.3} (middle), and \emph{J164100.10$+$345452.7} (bottom), respectively. All these sources did not show a significant trend from their $\emph{r-i}$ versus $\emph{r+i}$ color-magnitude diagrams.}
    \label{fig: OP_color_variability_ri_2_RQ-NLSy1}
\end{figure*}

\begin{figure*}
    \begin{minipage}[]{1.0\textwidth}
\includegraphics[width=1.0\textwidth,height=0.239\textheight,angle=00]{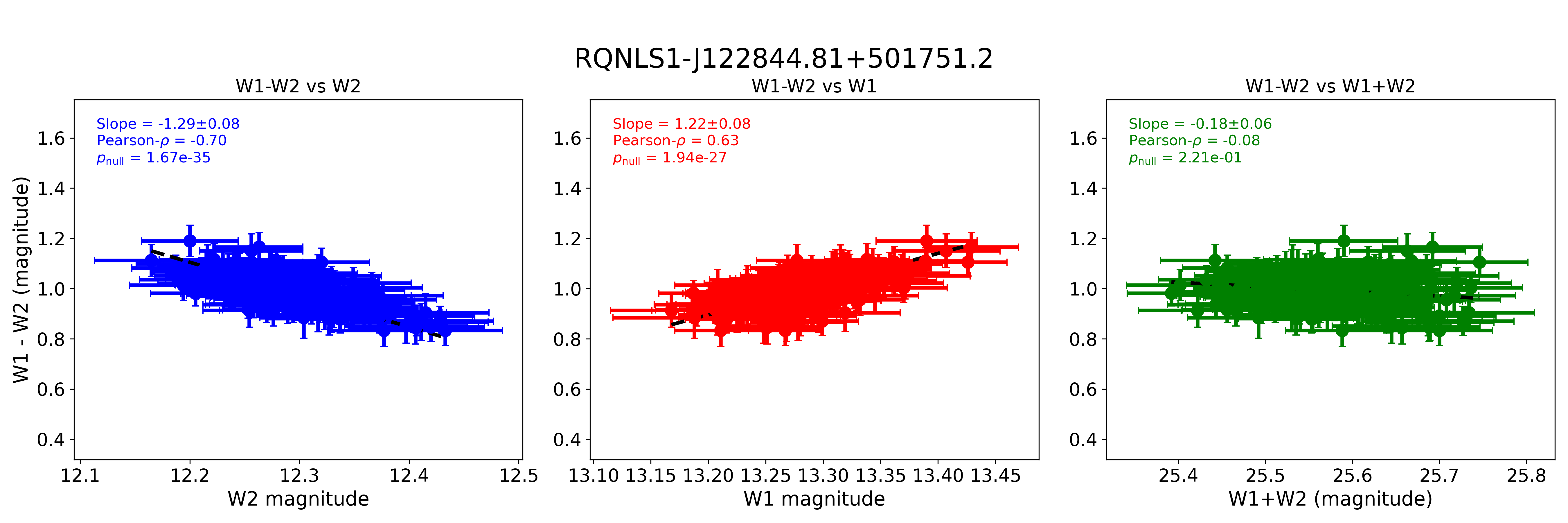}  
  \includegraphics[width=1.0\textwidth,height=0.239\textheight,angle=00]{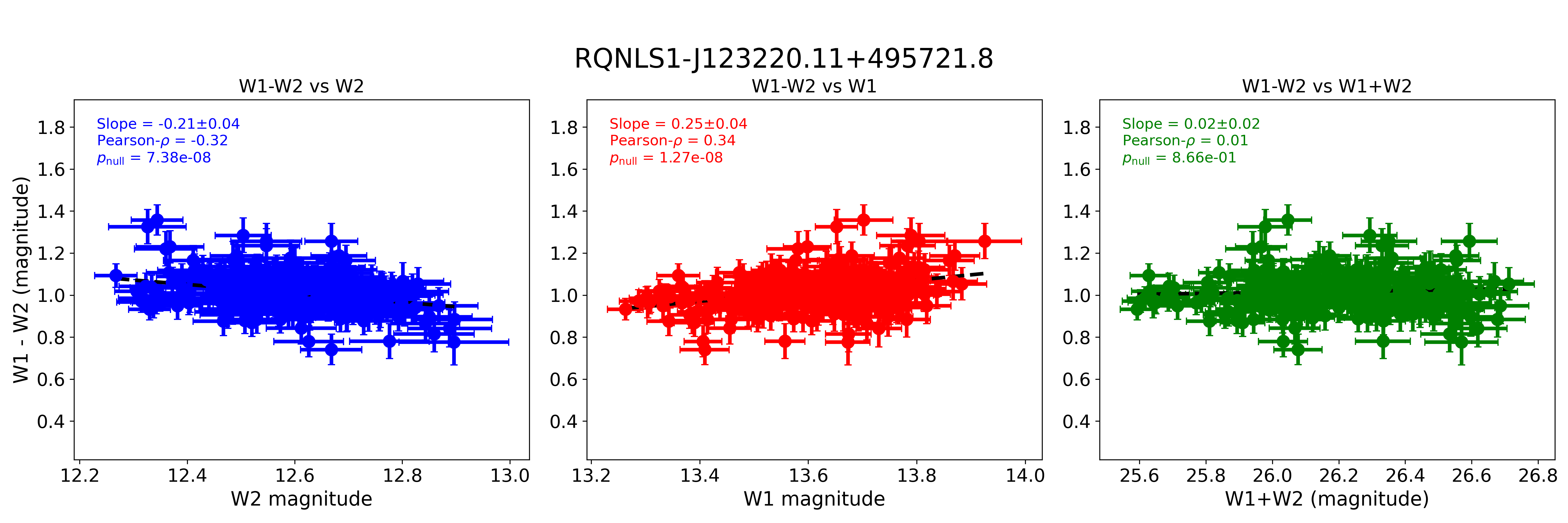}  
   \includegraphics[width=1.0\textwidth,height=0.239\textheight,angle=00]{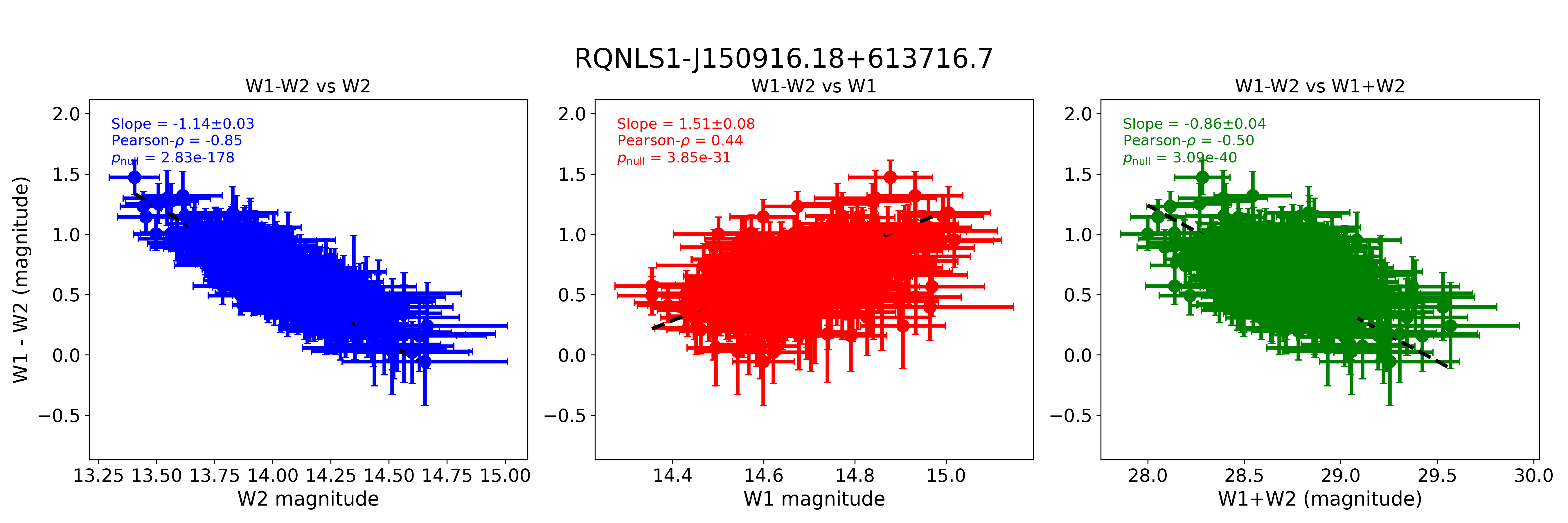} 
   \includegraphics[width=1.0\textwidth,height=0.239\textheight,angle=00]{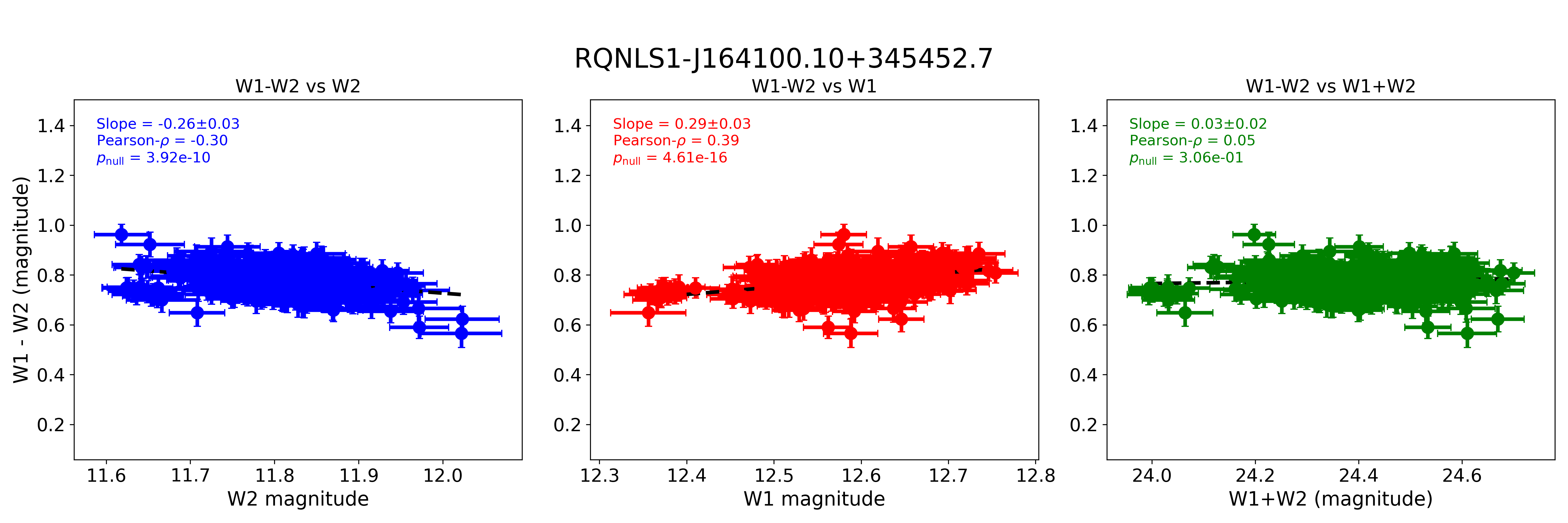} 
    \end{minipage}
    \caption{Same as Fig.~\ref{fig: OP_color_variability_gr_1_RQ-NLSy1} but for the long-term ($\emph{W1-W2}$) color variations versus $\emph{W2}$-magnitude, $\emph{W1}$-magnitude, and $\emph{(W1+W2)}$ for the RQ-NLSy1s \emph{J122844.81$+$501751.2}, \emph{J123220.11$+$495721.8}, \emph{J150916.18$+$613716.7}, and \emph{J164100.10$+$345452.7}, respectively, presented from the top to bottom. Out of four sources, only \emph{J150916.18$+$613716.7} showed a significant negative trend from its $\emph{W1-W2}$ versus $\emph{W1+W2}$ color-magnitude diagram.}
    \label{fig: IR_color_variability_w1w2_RQ-NLSy1}
\end{figure*}

\begin{figure*}
\begin{minipage}[]{1.0\textwidth}
\includegraphics[width=1.0\textwidth,height=0.45\textheight,angle=00]{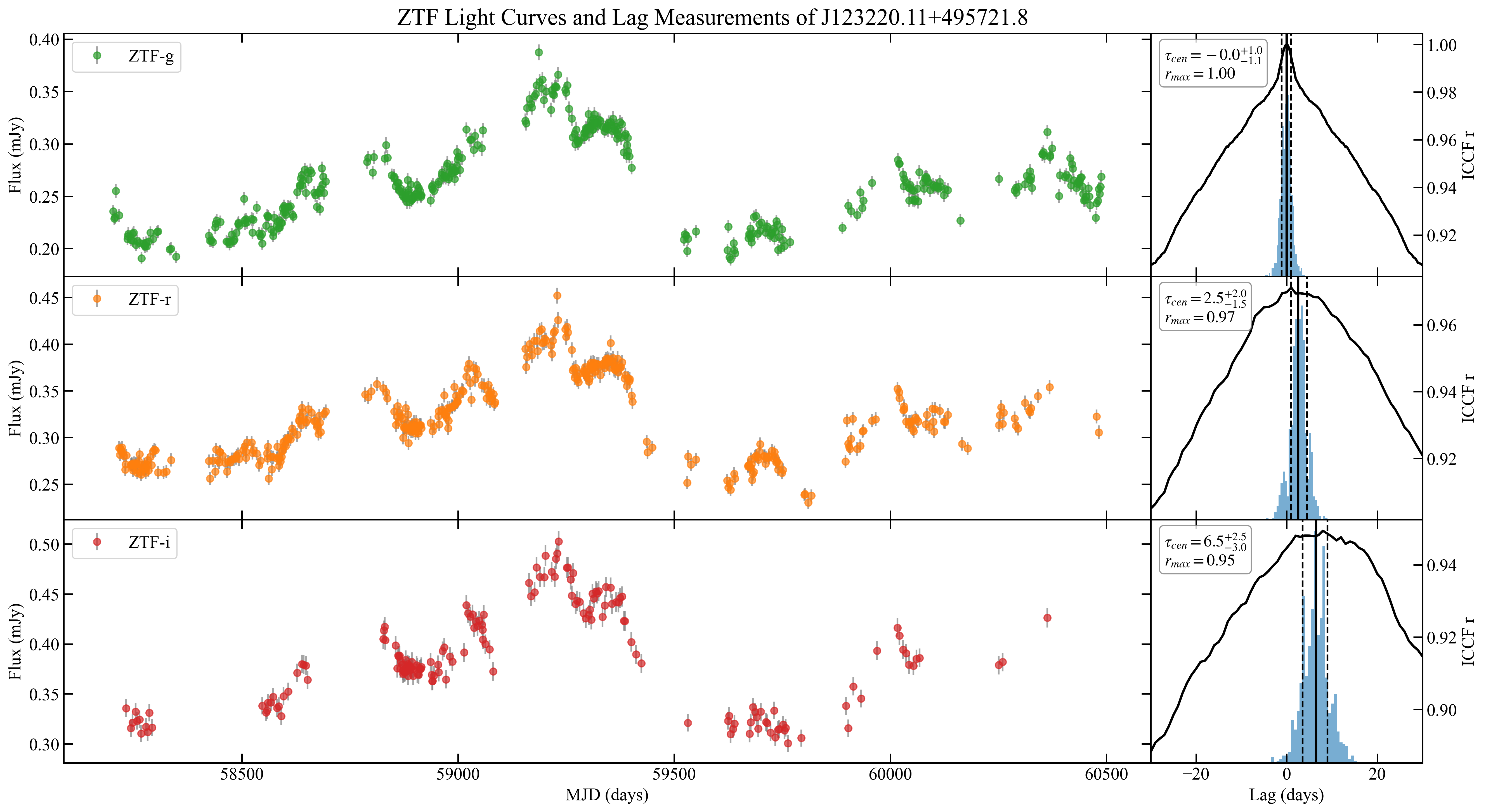} 
    \end{minipage}
    \caption{The \emph{ZTF} light curves and lag measurements for the source J123220.11$+$495721.8. Left panel: The \emph{ZTF} \emph{g-}, \emph{r-}, and \emph{i-}band light curves. Right panel: The cross-correlation functions (in black) and centroid distributions (in blue) for each band relative to the \emph{ZTF} \emph{g} band.}
    \label{fig: Lag_optical_bands_J1232}
\end{figure*}

\begin{figure*}
    \begin{minipage}[]{1.0\textwidth}
\includegraphics[width=1.0\textwidth,height=0.45\textheight,angle=00]{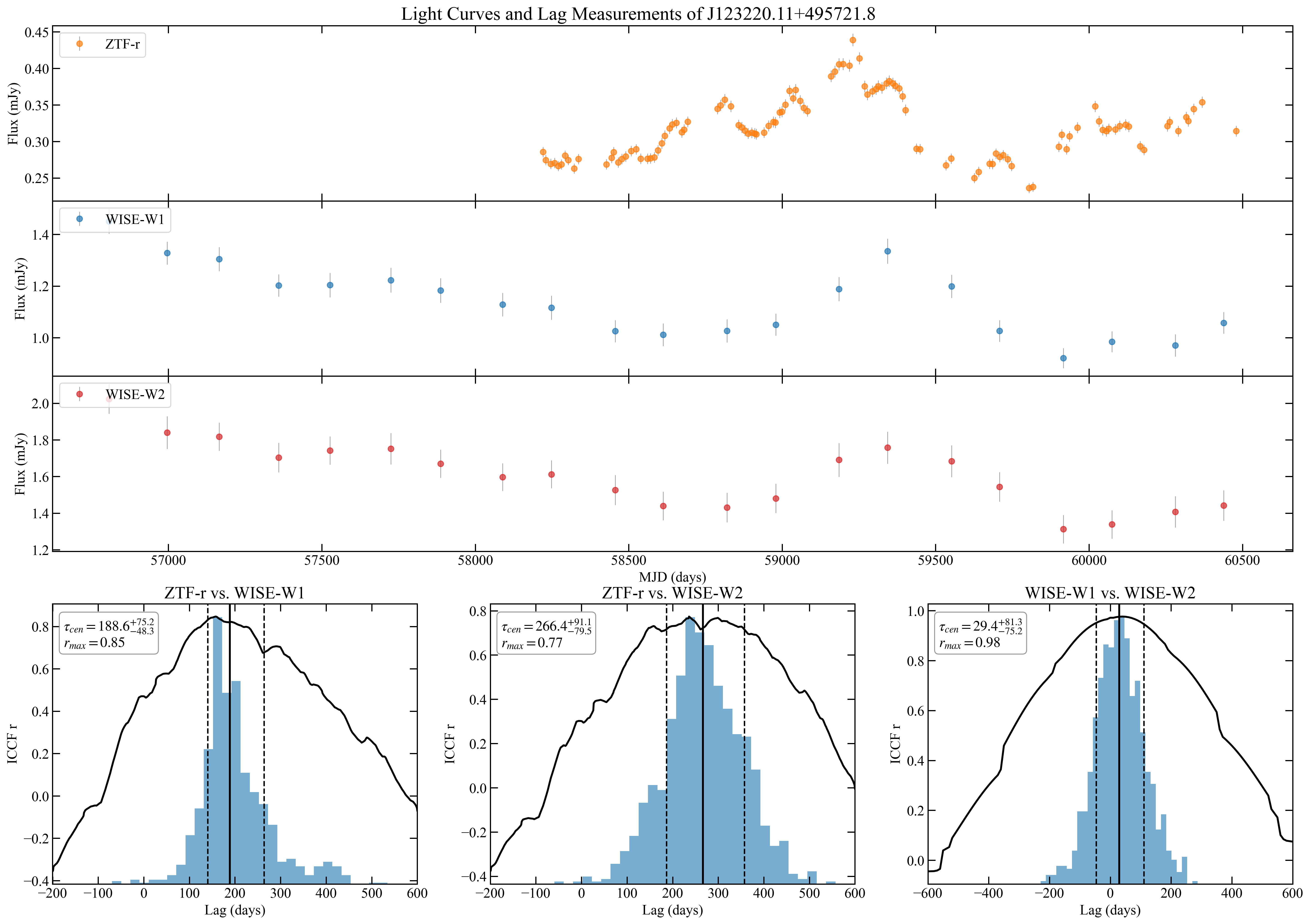} 
    \end{minipage}
    \caption{The \emph{ZTF} and \emph{WISE} binned light curves and lag measurements for the source J123220.11$+$495721.8. The upper three panels show the binned light curves of \emph{ZTF} \emph{r} band and \emph{WISE} \emph{W1}, \emph{W2} bands, while the bottom panel shows the corresponding lag measurements. Panels from left to right correspond to \emph{ZTF} \emph{r} vs. \emph{WISE} \emph{W1}, \emph{ZTF} \emph{r} vs. \emph{WISE} \emph{W2}, and \emph{WISE} \emph{W1} vs. \emph{WISE} \emph{W2}.}
    \label{fig: Lag_optical_MIR_bands_J1232}
\end{figure*}

\begin{figure*}
\begin{minipage}[]{1.0\textwidth}
\includegraphics[width=0.33\textwidth,height=0.24\textheight,angle=00]{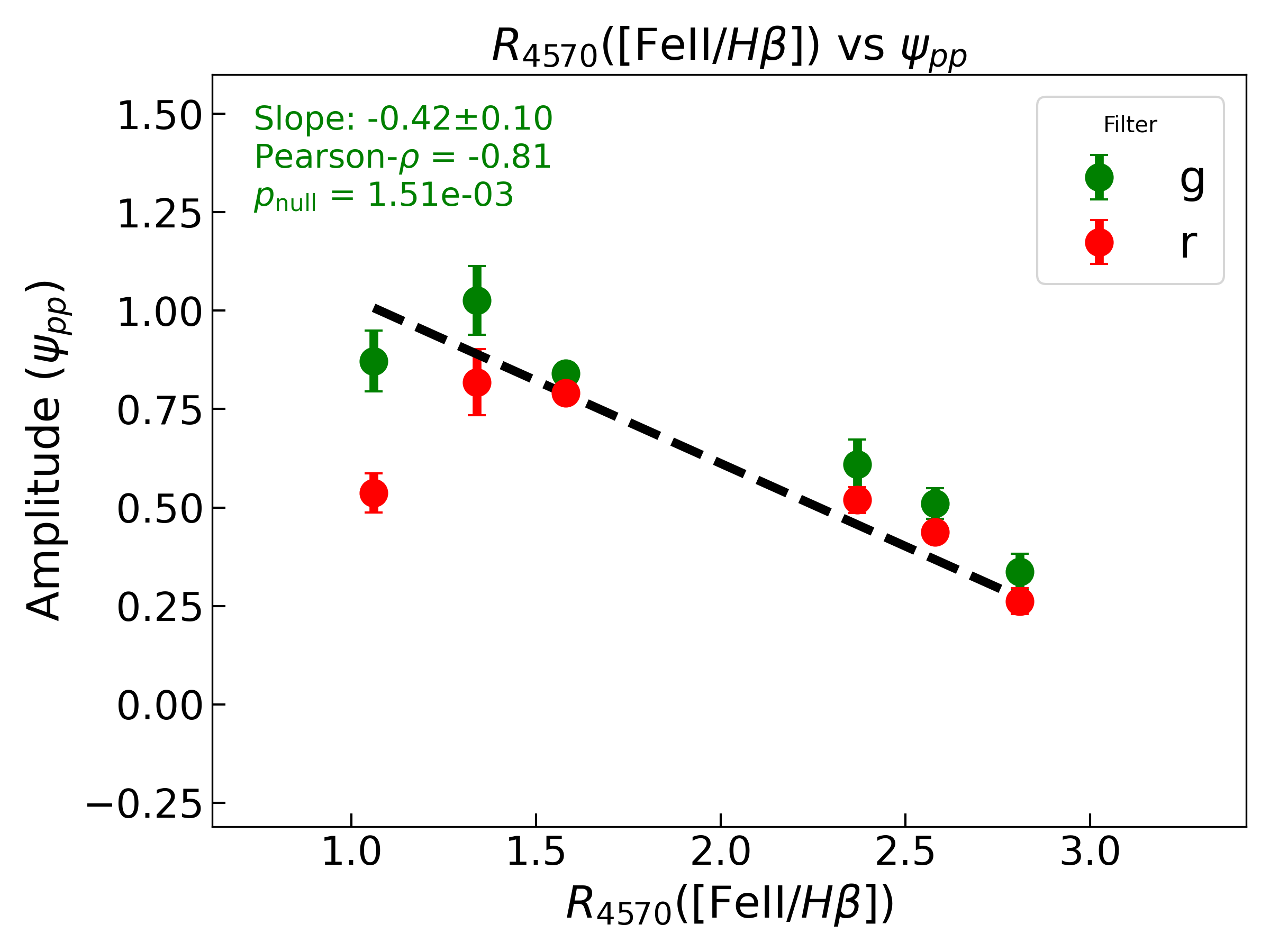}  
\includegraphics[width=0.33\textwidth,height=0.24\textheight,angle=00]{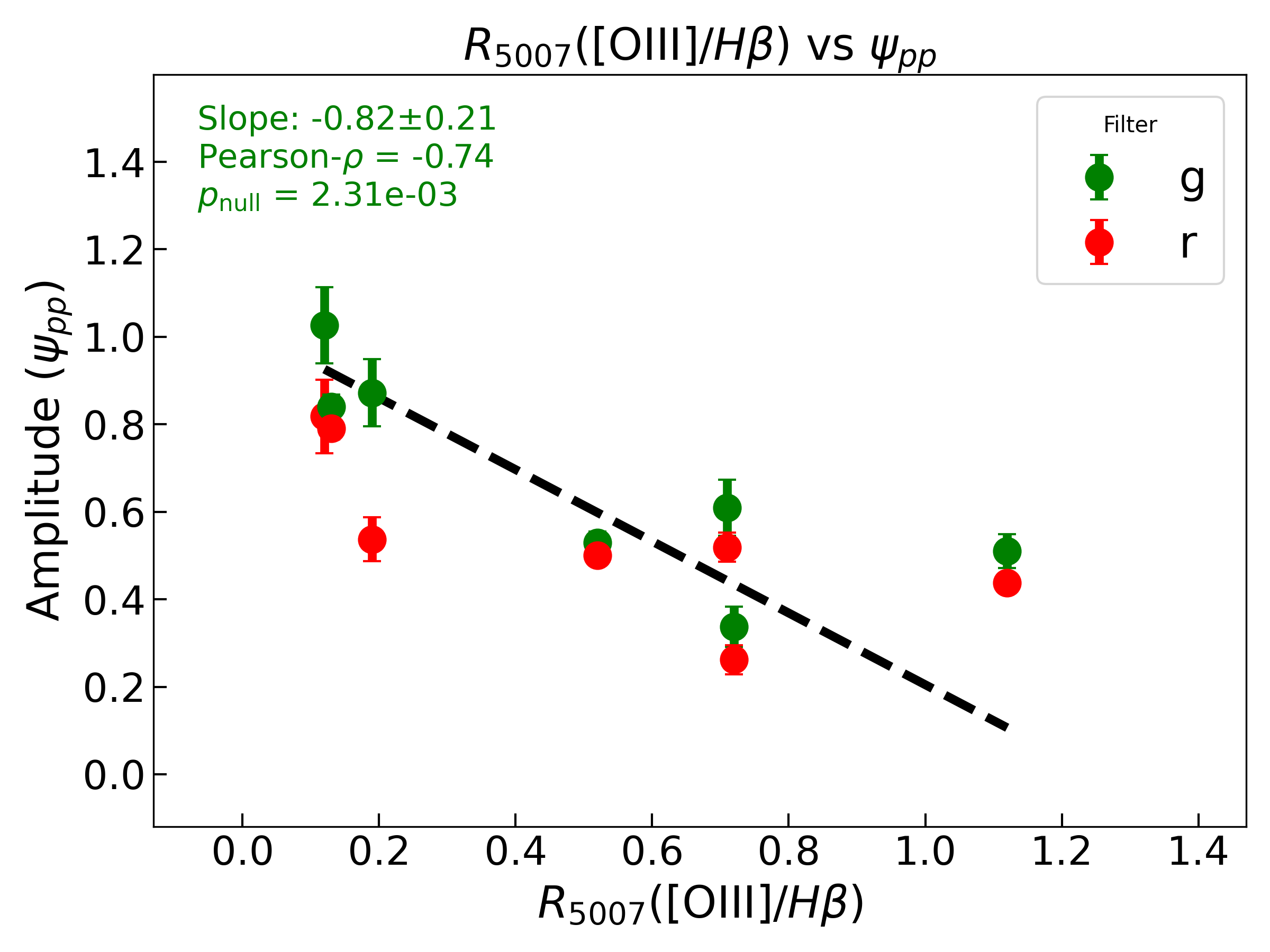} 
\includegraphics[width=0.33\textwidth,height=0.24\textheight,angle=00]{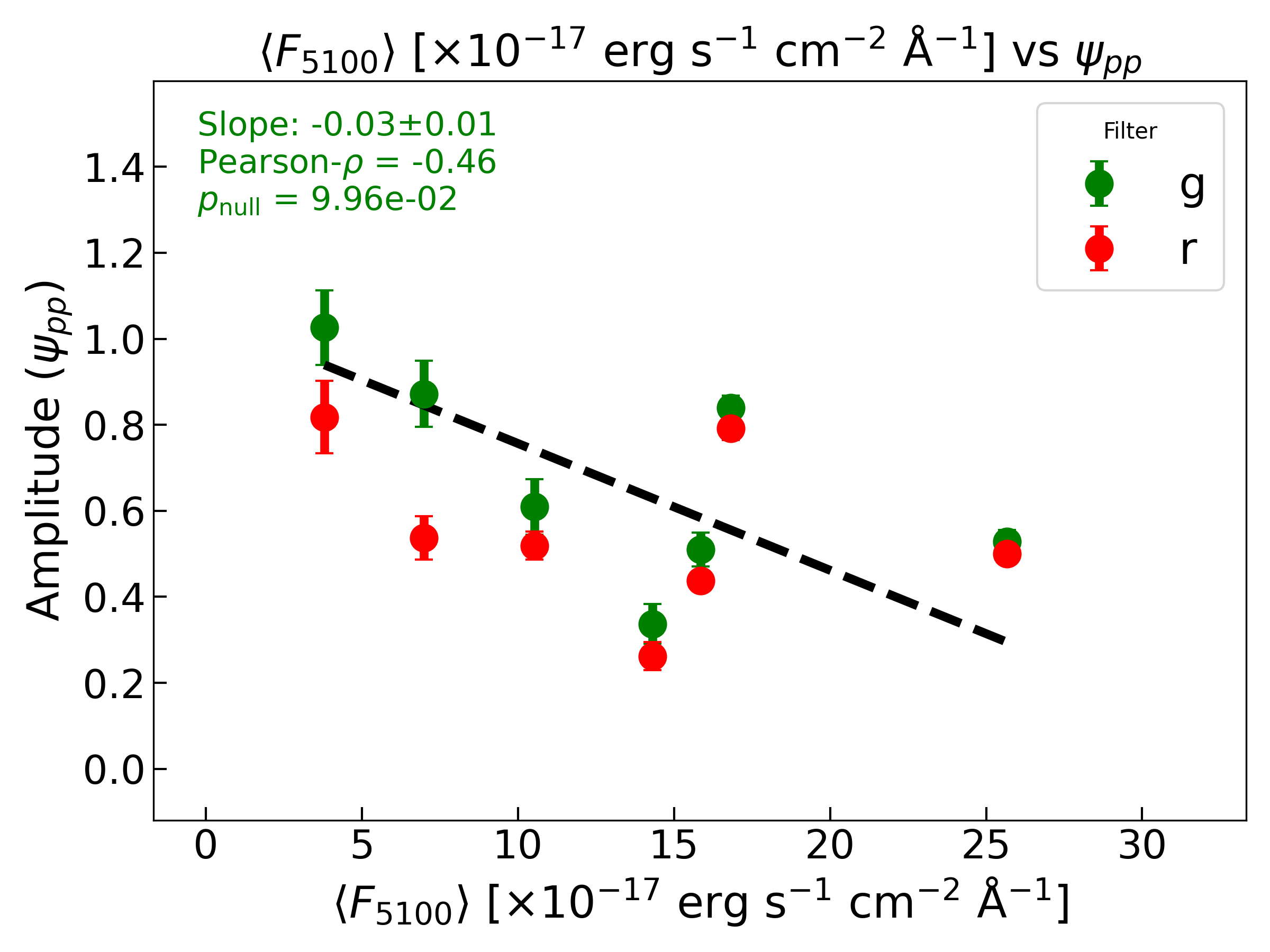}  
\includegraphics[width=0.33\textwidth,height=0.24\textheight,angle=00]{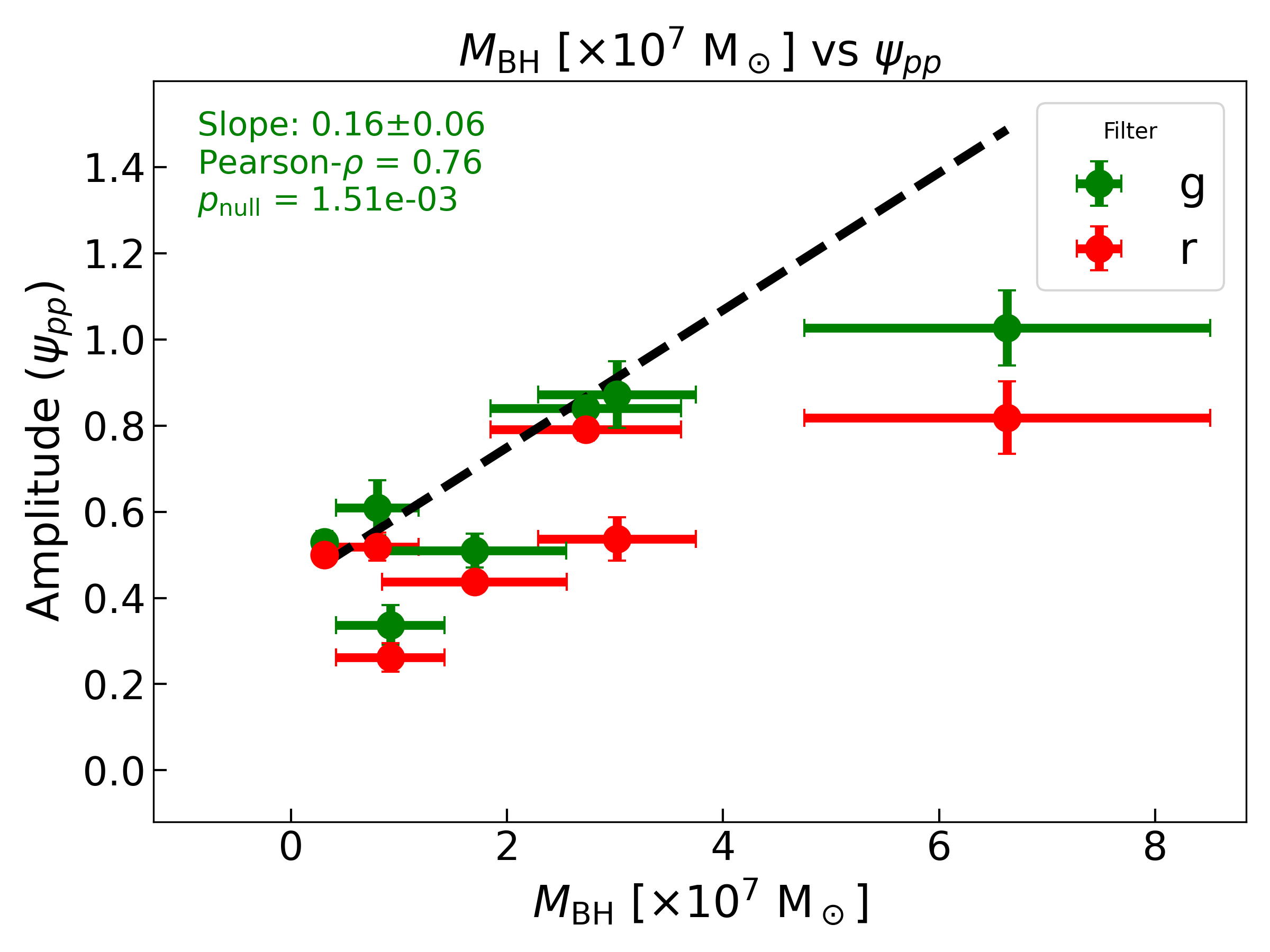}  
\includegraphics[width=0.33\textwidth,height=0.24\textheight,angle=00]{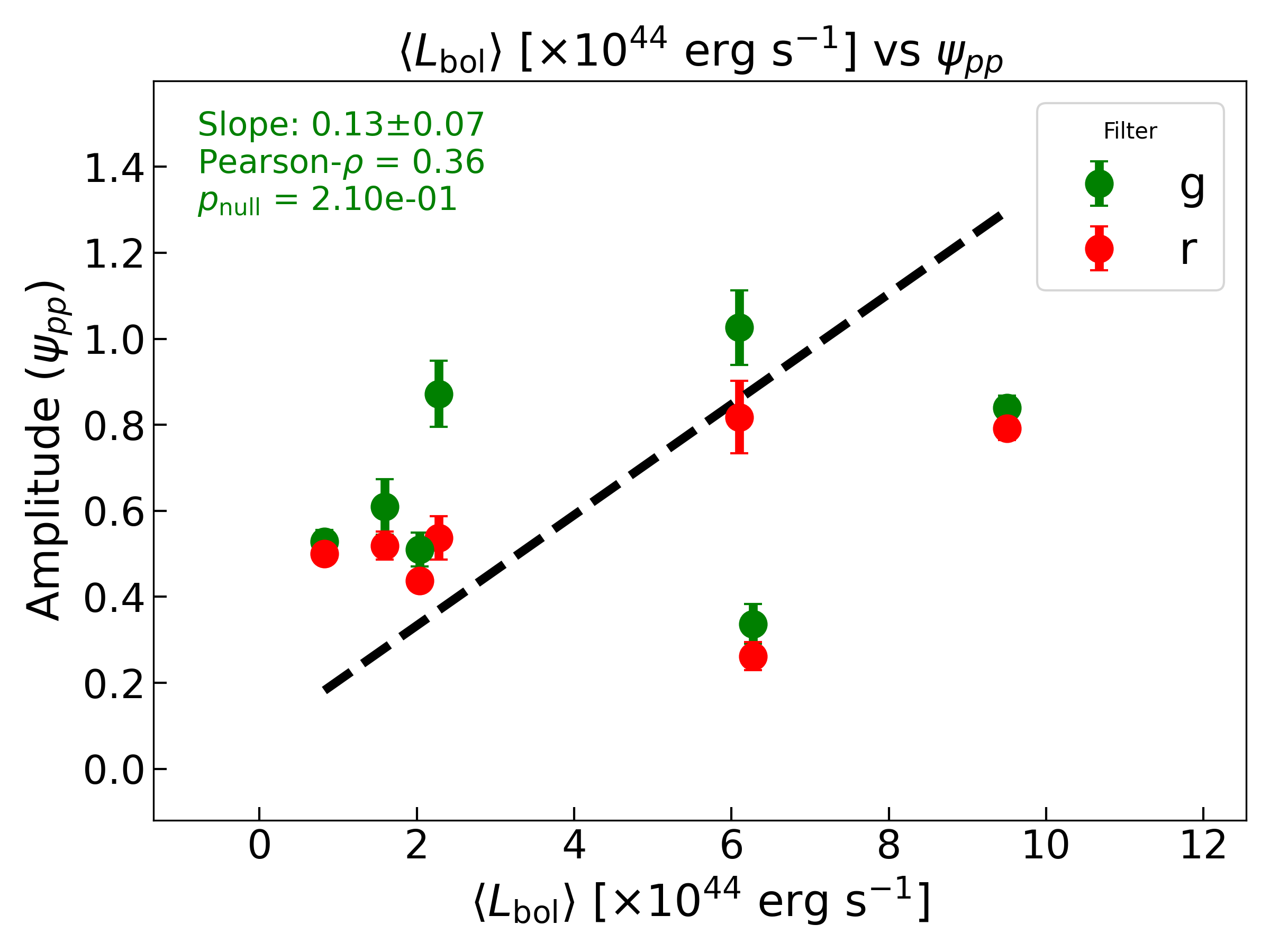} 
\includegraphics[width=0.33\textwidth,height=0.24\textheight,angle=00]{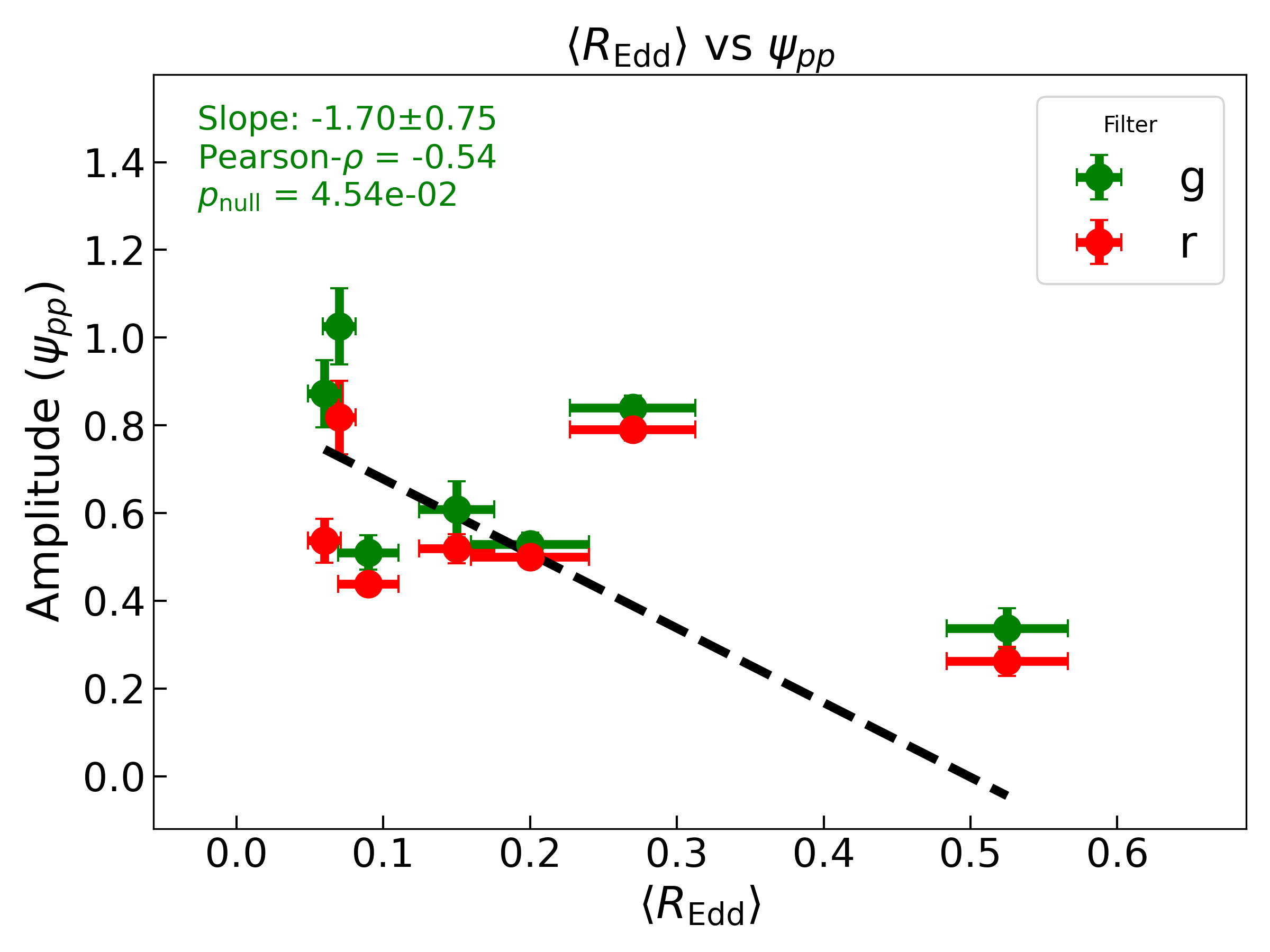}  
   \end{minipage}
    \caption{Figure showing correlation of variability amplitude ($\psi_{pp}$) with physical parameters of AGN, $\mathrm{R_{4570}}$, $\mathrm{R_{5007}}$, $\mathrm{\langle f_{5100\,\text{\AA}} \rangle}$, $\mathrm{M_{BH}}$, $\mathrm{\langle L_{bol} \rangle}$, and $\mathrm{\langle R_{Edd} \rangle}$. The black dashed line represents the best fit obtained using orthogonal distance regression, accounting for uncertainties in both axes parameters. The best-fit linear regression parameters, including the slope and Pearson-$\rho$, are computed including joined data from \emph{ZTF} \emph{g}, and \emph{r} bands of the seven RQ-NLSy1s and are displayed in the upper-left corner of each panel.}
    \label{fig: Correlation_bt_amp_agn_param}
\end{figure*}

\section{Results}
 \label{sec_4.0}

 \subsection{Multi-band optical and mid-infrared variability}
 \label{results_op_mir_var}
 We investigated flux and color variability in a sample of seven RQ-NLSy1s using optical and mid-infrared light curves. The optical analysis is based on high-cadence \emph{ZTF} observations in the \emph{g}, \emph{r}, and \emph{i} bands, complemented by \emph{WISE} \emph{W1} and \emph{W2} band data at the MIR wavelength. The presence of intrinsic optical variability was first assessed using the $F_{\mathrm{AGN}}$-test (Equation~\ref{eq:Ftest_new}), applied independently to the light curves of \emph{g-}, \emph{r-} and \emph{i-} band. All sources show statistically significant variability on long timescales. The variability amplitude was then quantified using $\psi_{\mathrm{pp}}$ and $F_{\mathrm{var}}$, calculated using Equations~\ref{eq:Amp_pp_new} and~\ref{eq:Fvar_new}, respectively. The median $\psi_{\mathrm{pp}}$ values are  \emph{0.609}, \emph{0.519} and \emph{0.438} in the \emph{g}, \emph{r} and \emph{i} bands, while the corresponding median $F_{\mathrm{var}}$ values are  \emph{0.422}, \emph{0.435} and \emph{0.411}  (see  Table~\ref{tab:OP_IR_variability}). For all sources, $\psi_{\mathrm{pp}}$ decreases systematically toward longer wavelengths, with $\psi_{\mathrm{pp}}^{g} > \psi_{\mathrm{pp}}^{r} > \psi_{\mathrm{pp}}^{i}$, except for J152205.41$+$393441.3, where $\psi_{\mathrm{pp}}^{r} < \psi_{\mathrm{pp}}^{i}$. A similar wavelength-dependent trend in $F_{\mathrm{var}}$ is observed for three objects, J102906.69$+$555625.2, J123220.11$+$495721.8, and J150916.18$+$613716.7 (see, Table~\ref{tab:OP_IR_variability}).\par
Given the high cadence of \emph{ZTF} observations in the \emph{g}, \emph{r}, and \emph{i} bands, we searched for flux variability in the seven RQ-NLSy1s on day-like (intra-night), week-like, and month-like timescales. 
The \emph{g-}, \emph{r-} and \emph{i}-band light curves were binned at 2-minute, half-day, and one-day intervals and subdivided into independent sub-epochs spanning half a day, seven days, and 30 days, respectively. This approach yielded multiple sub-epochs for each source on all investigated timescales. To ensure statistical robustness, only sub-epochs containing at least five data points were retained for further analysis; a criterion satisfied by several sub-epochs in each category is tabulated in Table~\ref{var_diff_time}. The variability statistics reported below are based on the complete set of retained sub-epochs, rather than on the selected light curves shown for illustrative purposes. Compared with the \emph{i}-band light curves, this condition is met more frequently in the \emph{g} and \emph{r} bands, likely reflecting the lower observational cadence of the \emph{ZTF} \emph{i}-band data. We, therefore, restrict the short- and intermediate-timescale variability analysis to the \emph{g-} and \emph{r}-band light curves.\par
All retained sub-epochs were subjected to the variability tests described in Sect.~\ref{section_3.1}, and the number of variable and total sub-epochs was determined for each source and investigated timescales. The corresponding statistics are listed in Table~\ref{var_diff_time}. Using these results, we compute the DC for each source. In most cases, the DC is higher in the \emph{r} band than in the \emph{g} band, except for J164100.10$+$345452.7 on intra-night timescales and J102906.69$+$555625.2 and J123220.11$+$495721.8 on month-like timescales. Despite this, the mean peak-to-peak variability amplitude ($\overline{\psi}_{\mathrm{pp}}$) consistently satisfies $\overline{\psi}_{\mathrm{pp}}^{g} > \overline{\psi}_{\mathrm{pp}}^{r}$ on all examined timescales. DC and $\overline{\psi}_{\mathrm{pp}}$ are summarized for each source in Table~\ref{DC_diff_time}.

Unlike optical analysis, MIR variability could be examined for only four of the seven sources, as these sources retain reliable \emph{WISE} \emph{W1} and \emph{W2} band measurements after applying the selection criteria described in Sect.~\ref{MIR_data}. The presence of intrinsic MIR variability was assessed by computing the redshift-corrected rest-frame intrinsic variability amplitude ($V_{\mathrm{mz}}$) using Equation~\ref{eq:mir_variability}. Of the four sources, three satisfy our adopted variability criterion of $V_{\mathrm{mz}} \gtrsim$ 0.1 within the associated uncertainties in both \emph{W1} and \emph{W2} bands, indicating statistically significant MIR variability. The corresponding values of $V_{\mathrm{mz}}$ for individual sources are reported in Table~\ref{tab:OP_IR_variability}.

\subsection{Optical and mid-infrared color variability}
\label{result_color_variability}
As discussed in Sect.~\ref{Color_variability}, the inferred color–magnitude trend can reverse when the shorter-wavelength (bluer) magnitude is placed on the horizontal axis therefore, following~\citet{Ojha2025ApJ...994...84O}, we base our analysis on the color–magnitude diagrams $(m_1 - m_2)$ versus $(m_1 + m_2)$, where this effect is properly mitigated. Based on the optical color–magnitude diagrams, a statistically significant BWB trend is detected in three of the seven RQ-NLSy1s, namely J102906.69$+$555625.2, J123220.11$+$495721.8, and J150916.18$+$613716.7 (see Fig.~\ref{fig: OP_color_variability_gr_1_RQ-NLSy1}). The remaining four sources do not show significant correlations and are therefore classified as NOT. Consistent results are obtained when the analysis is repeated using \emph{ZTF} \emph{r} and \emph{i} band data (see Fig.~\ref{fig: OP_color_variability_ri_1_RQ-NLSy1}). We applied the same methodology to investigate the color variability at MIR wavelengths. Among the four sources with reliable \emph{WISE} data, only J150916.18$+$613716.7 exhibits a statistically significant RWB trend, while the remaining sources do not (see Fig.~\ref{fig: IR_color_variability_w1w2_RQ-NLSy1}).

\subsection{Intra- and inter-band lag measurements}
Motivated by the pronounced continuum variability and color behavior discussed in Sects.~\ref{results_op_mir_var}~\&~\ref{result_color_variability}, we used the high-cadence \emph{ZTF} observations to investigate wavelength-dependent time delays within the optical regime. The ICCF method, described in Sect.~\ref{Lag_measurement}, was applied to each RQ-NLSy1 in the sample, adopting the \emph{g} band as the reference continuum. Optical intra-band lags were measured by cross-correlating the \emph{r-} and \emph{i-}band light curves with the \emph{g-}band light curve. The ICCF results for J123220.11$+$495721.8 are shown in Fig.~\ref{fig: Lag_optical_bands_J1232}, while those for the remaining six sources are presented in Figs.~\ref{fig: Lag_optical_bands_A1} and \ref{fig: Lag_optical_bands_A2}. Adopting a conservative threshold of the maximum correlation coefficient ($R_{\rm max}$) $> 0.6$ to identify reliable correlations, we find no statistically significant optical intra-band lags within the associated uncertainties for any of the targets except for J123220.11$+$495721.8, for which a systematic increase in the intra-band lags towards longer wavelength was found (see Fig.~\ref{fig: Lag_optical_bands_J1232}).\par
On the other hand, we performed an ICCF analysis for the four sources with available MIR coverage to measure the time lags between the optical \emph{r} band and the MIR \emph{W1} and \emph{W2} bands, as well as between the MIR \emph{W1} and \emph{W2} bands themselves. Among these four sources, three exhibit strong correlations with $r_{\rm max} > 0.6$. The ICCF results for J123220.11$+$495721.8 are presented in Fig.~\ref{fig: Lag_optical_MIR_bands_J1232}, while those for the remaining two sources are shown in Fig.~\ref{fig: Lag_optical_bands_A3}. Throughout this work, a positive lag indicates that variability in the shorter-wavelength band leads to the longer-wavelength band. In the following, we describe the results of the continuum optical to mid-infrared inter-band and MIR intra-band lag measurements for these three RQ-NLSy1s.\par

{\textbf{RQ-NLSy1 J123220.11$+$495721.8:}} For this source, the ICCF analysis yields highly significant centroid inter-band lags between the optical and MIR light curves. The \emph{ZTF} \emph{r-}band variations lead those in the \emph{WISE} bands by $\tau_{\rm centroid}^{r\text{–}W1} = 188.6^{+75.2}_{-48.3}$ days and $\tau_{\rm centroid}^{r\text{–}W2} = 266.4^{+91.1}_{-79.5}$ days, with corresponding  $R_{\rm max} = 0.85$ and $0.77$, respectively, indicating robust optical–MIR correlations. A strong MIR intra-band correlation is also detected between the \emph{WISE} \emph{W1} and \emph{W2} light curves, with a centroid lag of $\tau_{\rm centroid}^{W1\text{–}W2} = 29.4^{+81.3}_{-75.2}$ days and $R{\rm max} = 0.98$, representing the most significant MIR intra-band lag among the three sources with reliable measurements. The ICCF lag analysis for this source is shown in Fig.~\ref{fig: Lag_optical_MIR_bands_J1232}. Using these measured lags, we estimate that the upper limit of the characteristic distance of the MIR-emitting region from the accretion disk is 4.87$\times$10$^{17}$ cm and 6.89$\times$10$^{17}$ cm, and that the characteristic size of the emitting region at MIR wavelengths is 7.51$\times$10$^{16}$ cm following the relation, $R \sim c\tau_{\rm centroid}$. 
\par

{\textbf{RQ-NLSy1 J150916.18$+$613716.7:}} For this source, the ICCF analysis reveals correlated optical-mid-infrared variability, although the inferred centroid lags are weakly constrained due to large uncertainties. The \emph{ZTF} \emph{r-}band light curve leads the \emph{WISE} \emph{W1} and \emph{W2} light curves by $\tau_{\rm centroid}^{r\text{–}W1} = 69.3^{+239.4}_{-114.0}$ days and $\tau_{\rm centroid}^{r\text{–}W2} = 220.1^{+260.6}_{-283.5}$ days, with $R_{\rm max} = 0.84$ and $0.73$, respectively. While the correlation coefficients indicate statistically meaningful correlations, the large uncertainties imply that the inter-band delays are not well constrained. The MIR intra-band analysis between the \emph{WISE} \emph{W1} and \emph{W2} light curves yields a centroid lag of $\tau_{\rm centroid}^{W1\text{–}W2} = -9.6^{+350.0}_{-356.7}$ days with $R_{\rm max} = 0.66$. This lag is consistent with near-simultaneous variability between the two MIR bands within the uncertainties. The ICCF results for this source are presented in Fig.~\ref{fig: Lag_optical_bands_A3}. Using these measured lags, we estimate that the upper limit of characteristic distance of the MIR emitting region from the accretion disk is 1.79$\times$10$^{17}$ cm and 5.70$\times$10$^{17}$ cm, and that the characteristic size of the emitting region at MIR wavelengths is 2.49$\times$10$^{16}$ cm.\par

{\textbf{RQ-NLSy1 J164100.10$+$345452.7:}} For J164100.10$+$345452.7, the ICCF analysis indicates the presence of optical-to-mid-infrared time delays, although the correlations between the optical and MIR bands are comparatively weak. The \emph{ZTF} \emph{r-}band variations lead those in the \emph{WISE} bands by $\tau_{\rm centroid}^{r\text{–}W1} = 289.1^{+163.2}_{-83.7}$ days and $\tau_{\rm centroid}^{r\text{–}W2} = 330.8^{+146.8}_{-116.0}$ days, with $R_{\rm max} = 0.70$ and $0.60$, respectively. These values suggest delayed MIR variability relative to the optical emission, but with comparatively marginal correlation strengths. In contrast, a highly significant MIR intra-band lag is detected between the \emph{WISE} \emph{W1} and \emph{W2} light curves, with a centroid lag of $\tau_{\rm centroid}^{W1\text{–}W2} = 20^{+61.2}_{-40.7}$ days with a strong $R_{\rm max} = 0.96$. The ICCF lag analysis for this source is shown in Fig.~\ref{fig: Lag_optical_bands_A3}. Using these measured lags, we estimate that the upper limit of characteristic distance of the MIR emitting region from the accretion disk is 7.49$\times$10$^{17}$ cm and 8.55$\times$10$^{17}$ cm, and that the characteristic size of the emitting region at MIR wavelengths is 5.18$\times$10$^{16}$ cm.\par

\textcolor{black}{In all three sources, the characteristic size of the MIR-emitting region, inferred from the lag between the $W1$ and $W2$ bands, is of the order of $10^{16}$ cm, whereas its distance from the central engine is estimated to be of the order of $10^{17}$-$10^{18}$ cm ($\approx0.06$-$0.28$ pc). This scale is larger than the BLR radius adopted in the SED modelling, $R_{\rm BLR}\sim10^{17}$ cm ($\sim0.03$ pc), indicating that the MIR-emitting region is located outside the BLR. Such a location is naturally associated with thermal emission from the dusty torus.}

\subsection{Dependence of long-term variability amplitude on AGN parameters}
\label{results_corr_amp_agn_param}
To examine the dependence of $\psi_{\rm pp}$ on fundamental AGN properties, we compiled the spectroscopic and physical parameters of the seven RQ-NLSy1s from~\citet{Crepaldi2025A&A...696A..74C}. The adopted parameters include the Fe~{\sc ii} to H$\beta$ flux ratio ($R_{4570}$), the [O~{\sc iii}] to H$\beta$ flux ratio ($R_{5007}$), the average monochromatic flux density at 5100\AA\ ($\langle f_{5100,\text{\AA}} \rangle$), the black hole mass ($M_{\rm BH}$), the average bolometric luminosity ($\langle L_{\rm bol} \rangle$) and the average Eddington ratio ($\langle R_{\rm Edd} \rangle$). The underlying spectral analysis was carried out using Gran Telescopio Canarias spectra, as described in~\citet{Crepaldi2025A&A...696A..74C}.

In this work, we adopt $\langle f_{5100\text{\AA}} \rangle$ as the average of the flux densities measured from the Gran Telescopio Canarias and SDSS spectra, $\langle L_{\rm bol} \rangle$ as the average of the measured and derived bolometric luminosities, and $\langle R_{\rm Edd} \rangle$ as the average of the measured and derived Eddington ratios. The average black hole masses reported in~\citet{Crepaldi2025A&A...696A..74C} are used throughout. Since one source (J102906.69$+$555625.2) lacks \emph{ZTF} \emph{i-}band coverage, and because the \emph{i-}band light curves have substantially lower cadence than those in the \emph{g} and \emph{r} bands, we restrict this analysis to the \emph{g-} and \emph{r-}band variability amplitudes to maintain uniformity and statistical reliability.

The correlations between $\psi_{\rm pp}$ and the AGN parameters were quantified using the Pearson correlation coefficient, and the results are shown in Fig.~\ref{fig: Correlation_bt_amp_agn_param}. We find strong anti-correlations of $\psi_{\rm pp}$ with $R_{4570}$ (Pearson-$\rho$ $= -0.81$), $R_{5007}$ (Pearson-$\rho$ $= -0.74$), and $\langle R_{\rm Edd} \rangle$ (Pearson-$\rho$ $= -0.54$). A moderate negative correlation is also observed between $\psi_{\rm pp}$ and $\langle f_{5100,\text{\AA}} \rangle$, with Pearson-$\rho$ $= -0.46$. In contrast, a strong positive correlation is found between $\psi_{\rm pp}$ and $M_{\rm BH}$, with Pearson-$\rho$ $= +0.76$. A weak to mild positive correlation is present between $\psi_{\rm pp}$ and $\langle L_{\rm bol} \rangle$, with Pearson-$\rho$ $= +0.36$.

\section{Discussion}
\label{sec_5.0}

In this article, we investigate the flux and color variability properties of seven peculiar NLSy1 galaxies that exhibit extreme radio characteristics. Our analysis combines a detailed study of optical variability across multiple timescales with a long-term examination of flux and color variability in both the optical and mid-infrared wavelength regimes.

\subsection{Flux variability}
Table~\ref{tab:OP_IR_variability} shows that all seven RQ-NLSy1s in our sample exhibit statistically significant variability on long timescales. Such variability may arise from instabilities in the accretion disk~\citep[e.g.,][]{Wiita1991sepa.conf..557W, Chakrabarti1993ApJ...411..602C, Mangalam1993ApJ...406..420M} or from fluctuations associated with relativistic jets~\citep[e.g.,][]{Marscher1985ApJ...298..114M, Wagner1995ARA&A..33..163W, Marscher2014ApJ...780...87M}. To distinguish between these two possibilities, we applied the variability amplitude criterion proposed by~\citet{Bauer2009ApJ...699.1732B}, adopting a threshold of $\psi \geq 0.4$ mag for the \emph{ZTF} $g-$, $r-$ and $i-$band light curves of all seven sources. This threshold was motivated by the study of~\citet{Bauer2009ApJ...699.1732B}, who showed that variability amplitudes exceeding 0.4 mag are commonly observed in blazars and occur more frequently than in normal quasars. However, such large-amplitude variability alone does not uniquely establish the presence of a relativistic jet, as substantial optical variability can also arise from accretion-disk processes. Therefore, we use this criterion only as an empirical indicator of blazar-like variability and combine it with additional diagnostics. This criterion yields $\psi_{\rm pp}^{g,~r,~i} \geq 0.4$ for five sources, while only four sources also satisfy $F_{\rm var}^{g,~r,i} \geq 0.4$ (see Table~\ref{tab:OP_IR_variability}).  The simultaneous fulfillment of both conditions indicates that these four sources exhibit optical variability amplitudes comparable to those commonly seen in jet-dominated AGNs and are therefore plausible candidates for a significant jet contribution to their optical emission. To further test this interpretation, we examined the wavelength dependence of the variability amplitude for these four objects. We find that three sources, J102906.69$+$555625.2, J123220.11$+$495721.8, and J150916.18$+$613716.7, show a systematic decrease in variability amplitude with increasing wavelength, such that $\psi_{\rm pp}^{g} > \psi_{\rm pp}^{r} > \psi_{\rm pp}^{i}$ and $F_{\rm var}^{g} > F_{\rm var}^{r} > F_{\rm var}^{i}$ (see Table~\ref{tab:OP_IR_variability}). This behavior is consistent with expectations of the shock-in-jet model~\citep[see][]{2011MmSAI..82..104T}, in which particles are accelerated to high energies at the shock front. Because higher-energy electrons undergo more rapid radiative cooling than lower-energy ones, variability amplitudes are expected to be higher at higher observing frequencies~\citep{Kirk1998A&A...333..452K, Mastichiadis2002PASA...19..138M}.  Additionally, Table~\ref{tab:OP_IR_variability} shows that both the mean peak-to-peak variability $\overline{\psi_{\rm pp}^{g,~r,~i}}$ and the mean fractional variability $\overline{F_{\rm var}^{g,~r,~i}}$, for three RQ-NLSy1s, J102906.69$+$555625.2, J123220.11$+$495721.8 and J150916.18$+$613716.7, are higher than for the remaining four individual sources. This is consistent with previous findings that objects that host jets tend to display larger amplitudes of optical variability~\citep[][]{Rakshit2017ApJ...842...96R}.  Thus, based on these long-term optical variability properties, we therefore strongly suggest that jet-related processes most plausibly drive the optical variability in three of the seven RQ-NLSy1s in our sample. \par 
 However, we note that a similar wavelength dependence may also arise from intrinsic accretion-disk variability, therefore, while the observed trend supports the possibility of a jet contribution, it should not be regarded as definitive evidence for jet-dominated optical emission. Additionally, Table~\ref{tab:OP_IR_variability} shows that both mean peak-to-peak variability $\overline{\psi_{\rm pp}^{g,~r,~i}}$, and the mean fractional variability $\overline{F_{\rm var}^{g,~r,~i}}$, for three RQ-NLSy1s, J102906.69$+$555625.2, J123220.11$+$495721.8, and J150916.18$+$613716.7, are higher than those of the remaining four sources.  Although larger amplitudes of optical variability are commonly observed in jet-hosting AGNs~\citep[e.g.,][]{Rakshit2017ApJ...842...96R}, we note that enhanced variability may also be associated with lower Eddington ratios, as suggested by the trends shown in Fig.~\ref{fig: Correlation_bt_amp_agn_param}. Therefore, variability amplitude alone cannot be used as definitive evidence for jet activity, and additional diagnostics are required to assess the contribution of jet-related emission.

On the other hand, among the four RQ-NLSy1s with reliable \emph{WISE} \emph{W1} and \emph{W2} band measurements, three sources, J123220.11$+$495721.8, J150916.18$+$613716.7, and J164100.10$+$345452.7, exhibit statistically significant MIR variability.  Two of the three sources identified as jet-dominated candidates from the optical analysis exhibit detectable MIR variability, with amplitudes of 0.139$\pm$0.006 mag (W1) and 0.130$\pm$0.006 mag (W2) for J123220.11$+$495721.8, and 0.072$\pm$0.006 mag (W1) and 0.080$\pm$0.018 mag (W2) for J150916.18$+$613716.7. While MIR variability is observed in both jetted and non-jetted AGN and may arise from thermal dust reprocessing of accretion-disk fluctuations, the presence of MIR variability in these optically selected jet-candidate sources is consistent with an additional non-thermal contribution. However, MIR variability alone cannot be regarded as conclusive evidence for the presence of relativistic jets.\par
To explore optical variability on sub-timescales, we analyzed high-cadence \emph{ZTF} \emph{g-}, \emph{r-}, and \emph{i-}band light curves of the seven RQ-NLSy1s, following the strategy described in Sect.~\ref{results_op_mir_var}. As summarized in Table~\ref{var_diff_time}, three sources, J102906.69$+$555625.2, J123220.11$+$495721.8, and J150916.18$+$613716.7, do not show statistically significant variability on intra-night or week-like timescales in any of the optical bands. The only exception is J150916.18$+$613716.7, which exhibits weak variability on week-like timescales, with duty cycles of $\sim$1\% in the \emph{g-}band and $\sim$5\% in the \emph{r-}band (see Table~\ref{DC_diff_time}). When these results are considered together with the recent intra-night optical variability (INOV) study of the same sample by~\citet{Ojha2024MNRAS.529L.108O}, a more nuanced picture emerges. In that work, J102906.69$+$555625.2 and J123220.11$+$495721.8 were found to exhibit significant INOV, with a DC of $\sim$31\% and $\sim$21\%, respectively. In contrast, INOV was not detected in J150916.18$+$613716.7 during three observing epochs, each lasting more than 3 hours. Taken together, the results of the present analysis and the earlier INOV study support a jet-related origin for the sub-timescale optical variability in J102906.69$+$555625.2, J123220.11$+$495721.8, and J150916.18$+$613716.7, a conclusion that is further strengthened by their long-term color-variability properties (e.g., see below).\par

In contrast, although two other sources in the sample, J152205.41$+$393441.3 and J164100.10$+$345452.7, exhibit significant variability across all investigated timescales in the present work (see Table~\ref{DC_diff_time}), they do not show the systematic wavelength dependence of variability amplitude decrease, namely $\psi_{\rm pp}^{g} > \psi_{\rm pp}^{r} > \psi_{\rm pp}^{i}$ and $F_{\rm var}^{g} > F_{\rm var}^{r} > F_{\rm var}^{i}$ (see Table~\ref{tab:OP_IR_variability}), which would strongly favor a jet-dominated origin. Consequently, the physical drivers of variability in these two RQ-NLSy1s cannot be firmly established based on sub-timescale analysis alone and are therefore further examined through their color-variability behavior, as discussed in the subsequent subsection.

\subsection{Color variability}
To further discriminate between jet-dominated and accretion disk–dominated emission in the seven RQ-NLSy1s, we place particular emphasis on the color variability results presented in Sect.~\ref{result_color_variability}. Following the approach adopted in Sect.~\ref{result_color_variability}, a clear BWB was shown by three sources, J102906.69$+$555625.2, J123220.11$+$495721.8 and J150916.18$+$613716.7 from their diagrams $(g-r)$ versus $(g+r)$, with Pearson-$\rho$ exceeding 0.5 (see Fig.~\ref{fig: OP_color_variability_gr_1_RQ-NLSy1}). A similar strong BWB trend is observed for J123220.11$+$495721.8 and J150916.18$+$613716.7 in the $(r-i)$ versus $(r+i)$ diagrams (see Fig.~\ref{fig: OP_color_variability_ri_1_RQ-NLSy1}). The J102906.69$+$555625.2 could not be tested in this case due to the lack of data from the \emph{ZTF} $i-$band.

Together with the flux variability results, these color trends strongly indicate that, in three out of the seven RQ-NLSy1s, namely J102906.69$+$555625.2, J123220.11$+$495721.8, and J150916.18$+$613716.7, the optical emission is predominantly driven by jet-related processes. In contrast, the MIR color–magnitude analysis reveals a statistically significant trend only for J150916.18$+$613716.7, which shows a pronounced RWB behavior in its diagram $(W1-W2)$ versus $(W1+W2)$ ( see Fig.~\ref{fig: IR_color_variability_w1w2_RQ-NLSy1}). This RWB trend suggests an increased contribution from thermal components, such as dust reprocessing and the accretion disk, at MIR wavelengths.
The presence of strong optical BWB trends in these three RQ-NLSy1s, together with a systematic decrease in variability amplitude towards longer wavelengths, that is, $\psi_{\rm pp}^{g} > \psi_{\rm pp}^{r} > \psi_{\rm pp}^{i}$ and $F_{\rm var}^{g} > F_{\rm var}^{r} > F_{\rm var}^{i}$, is consistent with trends reported for BL Lac objects ~\citep[e.g., see][]{Negi2022MNRAS.510.1791N}.  However, we note that such behavior is not unique to jet-dominated systems and may also arise from accretion-disk variability, as shorter wavelengths are expected to exhibit larger variability amplitudes in a thermally fluctuating disk. Therefore, the observed BWB trend alone cannot unambiguously distinguish between disk- and jet-driven variability.

Nevertheless, several additional observational characteristics suggest that a significant jet contribution is present in these sources. Although no statistically significant intra-night variability was detected in the ZTF light curves, variability on timescales of weeks to months was observed in all three objects, while J150916.17+613716.6 also exhibited variability on week-like timescales during several epochs. More importantly, dedicated high-cadence optical monitoring revealed strong intra-night optical variability in J102906.69+555625.2 during two out of three observing sessions, each lasting more than 4 h, and in J123220.11+495721.8 during two out of five sessions, each lasting at least 3 h~\citep[][]{Ojha2024MNRAS.529L.108O}. The resulting duty cycles are comparable to those observed in blazar populations, where relativistic jets are known to dominate the optical variability. When these variability properties are considered together with the observed BWB behavior and the wavelength-dependent variability amplitudes, a jet-related origin for a substantial fraction of the optical variability appears more plausible than a scenario driven solely by accretion-disk fluctuations. Nevertheless, a contribution from the accretion disk cannot be excluded, and both components may contribute to the observed optical emission.

\subsection{Dependence of variability amplitude on AGN parameters}
To explore how the long-term optical variability amplitude depends on the physical characteristics of AGNs, we examined the correlations between $\psi_{\rm pp}^{g,r}$ and several key AGN parameters. The results of these tests are presented in Sect.~\ref{results_corr_amp_agn_param}. We find a strong anti-correlation between $\psi_{\rm pp}$ and both $R_{4570}$ and $R_{5007}$, in agreement with earlier studies of NLSy1s~\citep[e.g., see][]{Ai2010ApJ...716L..31A,Rakshit2017ApJ...842...96R}. This indicates that sources with stronger Fe{\sc ii} and [O{\sc iii}] emission tend to show lower amplitudes of optical variability.

We also detect a significant positive correlation between $\psi_{\rm pp}$ and black hole mass, $M_{\rm BH}$, consistent with similar trends reported earlier for Quasars~\citep[e.g. see][]{Wold2007MNRAS.375..989W, Wilhite2008MNRAS.383.1232W}. Such a relation can be understood within the framework of accretion disk models in which the mean accretion rate is of the order $\dot{m}_0 \sim 0.1$, with stochastic variations at the level of $0.1$–$0.5~\dot{m}_0$\citep[][]{Li2008MNRAS.387L..41L}. However,~\citet{Ai2010ApJ...716L..31A} noted that this correlation weakens once the dependence on the Eddington ratio is taken into account.

Nevertheless, we find a strong and fundamental anti-correlation between $\psi_{\rm pp}$ and $\langle R_{\rm Edd}\rangle$ for our sample of seven RQ-NLSy1s, which is largely independent of the black hole mass~\citep[see][]{Ai2010ApJ...716L..31A}. This behavior is consistent with previous results for larger samples of NLSy1s~\citep[see][]{Rakshit2017ApJ...842...96R, Rakshit2019MNRAS.483.2362R}. This can be understood within the standard Shakura–Sunyaev accretion disk framework~\citep{Shakura1973A&A....24..337S}, where the characteristic radius of the disk region emitting at a given wavelength $\lambda$ scales as
$R \propto T^{-4/3} \propto (\dot{m}/M_{\rm BH})^{1/3}\lambda^{4/3}$,
where $R$ is in units of the Schwarzschild radius, $T$ is the temperature of the disk, and $\dot{m}$ is the mass accretion rate in Eddington units. At a fixed wavelength, systems with higher Eddington ratios emit predominantly from larger disk radii, whereas lower Eddington ratio sources are dominated by emission from the inner disk. Since inner disk regions are expected to vary more strongly and on shorter timescales, this naturally leads to an anti-correlation between optical variability amplitude and Eddington ratio.

The above correlations indicate that, for the sample as a whole, the observed long-term optical variability is largely consistent with an accretion-disk origin. However, our suggestion of a possible jet contribution in a few individual sources is based on independent variability diagnostics, including strong intra-night optical variability, enhanced MIR variability, and blazar-like optical/MIR colour variability behaviour. These signatures probe variability on different timescales and do not need to dominate the long-term ensemble correlations discussed above. We further emphasize that the presence of a jet is established independently of these variability diagnostics through the ${\rm TS}>25$ GeV detections and the broadband SED modelling, neither of which can be explained by thermal AGN components alone. The observed colour and variability-amplitude trends, which are individually degenerate between disk- and jet-driven scenarios, are therefore used only to assess the relative contribution of the jet to the optical and MIR emission, rather than as primary evidence for the existence of the jet itself. We stress that the independent evidence for the presence of a jet comes from the GeV detections for which we consider a relativistic jet to be the most plausible, though not the only possible, origin.

To examine this further, we repeated the analysis of the dependence of $\psi_{\rm pp}$ on $\langle R_{\rm Edd}\rangle$ using only the three sources (J102906.69+555625.2, J123220.11+495721.8, and J150916.18+613716.7) for which we infer a possible jet contribution. In this case, no statistically significant correlation is found (slope $=0.49\pm0.57$, Pearson coefficient $=0.03$, and $P_{\rm null}=0.95$). Thus, the anti-correlation between $\psi_{\rm pp}$ and $\langle R_{\rm Edd}\rangle$ is primarily driven by the full sample and is not evident among the sources exhibiting the strongest signatures of jet-related activity. Therefore, while the long-term variability properties of the overall sample are well explained within the framework of accretion-disk variability, the presence of a jet contribution in a few individual sources cannot be excluded and is supported by their additional variability characteristics.

 The blazar-like SED parameters (Table~\ref{tab_SED_Modeling_data}) indicate relativistic emitting plasma; by ``weak or intermittent’’, we refer to low jet power and/or duty cycle, as suggested by the absence of these sources from 4FGL-DR4 and their detection only over the full $\sim$17.9 yrs \emph{Fermi}-LAT baseline. This interpretation is consistent with the radio observations, which show no persistent relativistic jets\citep{Berton2020A&A...636A..64B,Jarvela2024MNRAS.532.3069J}. The compact emitting regions ($R\sim10^{15}$cm, $R_H\lesssim10^{18}$cm) are self-absorbed below $\sim10^{11}$Hz and under-predict the archival radio fluxes (Fig.\ref{fig: sed_figures}), suggesting that the 37GHz flares may arise from episodic, short-lived dissipation regions\citep{2018A&A...614L...1L} rather than from a persistent extended jet, as expected for intermittent activity at high Eddington ratios~\citep{Czerny2009ApJ...698..840C}. Because \texttt{JETSET} is a steady-state one-zone model fitted to non-simultaneous data, it represents only a time-averaged state and cannot constrain the duty cycle; time-dependent modelling of simultaneous multi-wavelength data would therefore be required to describe such intermittent activity. Magnetic reconnection near the black hole could also produce faint, episodic $\gamma$-ray emission without a persistent jet~\citep{Kadowaki2015ApJ...802..113K,Kimura2022ApJ...937L..34K,Ojha2024MNRAS.529L.108O}, whereas coronal GeV emission is suppressed by $\gamma\gamma$ absorption~\citep{Inoue2020ApJ...891L..33I} and star formation or AGN-wind shocks are expected to yield $L_\gamma\lesssim10^{42}$~erg~s$^{-1}$~\citep{2012ApJ...755..164A, 2016A&A...596A..68L}, well below the observed $\nu L_\nu\sim10^{44}$–$10^{45}$~erg~s$^{-1}$. A relativistic jet therefore remains the most plausible origin, although a reconnection-powered origin cannot be excluded with the present data. \par

 An alternative to the relativistic-jet interpretation is magnetic reconnection in the magnetosphere of the central black hole, which can operate without requiring a persistent relativistic jet~\citep{Lazarian1999ApJ...517..700L, 2010A&A...518A...5D, Kadowaki2015ApJ...802..113K, Ripperda2022ApJ...924L..32R, Kimura2022ApJ...937L..34K}. In this scenario, magnetic flux emerging from the accretion flow can reconnect with field lines anchored to the black-hole magnetosphere, accelerating particles to relativistic energies and producing non-thermal emission~\citep{Blandford1977MNRAS.179..433B}. Such events can generate synchrotron emission and, through synchrotron self-Compton and/or inverse-Compton processes, potentially extend to the $\gamma$-ray band~\citep{Kadowaki2015ApJ...802..113K}. Turbulence-driven reconnection in efficiently accreting, low-luminosity AGNs has been shown to release $\sim10^{39}$–$10^{43}$~erg~s$^{-1}$ of magnetic power~\citep{2010A&A...518A...5D,Kadowaki2015ApJ...802..113K}, demonstrating that reconnection can produce appreciable non-thermal variability without a persistent jet. It could therefore provide an alternative explanation for the faint, episodic $\gamma$-ray emission detected in our sources after integrating over the full $\sim$17.9 yrs \emph{Fermi}-LAT baseline, particularly in view of the absence of persistent radio jets. Nevertheless, whether such a mechanism can account for the observed $\gamma$-ray luminosities of $\sim10^{44}$–$10^{45}$~erg~s$^{-1}$ remains uncertain. Coordinated radio, X-ray, and $\gamma$-ray observations would therefore be valuable for distinguishing magnetic-reconnection-powered emission from a weak or intermittent relativistic jet~\citep{2010A&A...518A...5D, Kimura2022ApJ...937L..34K}.

\section{Conclusions}
\label{sect_6.0}
 We have presented a comprehensive optical and mid-infrared variability study of seven radio-quiet narrow-line Seyfert 1 galaxies using high-cadence optical data from \emph{ZTF} \emph{g}, \emph{r}, and \emph{i} bands and long-term MIR observations in \emph{WISE} \emph{W1} and \emph{W2} bands. Our main conclusions are summarized below.

\begin{enumerate}
\item All seven RQ-NLSy1s exhibit statistically significant long-term optical variability in the \emph{ZTF} \emph{g}, \emph{r}, and \emph{i} bands. The variability amplitudes show a clear wavelength dependence in most cases, with larger amplitudes at shorter wavelengths. In three sources, J102906.69$+$555625.2, J123220.11$+$495721.8, and J150916.18$+$613716.7, both $\psi_{\rm pp}$ and $F_{\rm var}$ decrease systematically from the \emph{g} band to the \emph{i} band, a behavior often associated with a stronger contribution from non-thermal emission processes.

\item Optical variability on sub-timescales was explored using high-cadence \emph{ZTF} data. Although most sources show weak or absent intra-night and week-like variability in our analysis, the results are broadly consistent with previous INOV studies of the same sample. Although we do not detect significant intra-night variability in most cases, the combination of long-term optical variability characteristics and previously reported INOV detections suggests the possibility of jet-related activity in three RQ-NLSy1s, namely J102906.69$+$555625.2, J123220.11$+$495721.8, and J150916.18$+$613716.7.

\item The mid-infrared variability could be robustly examined for four sources with reliable \emph{WISE} \emph{W1} and \emph{W2} data. Three of these objects, namely J123220.11$+$495721.8, J150916.18$+$613716.7, and J164100.10$+$345452.7, exhibit statistically significant intrinsic MIR variability in both \emph{WISE} \emph{W1} and \emph{W2} bands, suggesting that non-thermal processes contribute to MIR emission, in addition to thermal radiation from the accretion disk and dusty torus.

\item Optical color–magnitude analysis, performed using bias-mitigated $(m_1-m_2)$ versus $(m_1+m_2)$ diagrams, reveals clear bluer-when-brighter trends in three RQ-NLSy1s, namely J102906.69$+$555625.2, J123220.11$+$495721.8 and J150916.18$+$613716.7. These trends are consistent with the presence of a variable non-thermal component and provide additional evidence for possible jet-related activity in these sources. In contrast, MIR color variability is detected only in J150916.18$+$613716.7, which shows a redder-when-brighter trend, likely reflecting an increased contribution from thermal dust emission at longer wavelengths.

\item Cross-correlation analysis using the ICCF method reveals no significant intra-optical inter-band lags, implying nearly simultaneous emission across optical bands. In contrast, significant optical–MIR and MIR intra-band lags are detected in three sources, with MIR emission lag the optical by tens to hundreds of days. These delays are consistent with reprocessing of optical/UV radiation by dust located at parsec-scale distances from the central engine.

\item Based on broadband SED modeling of six of seven RQ-NLSy1s, we find that the observed broadband emission can be reproduced by models that include relativistic jet components too. A targeted full-baseline ($\sim$17.9 yrs) \emph{Fermi}-LAT analysis carried out at the optical positions yields detections with ${\rm TS}>25$ for all six sources. The absence of these sources from 4FGL-DR4 can be attributed to the shorter catalogue baseline and the trials factor associated with a blind all-sky search, and is discussed in Sect.~\ref{LAT_analysis}, where we also describe the limitations of the analysis, in particular that the target positions were fixed at the optical coordinates rather than localised. As GeV emission cannot be produced by the standard thermal AGN components, these detections independently establish the presence of a non-thermal component in all six sources, most plausibly associated with relativistic jets, although a reconnection-powered origin without a persistent jet cannot be excluded. The fitted jet parameters are similar to those of blazars, but the faint GeV emission and absence of persistent radio jets suggest low-power and/or low-duty-cycle jets. The compact, self-absorbed emitting regions in our models are consistent with the lack of persistent radio emission, while the steady-state one-zone \texttt{JETSET} models represent only a time-averaged state of such intermittent activity. However, we emphasise that the GeV detections provide the independent evidence for the presence of a jet; the optical and MIR variability diagnostics are instead used to assess whether the jet may contribute to the emission at these wavelengths. The final SED models contain no dusty-torus or X-ray coronal component, since the available data cannot constrain either (Sect.~\ref{jetset_modelling}); consequently, the X-ray points in Fig.~\ref{fig: sed_figures} are traced by the SSC continuum by construction, and the SSC parameters should be regarded as upper limits on the jet contribution in that band. When combined with the optical and MIR variability characteristics, the SED results suggest that the jet contributes additionally to the optical and MIR variability in J102906.69$+$555625.2, J123220.11$+$495721.8, and J150916.18$+$613716.7.

\item The long-term optical variability amplitude shows strong correlations with the fundamental AGN parameters. In particular, $\psi_{\rm pp}$ anti-correlates with the Eddington ratio and emission-line ratios ($R_{4570}$ and $R_{5007}$), while positively correlating with black hole mass. These trends are consistent with expectations from standard accretion disk models, where higher Eddington ratio systems emit predominantly from larger, less variable disk radii.  Furthermore, the anti-correlation between variability amplitude and Eddington ratio supports an accretion-disk origin for the long-term optical variability of the sample as a whole. However, this correlation is not evident among the three sources showing the strongest indications of jet-related activity, suggesting that additional processes may contribute to their variability.

\item  We also searched for publicly available radio monitoring data for our RQNLS1 sample. Although sparse archival radio measurements are available for a few sources, densely sampled and contemporaneous radio light curves are generally lacking, preventing a meaningful optical-radio correlation analysis. Future coordinated optical and radio monitoring campaigns will be crucial for assessing the contribution of relativistic jets to the observed optical variability.
\end{enumerate}

Overall, our results indicate that the long-term optical variability of the RQ-NLSy1 galaxies is largely consistent with accretion-disk-driven variability. At the same time, a small subset of sources exhibits additional characteristics, including strong optical and MIR variability, variability on shorter timescales, BWB trends, and broadband SED properties, which are suggestive of a non-thermal jet contribution. The evidence for the presence of jets is provided by the GeV detections; the optical and MIR data are used to assess the possible contribution of the jets at these wavelengths, while the X-ray data provide additional constraints on the broadband emission. These findings imply that although accretion-disk processes dominate the variability properties of the population as a whole, weak or intermittently active jets (i.e. relativistic jets with blazar-like emitting regions but low power and/or low duty cycle) may be present in a subset of RQ-NLSy1s and contribute to their observed multi-wavelength variability. Future coordinated multi-wavelength monitoring, particularly including X-ray and radio observations, will be crucial for establishing the prevalence and physical nature of such jet activity in this intriguing class of AGNs.

\begin{acknowledgements}
 The authors sincerely thank the anonymous referee for careful reading of our manuscript and for constructive and insightful comments. The suggestions of the referee have helped us clarify several aspects of the analysis and strengthen the presentation and interpretation of our results. We greatly appreciate the time and effort devoted to reviewing our work. We also acknowledge the support of the National Key R\&D Program of China (2025YFA1614101) and the National Science Foundation of China (12133001). The work of LCH was supported by the National Science Foundation of China (12233001), the National Key R\&D Program of China (2022YFF0503401), and China Manned Space Program (CMS-CSST-2025-A09). H.C. expresses gratitude to the Inter-University Center for Astronomy and Astrophysics (IUCAA), India, for their hospitality under the IUCAA Associate Program. This research utilizes observational data collected with the 48-inch Samuel Oschin Telescope, as part of the \emph{Zwicky Transient Facility (ZTF)} initiative. The National Science Foundation funds the \emph{ZTF} project through Grant No. AST-2034437 is a collaborative effort involving institutions such as Caltech, IPAC, the Weizmann Institute of Science, the Oskar Klein Center at Stockholm University, the University of Maryland, DESY and Humboldt University, the TANGO Consortium in Taiwan, the University of Wisconsin–Milwaukee, Trinity College Dublin, Lawrence Livermore National Laboratory, and IN2P3 in France. The operational management of \emph{ZTF} is handled by Caltech Optical Observatories (COO), IPAC, and the University of Washington. This study also incorporates data from the \emph{Wide Field Infrared Survey Explorer} (\emph{WISE}), a collaborative mission between the University of California, Los Angeles, and NASA’s Jet Propulsion Laboratory at Caltech, supported by the National Aeronautics and Space Administration.
\end{acknowledgements}

\bibliographystyle{aa}
\bibliography{ref}

\begin{appendix}
\section{Log of the \emph{Swift}-XRT observations}
\label{app-xrt}

\begin{table*}
\centering
\caption{Log of the \emph{Swift}-XRT observations of the seven RQ-NLSy1s, together with the results of the default spectral fit returned by the UKSSDC products generator~\citep{Evans2007A&A...469..379E, 2009MNRAS.397.1177E}. $N_{\rm obs}$ is the number of observations stacked into the time-averaged spectrum, and $T_{\rm exp}$ the corresponding total on-source exposure. $\Gamma_{\rm X}$ is the photon index of an absorbed power law with the column density fixed to the Galactic value along the line of sight, and $F_{0.3-10}$ the observed 0.3--10~keV flux. Only the unfolded spectral points, not these fitted parameters, are used in the broadband SED modelling (Sect.~\ref{Xray_data}). Acronyms: UC - Unconstrained; ND - No data in Swift.}
\label{tab_xrt_log}
\resizebox{\textwidth}{!}{
\begin{tabular}{lcccccc} \hline
Object & ObsID(s) & Obs. date(s) & $N_{\rm obs}$ & $T_{\rm exp}$ & $\Gamma_{\rm X}$ & $F_{0.3-10}$ \\
       &          & (yyyy-mm-dd) &               & (ks)          &                  & ($10^{-13}$ erg~cm$^{-2}$ s$^{-1}$) \\ \hline
J102906.69$+$555625.2 & 00016005001 & 2023-05-04 & 1 & 4.5 & 2.2$^{+0.5}_{-0.4}$ & 3.07$^{+1.03}_{-0.85}$\\ \\
J122844.81$+$501751.2 & 00016062001 & 2023-05-30 & 13 & 35.8 & 2.31$^{+0.19}_{-0.16}$ & 7.4$^{+0.7}_{-0.6}$ \\
& 00016062002 & 2023-06-02 & & & &  \\
& 00016062003 & 2023-06-03 & & & &  \\
& 00016062005 & 2023-06-09 & & & &  \\
& 00016062012 & 2023-07-01 & & & &  \\
& 00016062011 & 2023-06-28 & & & &  \\
& 00016062008 & 2023-06-15 & & & &  \\
& 00016062013 & 2023-06-08 & & & &  \\
& 00016062007 & 2023-06-13 & & & &  \\
& 00016062006 & 2023-06-11 & & & &  \\
& 00016062010 & 2023-06-25 & & & &  \\
& 00016062009 & 2023-06-18 & & & &  \\ \\
J123220.11$+$495721.8 & 00045962009 & 2023-09-10 & 9 & 9.3 & 2.7$^{+0.5}_{-0.4}$ & 6.5$^{+1.4}_{-1.3}$ \\
& 00045962008 & 2023-09-09 & & & &  \\
& 00045962006 & 2023-09-02 & & & &  \\
& 00045962005 & 2023-08-29 & & & &  \\
& 00045962007 & 2023-09-04 & & & &  \\
& 00045962002 & 2018-01-28 & & & &  \\
& 00045962004 & 2021-09-10 & & & &  \\
& 00045962003 & 2021-09-09 & & & &  \\
& 00045962001 & 2013-05-30 & & & &  \\ \\
J150916.18$+$613716.7 & ND & ND & ND & ND & ND & ND \\ \\
J151020.06$+$554722.0 & 00046517001 & 2012-07-07 & 6 & 5.2 & UC & UC \\
& 00046517002 & 2013-04-12 & & & &  \\
& 00046517005 & 2021-05-18 & & & &  \\
& 00046517007 & 2021-05-26 & & & &  \\
& 00046517006 & 2021-05-22 & & & &  \\
& 00046517003 & 2021-05-17 & & & &  \\ \\
J152205.41$+$393441.3 & 00016006001 & 2023-05-04 & 1 & 4.8 & UC & UC \\ \\
J164100.10$+$345452.7 & 00011395096 & 2022-03-27 & 91 & 212.9 & 1.68$^{+0.09}_{-0.08}$ & 1.15$^{+0.05}_{-0.04}$ \\
& .... & .... & & & &  \\ 
& 00011395027 & 2020-05-11 & & & &  \\ 
& .... & .... & & & &  \\ 
& 00011395002 & 2019-08-29 & & & &  \\
& 00011395001 & 2019-05-25 & & & &  \\ \hline
\end{tabular}
}
\end{table*}

\section{Light curves and lag measurements}

\begin{figure*}
\begin{minipage}[]{1.0\textwidth}
\includegraphics[width=1.0\textwidth,height=0.30\textheight,angle=00]{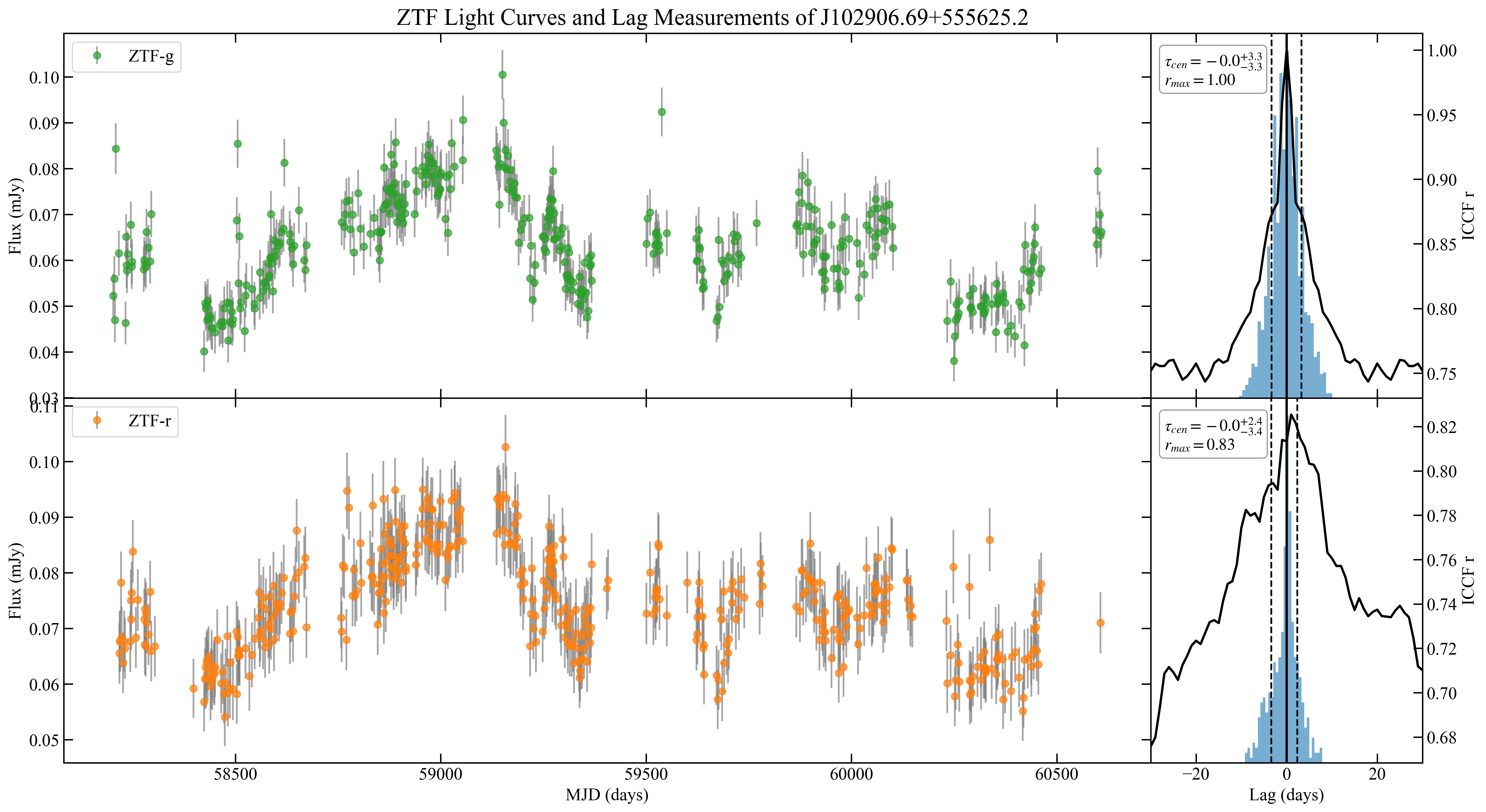}  
\includegraphics[width=1.0\textwidth,height=0.30\textheight,angle=00]{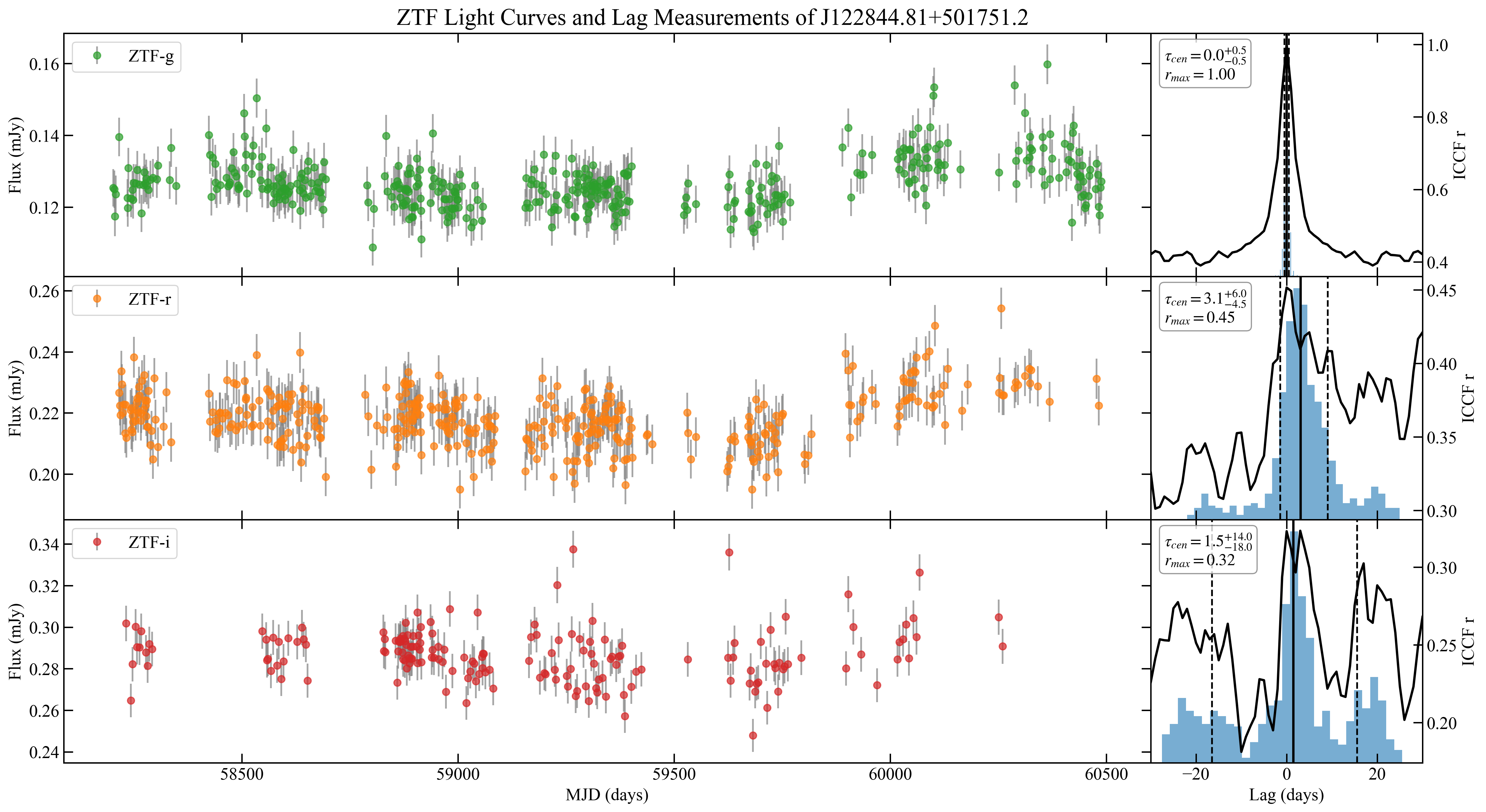}  
\includegraphics[width=1.0\textwidth,height=0.30\textheight,angle=00]{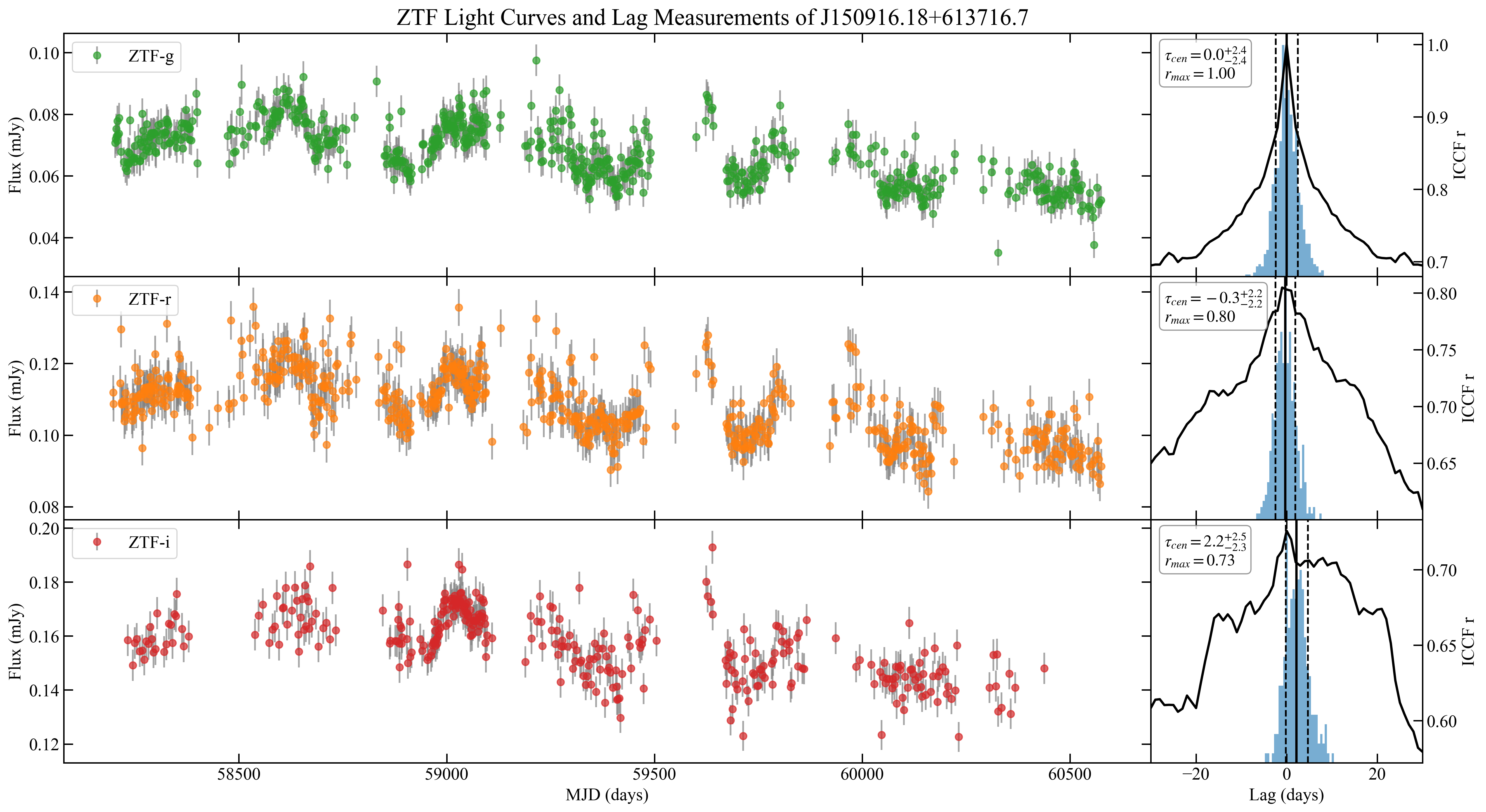}  
    \end{minipage}
    \caption{same as Fig.~\ref{fig: Lag_optical_bands_J1232} but for the source J102906.69$+$555625.2, J122844.81$+$501751.2, and J151020.06$+$554722.0.}
    \label{fig: Lag_optical_bands_A1}
\end{figure*}

\begin{figure*}
\begin{minipage}[]{1.0\textwidth}
\includegraphics[width=1.0\textwidth,height=0.32\textheight,angle=00]{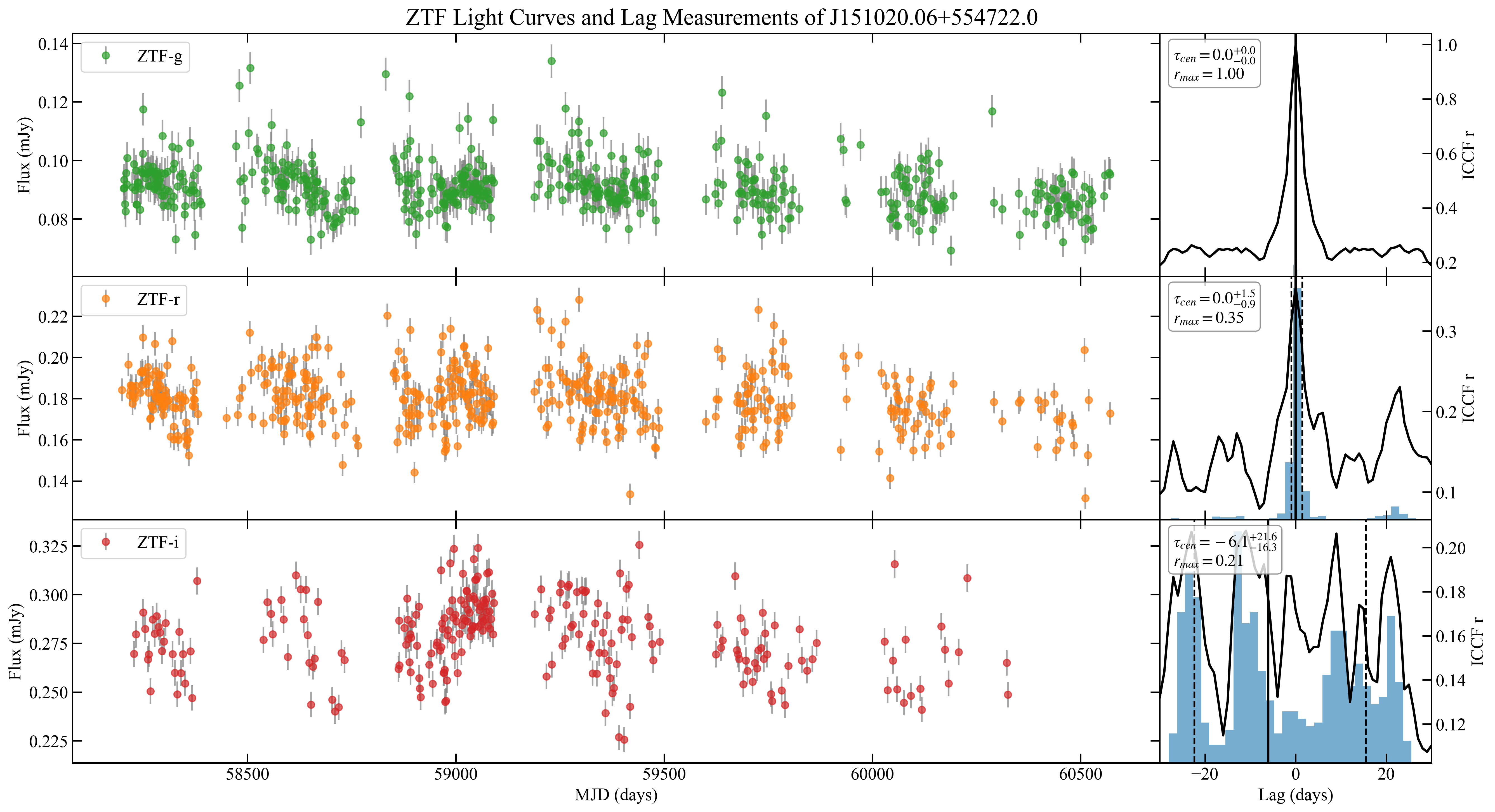}  
\includegraphics[width=1.0\textwidth,height=0.32\textheight,angle=00]{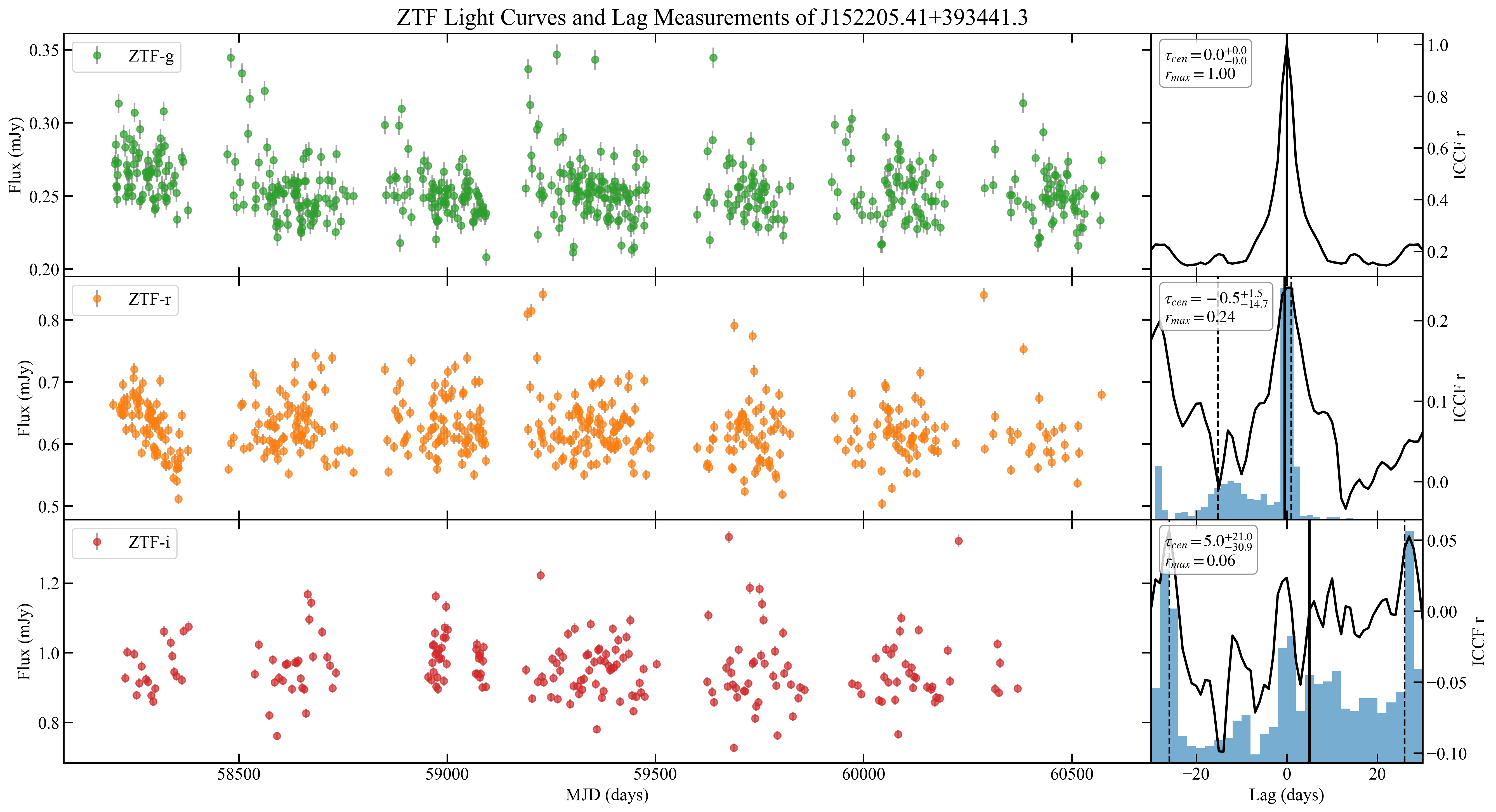} 
\includegraphics[width=1.0\textwidth,height=0.32\textheight,angle=00]{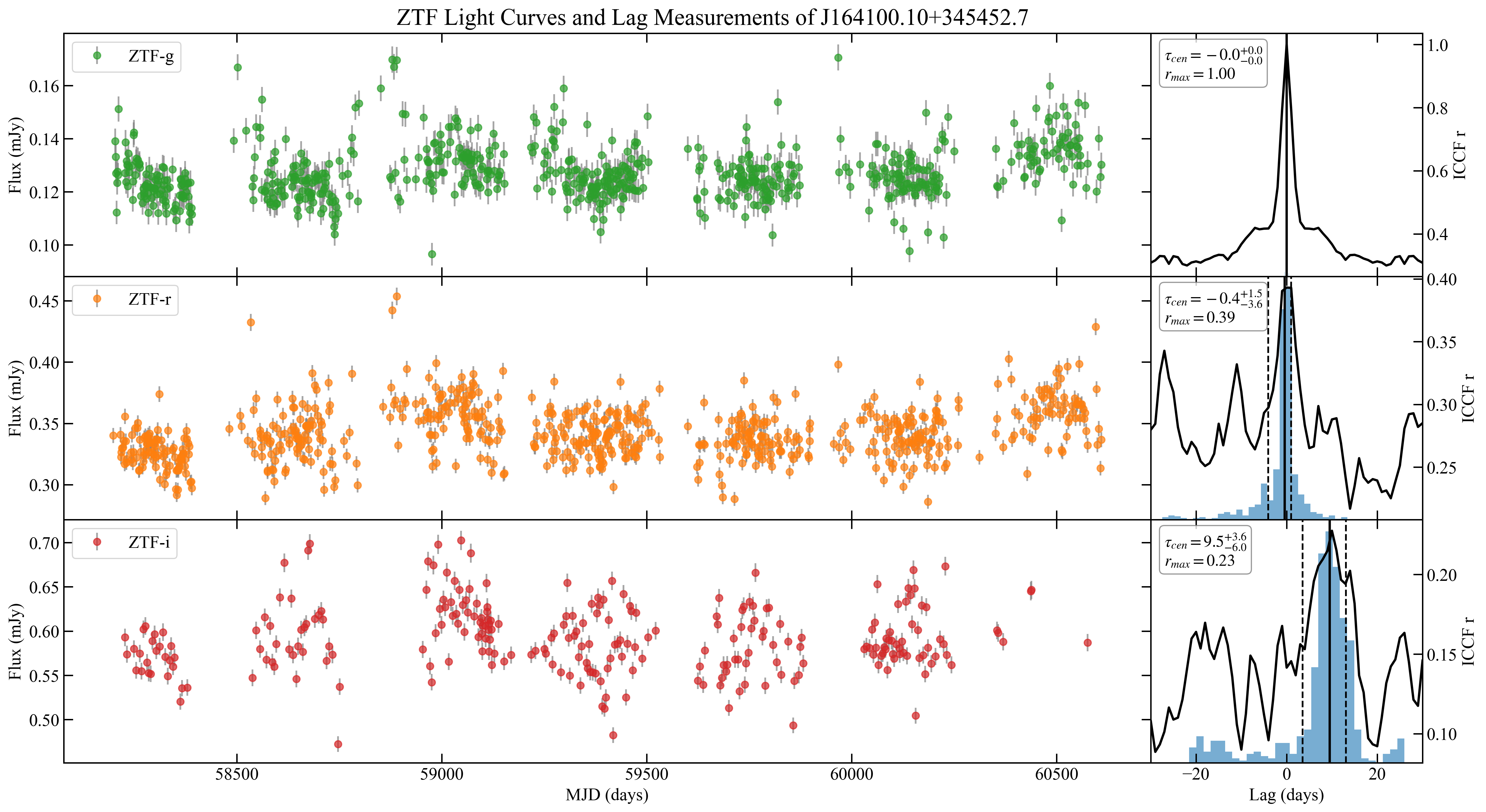}  
\end{minipage}
 \caption{same as Fig.~\ref{fig: Lag_optical_bands_J1232} but for the sources J151020.06$+$554722.0, J152205.41$+$393441.3, and J164100.10$+$345452.7.}
\label{fig: Lag_optical_bands_A2}
\end{figure*}

\begin{figure*}
\begin{minipage}[]{1.0\textwidth}
\includegraphics[width=1.0\textwidth,height=0.48\textheight,angle=00]{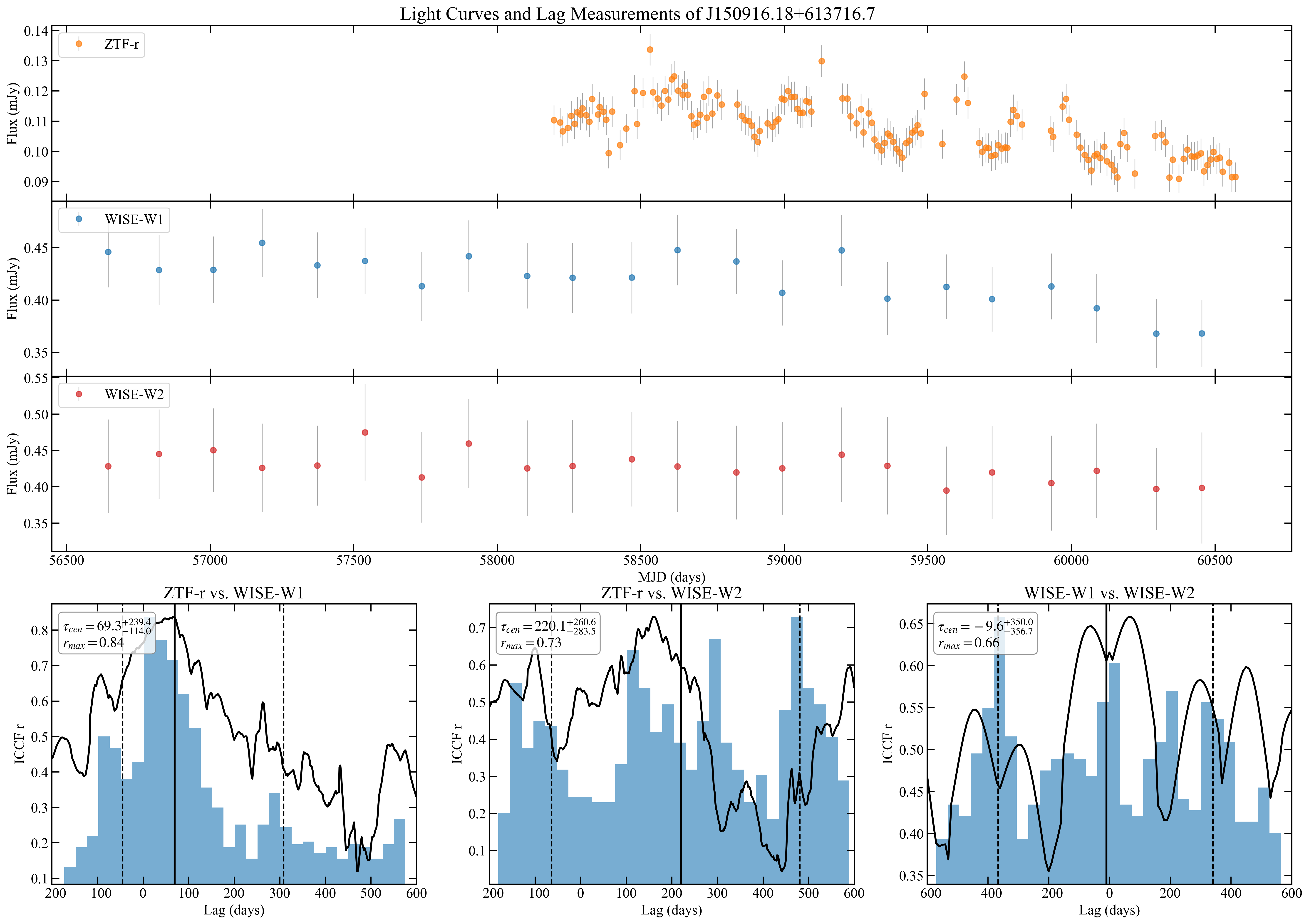}  
\includegraphics[width=1.0\textwidth,height=0.48\textheight,angle=00]{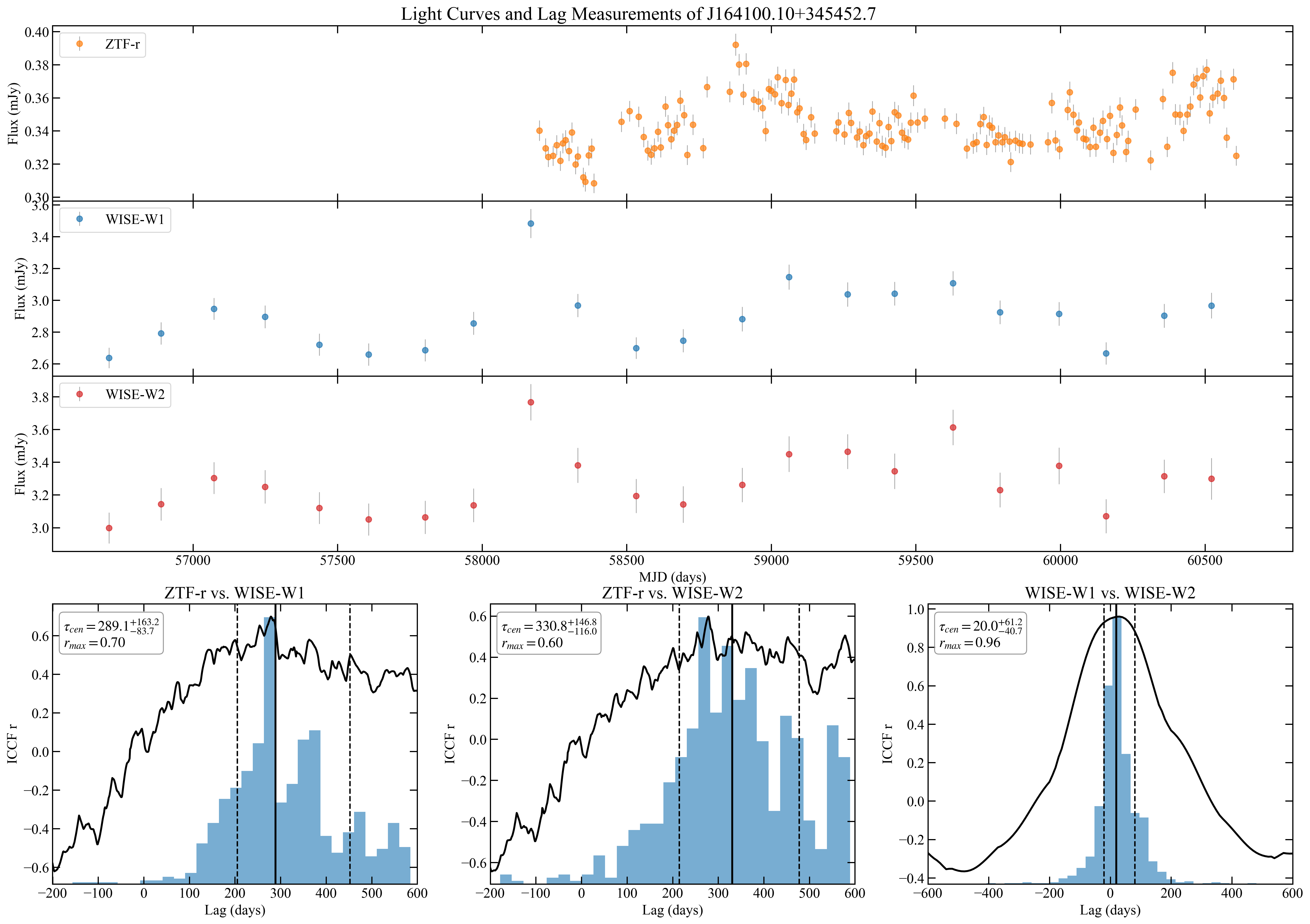} 
\end{minipage}
 \caption{Same as Fig.~\ref{fig: Lag_optical_MIR_bands_J1232} but for the sources J150916.18+613716.7 and J164100.10+345452.7.}
\label{fig: Lag_optical_bands_A3}
\end{figure*} 

\end{appendix}
\end{document}